\documentclass[11pt,a4paper]{article}
\pdfoutput=1
\usepackage{graphicx,caption,array}
\usepackage{ifpdf}
\usepackage{xspace}
\usepackage{jheppub}
\usepackage{dsfont}
\usepackage{amsfonts}
\usepackage{amsmath}
\usepackage{euscript}
\usepackage{amssymb}
\usepackage{bbold}
\usepackage{xurl}
\usepackage{hyperref}
\usepackage{axodraw2}
\usepackage{multirow}
\usepackage{pstricks}
\usepackage{relsize,exscale,scalefnt,anyfontsize,cases}
\usepackage[utf8]{inputenc}
\usepackage{color}

\def\bc{\begin{center}}
\def\ec{\end{center}}
\def\bi{\begin{itemize}}
\def\ei{\end{itemize}}
\def\be{\begin{equation}}
\def\ee{\end{equation}}
\def\bea{\begin{eqnarray}}
\def\eea{\end{eqnarray}}
\def\bt{\begin{tabular}}
\def\et{\end{tabular}}
\def\ba{\begin{array}}
\def\ea{\end{array}}

\newcommand{\nn}{\nonumber} 
\newcommand{\ie}{{i.e.}}  
\newcommand{\eg}{{e.g.}}
\newcommand{\diag}{{\rm diag}}

\def\wv#1{\widetilde{#1}} 
\newcommand{\lqcd}{\Lambda_{_{\rm QCD}}}
\def\xtwo{x_{_2}}
\def\sun{{\rm SU}(\Nc)}

\newcommand{\Nc}{\ensuremath{N}\xspace}
\newcommand{\CF}{C_F}
\newcommand{\xibar}{{\bar{\xi}}} 
\newcommand{\np}{n} 
\newcommand{\ns}{{\tilde n}} 

\newcommand{\morder}[1]{{\cal O}\left(#1 \right)}
\newcommand{\eq}[1]{(\ref{#1})}

\newcommand{\ave}[1]{\langle{#1}\rangle}

\newcommand{\tf}{t_{\mathrm{f}}}

\newcommand{\al}{\alpha}

\newcommand{\dd}{{\rm d}}

\newcommand{\lsim}{\lesssim} 
\newcommand{\gsim}{\gtrsim}

\def\bm#1{\mbox{\boldmath$#1$}}
\newcommand{\Tr}{\mathrm{Tr}\,}
\newcommand{\tr}{\mathrm{tr}\,}
\newcommand{\trc}{\tr_{_{\! \! \rm c}}}
\newcommand{\trd}{\tr_{_{\! \! \rm d}}}

\newcommand{\T}{\Theta}
\newcommand{\B}{\mathds{B}}
\newcommand{\Bbar}{\overline{\mathds{B}}}

\newcommand{\ADM}{{\cal Q}}

\newcommand{\ket}[1]{\vert{#1}\rangle}
\newcommand{\bra}[1]{\langle{#1}\vert}

\newcommand{\proj}{{\mathds{P}}} 
\newcommand{\trans}{{\mathds{T}}} 
\newcommand{\mathsize}[2]{\mbox{\fontsize{#1}{2}\selectfont $#2$}}
\newcommand{\Proj}[1]{{\mathds{P}}_{\mathsize{7}{#1}}} 
\newcommand{\Trans}[1]{{\mathds{T}}_{\mathsize{7}{#1}}} 
\newcommand{\Rho}[1]{\rho_{\mathsize{7}{#1}}}
\newcommand{\Cas}[1]{C_{\mathsize{7}{#1}}}

\def \drawqua (#1) {
\draw[postaction={decorate},decoration={markings,mark=at position #1 with {\large\arrow{>}}}] 
}
\def \drawanti (#1) {
\draw[postaction={decorate},decoration={markings,mark=at position #1 with {\large\arrow{<}}}] 
}

\newcommand{\tikeq}[1]{\vcenter{\hbox{\begin{tikzpicture}#1\end{tikzpicture}}}}
\newcommand{\tikeqbis}[1]{\hbox{\begin{tikzpicture}#1\end{tikzpicture}}}

\newcommand{\drawdash}{\draw[dash pattern=on 1pt off 1.3pt, line width=1.2]}

\def \gggqloopCCW {\tikeqbis{\begin{scope}[scale=0.8,yshift=3mm]
\draw[anti] (0.2,0) arc(180:0:.3); \draw (0.2,0) arc(-180:0:.3); 
\draw[glu] (0.5,-0.6) -- (0.5,-0.3); \draw[glu] (-0.2,0) -- (0.2,0);  \draw[glu] (0.8,0) -- (1.2,0); 
\draw (-0.2,0) node[left] {$c$}; \draw (1.2,0) node[right] {$b$}; \draw (0.5,-0.6) node[below] {$a$}; 
\end{scope}}}

\def \gggqloopCW  {\tikeqbis{\begin{scope}[scale=0.8,yshift=3mm]
\draw[qua] (0.2,0) arc(180:0:.3); \draw (0.2,0) arc(-180:0:.3); 
\draw[glu] (0.5,-0.6) -- (0.5,-0.3); \draw[glu] (-0.2,0) -- (0.2,0);  \draw[glu] (0.8,0) -- (1.2,0); 
\draw (-0.2,0) node[left] {$c$}; \draw (1.2,0) node[right] {$b$}; \draw (0.5,-0.6) node[below] {$a$}; 
\end{scope}}}

\def \qEdge (#1) {
\tikeq{\begin{scope}[scale=#1]
   \draw (0,0) node[left] {$i$};
  \draw[qua](0,0) -- (.8,0);
  \draw (.8,0) node[right] {$j$};
\end{scope}}}

\def \gEdge (#1) {
\tikeq{\begin{scope}[scale=#1]
   \draw (0,0) node[left] {$a$};
  \draw[glu](0,0) -- (.8,0);
  \draw (.8,0) node[right] {$b$};
\end{scope}}}

\def \gqqVert (#1) {
\tikeq{\begin{scope}[scale=#1]
   \draw (0,0) node[left] {$i$};
  \draw[qua](0,0) -- (.4,0);
  \draw[qua](.4,0) -- (.8,0);
  \draw (.8,0) node[right] {$j$};
  \draw[glu](.4,-0.4) -- (.4,0);
  \draw (.4,-0.4) node[below] {$a$};
    \end{scope}}}
  
 \def \gqbarqbarVert (#1) {
\tikeq{\begin{scope}[scale=#1]  
   \draw (0,0) node[left] {$i$};
  \draw[anti](0,0) -- (.4,0);
  \draw[anti](.4,0) -- (.8,0);
  \draw (.8,0) node[right] {$j$};
  \draw[glu](.4,-0.4) -- (.4,0);
  \draw (.4,-0.4) node[below] {$a$};
   \end{scope}}}
   
\def \gggVert (#1) {
\tikeq{\begin{scope}[scale=#1]   
   \draw (0,0) node[left] {$c$};
  \draw[glu](0,0) -- (.4,0)node{$\scriptstyle\bullet$};
  \draw[glu](.4,0) -- (.8,0);
  \draw (.8,0) node[right] {$b$};
  \draw[glu](.4,-0.4) -- (.4,0);
  \draw (.4,-0.4) node[below] {$a$};
 \end{scope}}}
 
\def \starVert (#1) {
\tikeq{\begin{scope}[scale=#1]  
   \draw (0,0) node[left] {$c$};
  \draw[glu](0,0) -- (.45,0);
  \draw[glu](.45,0) -- (.9,0);
  \draw (.9,0) node[right] {$b$};
  \draw (0.4,0) node {$\bigstar$};
  \draw[glu](.4,-0.4) -- (.4,-.05);
  \draw (.4,-0.4) node[below] {$a$};
\end{scope}}}

\def \Pqqbar (#1,#2) {
\tikeqbis{\begin{scope}[scale=#2]
	\coordinate (A) at (1.,0); 
	\coordinate (O) at (0,0); 
	\draw[anti] ($(A)+(0,-.3)$) -- ++ (.4,0) ;
	\draw[qua] ($(A)+(0,+.3)$) -- ++ (.4,0) ;
	\draw[anti] ($(A)+(1.,-.3)$) -- ++ (.4,0) ;
  \draw[qua] ($(A)+(1.,+.3)$) -- ++ (.4,0) ;
  \fill[white,draw=black] ($(O)+(1.7,.0)$) circle (.42);  
  \draw ($(O)+(1.7,0)$) node {$#1$};
  \end{scope}
}}
\def \Pgg (#1,#2) {
\tikeqbis{\begin{scope}[scale=#2]
	\coordinate (A) at (1.,0); 
	\coordinate (O) at (0,0); 
	\draw[glu] ($(A)+(0,-.3)$) -- ++ (.4,0) ;
	\draw[glu] ($(A)+(0,+.3)$) -- ++ (.4,0) ;
	\draw[glu] ($(A)+(1.,-.3)$) -- ++ (.4,0) ;
  \draw[glu] ($(A)+(1.,+.3)$) -- ++ (.4,0) ;
  \fill[white,draw=black] ($(O)+(1.7,.0)$) circle (.42);  
  \draw ($(O)+(1.7,0)$) node {$#1$};
    \end{scope}}}
\def \Pqqbaranti (#1,#2)  {
\tikeqbis{\begin{scope}[scale=#2]
	\coordinate (A) at (1.,0); 
	\coordinate (O) at (0,0); 
	\draw[anti] ($(A)+(0,-.25)$) -- ++ (.4,0) ;
	\draw[qua] ($(A)+(0,+.25)$) -- ++ (.4,0) ;
	\draw[qua] ($(A)+(.72,-.25)$) -- ++ (.4,0) ;
  \draw[anti] ($(A)+(.72,+.25)$) -- ++ (.4,0) ;
  \fill[white,draw=black] ($(O)+(1.55,.0)$) circle (.3);  
  \draw ($(O)+(1.55,0)$) node {$#1$};
 \end{scope}}}
\def \Pqqbaranticross (#1,#2)  {
\tikeqbis{\begin{scope}[scale=#2]
	\coordinate (A) at (1.,0); 
	\coordinate (O) at (0,0); 
	\draw[anti] ($(A)+(0,-.25)$) -- ++ (.4,0) ;
	\draw[qua] ($(A)+(0,+.25)$) -- ++ (.4,0) ;
	\draw[anti] ($(A)+(.72,-.25)$) -- ++ (.4,0) ;
  \draw[qua] ($(A)+(.72,+.25)$) -- ++ (.4,0) ;
  \fill[white,draw=black] ($(O)+(1.55,.0)$) circle (.3);  
  \draw ($(O)+(1.55,0)$) node {$#1$};
  \draw ($(A)+(1.12,+.25)$) -- ++(0.4,-.5);
	\draw ($(A)+(1.12,-.25)$) -- ++(0.4,0.5);
  \draw[anti] ($(A)+(1.52,+.25)$) -- ++(0.35,-.0);
  \draw[qua] ($(A)+(1.52,-.25)$) -- ++(0.35,0.0);
 \end{scope}}}
\def \Pthreebarqq  {\frac{1}{2} \left[ \ 
\tikeq{
		\coordinate (O) at (0,0);
		\draw[qua] ($(O)+(0,0)$) -- ++(0.7,0);
		\draw[qua] ($(O)+(0,.5)$) -- ++(0.7,0);
}
-
\tikeq{
		\coordinate (O) at (0,0);
		\draw (O) -- ++(0.7,.5);
		\draw ($(O)+(0,.5)$) -- ++(0.7,-0.5);
		\draw[qua] (O) -- ++(0.35,.25);
		\draw[qua] ($(O)+(0,.5)$) -- ++(0.35,-0.25);
} \ \right] }
\def \Psixqq  {\frac{1}{2} \left[ \ 
\tikeq{
		\coordinate (O) at (0,0);
		\draw[qua] ($(O)+(0,0)$) -- ++(0.7,0);
		\draw[qua] ($(O)+(0,.5)$) -- ++(0.7,0);
}
+
\tikeq{
		\coordinate (O) at (0,0);
		\draw (O) -- ++(0.7,.5);
		\draw ($(O)+(0,.5)$) -- ++(0.7,-0.5);
		\draw[qua] (O) -- ++(0.35,.25);
		\draw[qua] ($(O)+(0,.5)$) -- ++(0.35,-0.25);
} \ \right]}
\def \Poneqqbar { \displaystyle \frac1{\Nc} \
\tikeq{
    \coordinate (O) at (0,0);
    \draw[anti] (O) -- ++(0.1,0) -- ++(0,.5) -- ++(-0.1,0);  
    \draw[qua] ($(O)+(.5,0)$) -- ++(-0.1,0) -- ++(0,.5) -- ++(0.1,0);
}}
\def \Peightqqbar {2 \ 
\tikeq{
        \coordinate (O) at (0,0);
        \draw[qua] ($(O)+(0,.5)$) -- ++(.3,-.25);
        \draw[anti] (O) -- ++(.3,.25);
        \draw[glu] ($(O)+(.3,.25)$) -- ++(.7,0);
        \draw[qua] ($(O)+(1,.25)$) -- ++(.3,.25);
        \draw[anti] ($(O)+(1,.25)$) -- ++(.3,-.25);
}} 
\def \Pthreeqg  { \displaystyle \frac{1}{\CF} \, 
\tikeq{
		\coordinate (A) at (.3,.0);
		\coordinate (B) at (.7,.0);
    \draw[qua] ($(A)-(.3,-.35)$) -- (A);
    \draw[qua] (A) -- (B);
    \draw[qua] (B) -- ($(B)+(.3,.35)$) ;
    \draw[glu] ($(A)-(.3,.35)$) -- (A);
    \draw[glu] (B) -- ($(B)+(.3,-.35)$) ;
} }
\def \Psixqg  { \displaystyle \frac12 \ 
\tikeq{
		\coordinate (A) at (.3,.0);
		\coordinate (B) at (1.,.0);
    \draw[qua] ($(A)-(.3,-.25)$) -- ++(.7,0);
    \draw[glu] ($(A)-(.3,.25)$) -- ++(.7,0);
} 
-\frac{\Nc+1}2 \mathsize{11}{\Proj{\bm{3}} }
+ 
\tikeq{
		\coordinate (O) at (0,0);
		\draw[glu] ($(O)+(0,0)$) -- ++(0.5,.1) coordinate(A);
		\draw[glu] (A) -- ++(0.5,-.1);
		\draw[qua] ($(O)+(0,.7)$) -- ++(0.5,-.1)coordinate(B);
		\draw[qua] (B) -- ++(0.5,.1);
		\draw[glu] (A) --(B);
}}
\def \Pfifteenqg  { \displaystyle \frac12 \ 
\tikeq{
		\coordinate (A) at (.3,.0);
		\coordinate (B) at (1.,.0);
    \draw[qua] ($(A)-(.3,-.25)$) -- ++(.7,0);
    \draw[glu] ($(A)-(.3,.25)$) -- ++(.7,0);
} 
+\frac{\Nc-1}2 \mathsize{11}{\Proj{\bm{3}} }
-
\tikeq{
		\coordinate (O) at (0,0);
		\draw[glu] ($(O)+(0,0)$) -- ++(0.5,.1) coordinate(A);
		\draw[glu] (A) -- ++(0.5,-.1);
		\draw[qua] ($(O)+(0,.7)$) -- ++(0.5,-.1)coordinate(B);
		\draw[qua] (B) -- ++(0.5,.1);
		\draw[glu] (A) --(B);
}}
\def \Ponegg  { \displaystyle \frac1{\Nc^2-1} \ 
\tikeq{
		\coordinate (A) at (.0,.0);
		\coordinate (B) at (1,-.4);  
     \draw[glu] (A) .. controls +(0.5,0) and +(0.5,0) .. ($(A)+(0,-.4)$) ;            	
     \draw[glu] (B) .. controls +(-0.5,0) and +(-0.5,0) .. ($(B)+(0,.4)$); 
} 
}
\def \PeightAgg  { \displaystyle \frac1{\Nc} \ 
\tikeq{
        \coordinate (O) at (0,0);
        \draw[glu] ($(O)+(0,.5)$) -- ++(.3,-.25);
        \draw[glu] (O) -- ++(.3,.25);
        \draw[glu] ($(O)+(.3,.25)$) -- ++(.7,0);
        \draw[glu] ($(O)+(1,.25)$) -- ++(.3,.25);
        \draw[glu] ($(O)+(1,.25)$) -- ++(.3,-.25);
}
}
\def \PeightSgg  { \displaystyle \frac{\Nc}{\Nc^2-4} \ 
\tikeq{
        \coordinate (O) at (0,0);
        \draw[glu] ($(O)+(0,.5)$) -- ++(.3,-.25);
        \draw[glu] (O) -- ++(.3,.25);
        \draw[glu] ($(O)+(.3,.25)$) -- ++(.7,0);
        \draw[glu] ($(O)+(1,.25)$) -- ++(.3,.25);
        \draw[glu] ($(O)+(1,.25)$) -- ++(.3,-.25);
        \node [star, fill=black, star points=5, star point ratio=3.2, scale=.3] at ($(O)+(.3,.25)$) {};
        \node [star, fill=black, star points=5, star point ratio=3.2, scale=.3] at ($(O)+(1.,.25)$) {};
} 
}
\newcommand{\Ptengg}{ \displaystyle \frac12 \left[ \ 
\tikeq{
		\coordinate (A) at (.3,.0);
    \draw[glu] ($(A)-(.3,-.25)$) -- ++(.6,0);
    \draw[glu] ($(A)-(.3,.25)$) -- ++(.6,0);
}  -
\tikeq{
		\coordinate (A) at (.3,.0);
    \draw[glu] ($(A)-(.3,-.25)$) -- ++(.6,-.5);
    \draw[glu] ($(A)-(.3,.25)$) -- ++(.6,.5);
} \ \right]
- \mathsize{11}{\Proj{\bm{8}_{\rm a}}}
}

\newcommand{\Ptwentysevenggcut}{ \displaystyle 
\bigg( 
\frac12 \ \tikeq{
		\coordinate (A) at (.3,.0);
    \draw[glu] ($(A)-(.3,-.25)$) -- ++(.6,0);
    \draw[glu] ($(A)-(.3,.25)$) -- ++(.6,0);
}  + 2\
\tikeq{
		\coordinate (A) at (0,.25);
		\coordinate (B) at (0,-.25);
		\coordinate (C) at (.5,.25);
		\coordinate (D) at (.5,-.25);
    \draw[glu] ($(A)-(.3,.0)$) -- (A);
    \draw[glu] ($(B)-(.3,.0)$) -- (B);
    \draw[glu] (C) -- ($(C)+(.3,.0)$);
    \draw[glu] (D) -- ($(D)+(.3,.0)$);
    \draw[qua] (A) -- (C);
    \draw[qua] (B) -- (D);
    \draw (A) -- (D);
    \draw (B) -- (C);
}  
\bigg)
}
\newcommand{\Pzeroggcut}{ \displaystyle 
\bigg( 
\frac12 \ \tikeq{
		\coordinate (A) at (.3,.0);
    \draw[glu] ($(A)-(.3,-.25)$) -- ++(.6,0);
    \draw[glu] ($(A)-(.3,.25)$) -- ++(.6,0);
}  - 2\
\tikeq{
		\coordinate (A) at (0,.25);
		\coordinate (B) at (0,-.25);
		\coordinate (C) at (.5,.25);
		\coordinate (D) at (.5,-.25);
    \draw[glu] ($(A)-(.3,.0)$) -- (A);
    \draw[glu] ($(B)-(.3,.0)$) -- (B);
    \draw[glu] (C) -- ($(C)+(.3,.0)$);
    \draw[glu] (D) -- ($(D)+(.3,.0)$);
    \draw[qua] (A) -- (C);
    \draw[qua] (B) -- (D);
    \draw (A) -- (D);
    \draw (B) -- (C);
}  
\bigg)
}
\def \Q {\displaystyle 
\frac12 \left[ \ 
\tikeq{\begin{scope}[scale=0.8] \coordinate (A) at (.3,.0);
    \draw[glu] ($(A)-(.3,-.25)$) -- ++(.6,0);
    \draw[glu] ($(A)-(.3,.25)$) -- ++(.6,0);
\end{scope}}  +
\tikeq{\begin{scope}[scale=0.8] \coordinate (A) at (.3,.0);
    \draw[glu] ($(A)-(.3,-.25)$) -- ++(.6,-.5);
    \draw[glu] ($(A)-(.3,.25)$) -- ++(.6,.5);
\end{scope}} \ \right] 
- \mathsize{11}{\Proj{\bm{8}_{\rm s}}}
- \mathsize{11}{\Proj{\bm{1}}}}

\def \Wm (#1) {\tikeqbis{\begin{scope}[scale=#1]
		\coordinate (A) at (0,.25);
		\coordinate (B) at (0,-.25);
		\coordinate (C) at (.5,.25);
		\coordinate (D) at (.5,-.25);
    \draw[glu] ($(A)-(.3,.0)$) -- (A);
    \draw[glu] ($(B)-(.3,.0)$) -- (B);
    \draw[glu] (C) -- ($(C)+(.3,.0)$);
    \draw[glu] (D) -- ($(D)+(.3,.0)$);
    \draw[qua] (A) -- (C);
    \draw[qua] (B) -- (D);
    \draw (A) -- (D);
    \draw (B) -- (C);\end{scope}}}
    
    \def \Wp (#1) {\tikeqbis{\begin{scope}[scale=#1]
		\coordinate (A) at (0,.25);
		\coordinate (B) at (0,-.25);
		\coordinate (C) at (.5,.25);
		\coordinate (D) at (.5,-.25);
    \draw[glu] ($(A)-(.3,.0)$) -- (A);
    \draw[glu] ($(B)-(.3,.0)$) -- (B);
    \draw[glu] (C) -- ($(C)+(.3,.0)$);
    \draw[glu] (D) -- ($(D)+(.3,.0)$);
    \draw[anti] (A) -- (C);
    \draw[anti] (B) -- (D);
    \draw (A) -- (D);
    \draw (B) -- (C);\end{scope}}}

\def \Tonegg {
\tikeq{
		\coordinate (A) at (.0,.0);
		\coordinate (B) at (1,-.4);   
     \draw[glu] (A) .. controls +(0.5,0) and +(0.5,0) .. ($(A)+(0,-.4)$) ;            	
     \draw[qua] (B) .. controls +(-0.5,0) and +(-0.5,0) .. ($(B)+(0,.4)$); 
} }
\def \TeightAgg  {
\tikeq{
        \coordinate (O) at (0,0);
        \draw[glu] ($(O)+(0,.5)$) -- ++(.3,-.25);
        \draw[glu] (O) -- ++(.3,.25);
        \draw[glu] ($(O)+(.3,.25)$) -- ++(.7,0);
        \draw[qua] ($(O)+(1,.25)$) -- ++(.3,.25);
        \draw[anti] ($(O)+(1,.25)$) -- ++(.3,-.25);
}}
\def \TeightSgg  {
\tikeq{
        \coordinate (O) at (0,0);
        \draw[glu] ($(O)+(0,.5)$) -- ++(.3,-.25);
        \draw[glu] (O) -- ++(.3,.25);
        \draw[glu] ($(O)+(.3,.25)$) -- ++(.7,0);
        \draw[qua] ($(O)+(1,.25)$) -- ++(.3,.25);
        \draw[anti] ($(O)+(1,.25)$) -- ++(.3,-.25);
        \node [star, fill=black, star points=5, star point ratio=3.2, scale=.3] at ($(O)+(.3,.25)$) {};
      } }

\def \Mqoneqtwo {\tikeq{
		\coordinate (O) at (0,0);
		\draw[qua] ($(O)+(0,0)$) -- ++(0.4,0) coordinate(A);
		\draw[qua] (A) -- ++(0.4,0);
		\draw[qua] ($(O)+(0,.5)$) -- ++(0.4,0)coordinate(B);
		\draw[qua] (B) -- ++(0.4,0);
		\draw[glu] (A) --(B);
}}
\newcommand{\qqtchannel}[1]{\tikeq{\begin{scope}[scale=#1]
		\coordinate (O) at (0,0);
		\draw[qua] ($(O)+(0,0)$) -- ++(0.4,0) coordinate(A);
		\draw[qua] (A) -- ++(0.4,0);
		\draw[qua] ($(O)+(0,.5)$) -- ++(0.4,0)coordinate(B);
		\draw[qua] (B) -- ++(0.4,0);
		\draw[glu] (A) --(B);
\end{scope}
}}

\newcommand{\qquchannel}[1]{\tikeq{\begin{scope}[scale=#1]
		\coordinate (O) at (0,0);
		\draw[qua] ($(O)+(0,0)$) -- ++(0.4,0) coordinate(A);
		\draw (A) -- ++(0.4,0.5);
		\draw[qua] ($(O)+(0,.5)$) -- ++(0.4,0)coordinate(B);
		\draw(B) -- ++(0.4,-.5);
		\draw[glu] (A) --(B);
\end{scope}
}}

\def \Mqoneqone {\mathcal{B}_t \ \tikeq{
		\coordinate (O) at (0,0);
		\draw[qua] ($(O)+(0,0)$) -- ++(0.4,0) coordinate(A);
		\draw[qua] (A) -- ++(0.4,0);
		\draw[qua] ($(O)+(0,.5)$) -- ++(0.4,0)coordinate(B);
		\draw[qua] (B) -- ++(0.4,0);
		\draw[glu] (A) --(B);
} 
+ \mathcal{B}_u \  \tikeq{
		\coordinate (O) at (0,0);
		\draw[qua] ($(O)+(0,0)$) -- ++(0.4,0) coordinate(A);
		\draw (A) -- ++(0.4,0.5);
		\draw[qua] ($(O)+(0,.5)$) -- ++(0.4,0)coordinate(B);
		\draw(B) -- ++(0.4,-.5);
		\draw[glu] (A) --(B);
}}
\def \Mqoneqtwobar {\tikeq{
		\coordinate (O) at (0,0);
		\draw[anti] ($(O)+(0,0)$) -- ++(0.4,0) coordinate(A);
		\draw[anti] (A) -- ++(0.4,0);
		\draw[qua] ($(O)+(0,.5)$) -- ++(0.4,0)coordinate(B);
		\draw[qua] (B) -- ++(0.4,0);
		\draw[glu] (A) --(B);
}}
\def \Mqoneqonebar  {\tikeq{
        \coordinate (O) at (0,0);
        \draw[qua] ($(O)+(0,.5)$) -- ++(.3,-.25);
        \draw[anti] (O) -- ++(.3,.25);
        \draw[glu] ($(O)+(.3,.25)$) -- ++(.7,0);
        \draw[qua] ($(O)+(1,.25)$) -- ++(.3,.25);
        \draw[anti] ($(O)+(1,.25)$) -- ++(.3,-.25);
}}
\def \Mqoneqonebarbis { \mathcal{E}_s \ \tikeq{
        \coordinate (O) at (0,0);
        \draw[qua] ($(O)+(0,.5)$) -- ++(.3,-.25);
        \draw[anti] (O) -- ++(.3,.25);
        \draw[glu] ($(O)+(.3,.25)$) -- ++(.7,0);
        \draw[qua] ($(O)+(1,.25)$) -- ++(.3,.25);
        \draw[anti] ($(O)+(1,.25)$) -- ++(.3,-.25);
}+ \mathcal{E}_t \ \tikeq{
		\coordinate (O) at (0,0);
		\draw[anti] ($(O)+(0,0)$) -- ++(0.4,0) coordinate(A);
		\draw[anti] (A) -- ++(0.4,0);
		\draw[qua] ($(O)+(0,.5)$) -- ++(0.4,0)coordinate(B);
		\draw[qua] (B) -- ++(0.4,0);
		\draw[glu] (A) --(B);
}} 
\def \Mqg { \tikeq{
		\coordinate (O) at (0,0);
		\draw[glu] ($(O)+(0,0)$) -- ++(0.5,.1) coordinate(A);
		\draw[glu] (A) -- ++(0.5,-.1);
		\draw[qua] ($(O)+(0,.7)$) -- ++(0.5,-.1)coordinate(B);
		\draw[qua] (B) -- ++(0.5,.1);
		\draw[glu] (A) --(B);
}
 - \xi \ 
 \tikeq{
		\coordinate (A) at (.3,.0);
		\coordinate (B) at (.7,.0);
    \draw[qua] ($(A)-(.3,-.35)$) -- (A);
    \draw[qua] (A) -- (B);
    \draw[qua] (B) -- ($(B)+(.3,.35)$) ;
    \draw[glu] ($(A)-(.3,.35)$) -- (A);
    \draw[glu] (B) -- ($(B)+(.3,-.35)$) ;
}}

\def \Mggqq { 
\tikeq{
		\coordinate (O) at (0,0);
		\draw[glu] ($(O)+(0,0)$) -- ++(0.5,.1) coordinate(A);
		\draw[anti] (A) -- ++(0.5,-.1);
		\draw[glu] ($(O)+(0,.7)$) -- ++(0.5,-.1)coordinate(B);
		\draw[qua] (B) -- ++(0.5,.1);
		\draw[qua] (A) --(B);
}
 - \xi \ 
\tikeq{
        \coordinate (O) at (0,0);
        \draw[glu] ($(O)+(0,.5)$) -- ++(.3,-.25);
        \draw[glu] (O) -- ++(.3,.25);
        \draw[glu] ($(O)+(.3,.25)$) -- ++(.5,0);
        \draw[qua] ($(O)+(.8,.25)$) -- ++(.3,.25);
        \draw[anti] ($(O)+(.8,.25)$) -- ++(.3,-.25);
}
} 

\def \xibird (#1,#2) {\tikeqbis{\begin{scope}[scale=#1]
	        \coordinate (O) at (0,0); 
	        \coordinate (S) at   ($(O)+(.5,0)$); 
	        \coordinate (A2) at  ($(O)+(0,.5)$) ;  
	        \coordinate (A1) at  ($(A2)+(0,.6)$) ; 
	        \coordinate (A3) at  ($(A2)+(1.,0)$); 
	        \coordinate (A4) at  ($(A1)+(1.,0)$);  
	        \coordinate (B2) at  ($(O)-(0,.5)$) ;  
	        \coordinate (B1) at  ($(B2)-(0,.6)$) ;  
	        \coordinate (B3) at  ($(B2)+(1.,0)$);  
	        \coordinate (B4) at  ($(B1)+(1.,0)$);          
	        \drawdash  (A1) -- ++ (-1,0) arc (90:180:.3) -- ++ (0,-1.6) arc (-180:-90:.3) -- (B1);
	        \drawdash  (A2) -- ++ (-.7,0) arc (90:180:.2) -- ++ (0,-0.6) arc (-180:-90:.2) -- (B2);
	        \drawdash  (A4) -- ++ (1,0) arc (90:0:.3) -- ++ (0,-1.6) arc (0:-90:.3) -- (B4);
	        \drawdash  (A3) -- ++ (.7,0) arc (90:0:.2) -- ++ (0,-0.6) arc (0:-90:.2) -- (B3);
  \fill[color=gray!30!white,opacity=1] ($(A1)+(0,.1)$) rectangle ($(A3)+(0,-.1)$);  
  \draw ($(A1)+(.5,-.3)$) node {$\scriptstyle{{\cal M}}$};
  \fill[color=gray!30!white,opacity=1] ($(B2)+(0,.1)$) rectangle ($(B4)+(0,-.1)$); 
  \draw ($(B2)+(.5,-.3)$) node {$\scriptstyle{{\cal M}^*}$};
 \draw[glu] ($(A1)+(-0.6,0)$) -- ++ (0,.2) arc (180:90:.3) -- ++ (1.6,0) arc (90:0:.3) -- ++ (0,#2); 
\end{scope}}}

\def \xibirdstraight (#1,#2) {
\tikeqbis{  \begin{scope}[scale=#1]
	        \coordinate (O) at (0,0); 
	        \coordinate (S) at   ($(O)+(.5,0)$); 
	        \coordinate (A2) at  ($(O)+(0,.5)$) ;  
	        \coordinate (A1) at  ($(A2)+(0,.6)$) ; 
	        \coordinate (A3) at  ($(A2)+(1.,0)$); 
	        \coordinate (A4) at  ($(A1)+(1.,0)$);  
	        \coordinate (B2) at  ($(O)-(0,.5)$) ;  
	        \coordinate (B1) at  ($(B2)-(0,.6)$) ;  
	        \coordinate (B3) at  ($(B2)+(1.,0)$);  
	        \coordinate (B4) at  ($(B1)+(1.,0)$);          
    \draw ($(A1)+(-.2,.15)$) node {$1$};
    \draw ($(A2)+(-.2,-.15)$) node {$2$};
    \draw ($(A3)+(.2,-.15)$) node {$3$};
    \draw ($(A4)+(.2,.15)$) node {$4$};
	           
	        \draw[red]  (A1) -- ++ (-1,0) arc (90:180:.3) -- ++ (0,-1.6) arc (-180:-90:.3) -- (B1);
	        \draw[blue]  (A2) -- ++ (-.7,0) arc (90:180:.2) -- ++ (0,-0.6) arc (-180:-90:.2) -- (B2);
	        \draw[green]  (A4) -- ++ (1,0) arc (90:0:.3) -- ++ (0,-1.6) arc (0:-90:.3) -- (B4);
	        \draw[brown]  (A3) -- ++ (.7,0) arc (90:0:.2) -- ++ (0,-0.6) arc (0:-90:.2) -- (B3);

  \fill[color=gray!30!white,opacity=1] ($(A1)+(0,.1)$) rectangle ($(A3)+(0,-.1)$);  
  \draw ($(A1)+(.5,-.3)$) node {${\cal M}$};
  \fill[color=gray!30!white,opacity=1] ($(B2)+(0,.1)$) rectangle ($(B4)+(0,-.1)$); 
  \draw ($(B2)+(.5,-.3)$) node {${\cal M}^*$};
 
 \draw[glu] ($(O)+(-1.3,0)$) -- ($(O)+(#2,0)$) ; 
  \end{scope}
 }}

\def \xibirdalpha (#1,#2,#3,#4) {
\tikeqbis{  \begin{scope}[scale=#1]
	        \coordinate (O) at (0,0); 
	        \coordinate (S) at   ($(O)+(.5,0)$); 
	        \coordinate (A2) at  ($(O)+(0,.5)$) ;  
	        \coordinate (A1) at  ($(A2)+(0,.6)$) ; 
	        \coordinate (A3) at  ($(A2)+(1.,0)$); 
	        \coordinate (A4) at  ($(A1)+(1.,0)$);  
	        \coordinate (B2) at  ($(O)-(0,.5)$) ;  
	        \coordinate (B1) at  ($(B2)-(0,.6)$) ;  
	        \coordinate (B3) at  ($(B2)+(1.,0)$);  
	        \coordinate (B4) at  ($(B1)+(1.,0)$);          	           
	        \draw[red]   ($(A1)+(.4,0)$) -- (A1) -- ++ (-1,0) arc (90:180:.3) -- ++ (0,-1.6) arc (-180:-90:.3) -- ($(B1)+(.2,0)$) ;
	        \draw[blue]   ($(A2)+(.4,0)$) -- (A2) -- ++ (-.7,0) arc (90:180:.2) -- ++ (0,-0.6) arc (-180:-90:.2) -- ($(B2)+(.2,0)$) ;
	        \draw[green]   ($(A4)+(-.4,0)$) -- (A4) -- ++ (1,0) arc (90:0:.3) -- ++ (0,-1.6) arc (0:-90:.3) -- ($(B4)+(-.2,0)$);
	        \draw[brown]   ($(A3)+(-.4,0)$) -- (A3) -- ++ (.7,0) arc (90:0:.2) -- ++ (0,-0.6) arc (0:-90:.2) -- ($(B3)+(-.2,0)$);
  \fill[white,draw=black] ($(A1)+(.5,-.3)$) circle (.42);   
  \draw ($(A1)+(.5,-.3)$) node {$#3$};
  \fill[white,draw=black] ($(B2)+(.5,-.3)$) circle (.42);  
  \draw ($(B2)+(.5,-.3)$) node {$#4$};
 \draw[glu] ($(A1)+(-0.6,0)$) -- ++ (0,.2) arc (180:90:.3) -- ++ (1.6,0) arc (90:0:.3) -- ++ (0,#2); 
  \end{scope}
 }}
\def \xibirdalphastraight (#1,#2,#3,#4) {
\tikeqbis{  \begin{scope}[scale=#1]
	        \coordinate (O) at (0,0);
	        \coordinate (A2) at  ($(O)+(0,.5)$) ;  
	        \coordinate (A1) at  ($(A2)+(0,.6)$) ; 
	        \coordinate (A3) at  ($(A2)+(1.,0)$); 
	        \coordinate (A4) at  ($(A1)+(1.,0)$);  
	        \coordinate (B2) at  ($(O)-(0,.5)$) ;  
	        \coordinate (B1) at  ($(B2)-(0,.6)$) ;  
	        \coordinate (B3) at  ($(B2)+(1.,0)$);  
	        \coordinate (B4) at  ($(B1)+(1.,0)$);          
    \draw ($(A1)+(-.2,.15)$) node {$1$};
    \draw ($(A2)+(-.2,-.15)$) node {$2$};
    \draw ($(A3)+(.2,-.15)$) node {$3$};
    \draw ($(A4)+(.2,.15)$) node {$4$};       
	        \draw[red]  ($(A1)+(.4,0)$) -- (A1) -- ++ (-1.,0) arc (90:180:.3) -- ++ (0,-1.6) arc (-180:-90:.3) -- ($(B1)+(.2,0)$) ;
	        \draw[blue]  ($(A2)+(.4,0)$) -- (A2) -- ++ (-.7,0) arc (90:180:.2) -- ++ (0,-0.6) arc (-180:-90:.2) -- ($(B2)+(.2,0)$);
	        \draw[green]  ($(A4)+(-.4,0)$) -- (A4) -- ++ (1,0) arc (90:0:.3) -- ++ (0,-1.6) arc (0:-90:.3) -- ($(B4)+(-.2,0)$);
	        \draw[brown] ($(A3)+(-.4,0)$) -- (A3) -- ++ (.7,0) arc (90:0:.2) -- ++ (0,-0.6) arc (0:-90:.2) -- ($(B3)+(-.2,0)$);
   \fill[white,draw=black] ($(O)+(.5,.8)$) circle (.42);  
  \draw ($(O)+(.5,.8)$) node {$#3$};
    \fill[white,draw=black] ($(O)+(.5,-.8)$) circle (.42);  
  \draw ($(O)+(.5,-.8)$) node {$#4$};
 \draw[glu] ($(O)+(-1.3,0)$) -- ($(O)+(#2,0)$) ; 
  \end{scope}
 }}

\def \Msquaredbird (#1) {
\tikeqbis{ \begin{scope}[scale=#1]
	        \coordinate (O) at (0,0); 
	        \coordinate (S) at   ($(O)+(.5,0)$); 
	        \coordinate (A2) at  ($(O)+(0,.5)$) ;  
	        \coordinate (A1) at  ($(A2)+(0,.6)$) ; 
	        \coordinate (A3) at  ($(A2)+(1.,0)$); 
	        \coordinate (A4) at  ($(A1)+(1.,0)$);  
	        \coordinate (B2) at  ($(O)-(0,.5)$) ;  
	        \coordinate (B1) at  ($(B2)-(0,.6)$) ;  
	        \coordinate (B3) at  ($(B2)+(1.,0)$);  
	        \coordinate (B4) at  ($(B1)+(1.,0)$);          
	        \drawdash  (A1) -- ++ (-1,0) arc (90:180:.3) -- ++ (0,-1.6) arc (-180:-90:.3) -- (B1);
	        \drawdash  (A2) -- ++ (-.7,0) arc (90:180:.2) -- ++ (0,-0.6) arc (-180:-90:.2) -- (B2);
	        \drawdash  (A4) -- ++ (1,0) arc (90:0:.3) -- ++ (0,-1.6) arc (0:-90:.3) -- (B4);
	        \drawdash  (A3) -- ++ (.7,0) arc (90:0:.2) -- ++ (0,-0.6) arc (0:-90:.2) -- (B3);
  \fill[color=gray!30!white,opacity=1] ($(A1)+(0,.1)$) rectangle ($(A3)+(0,-.1)$);  
  \draw ($(A1)+(.5,-.3)$) node {$\scriptstyle{{\cal M}}$};
  \fill[color=gray!30!white,opacity=1] ($(B2)+(0,.1)$) rectangle ($(B4)+(0,-.1)$); 
  \draw ($(B2)+(.5,-.3)$) node {$\scriptstyle{{\cal M}^*}$};
 \end{scope}}}

\def \Operatorbird (#1) {
\tikeq{  \begin{scope}[scale=#1]
	        \coordinate (O) at (0,0); 
	        \coordinate (A2) at  ($(O)+(-.6,0)$) ;  
	        \coordinate (A1) at  ($(A2)+(-.6,0)$) ; 
	        \coordinate (A3) at  ($(A2)+(0,1)$); 
	        \coordinate (A4) at  ($(A1)+(0,1)$);  
	        \coordinate (B2) at  ($(O)-(-.6,0)$) ;  
	        \coordinate (B1) at  ($(B2)-(-.6,0)$) ;  
	        \coordinate (B3) at  ($(B2)+(0,1)$);  
	        \coordinate (B4) at  ($(B1)+(0,1)$);                
	        \drawdash  ($(A1)+(0,.2)$) -- (A1) -- ++ (0,-.5) arc (180:270:.3) -- ++ (1.8,0) arc (-90:0:.3) -- ($(B1)+(0,.2)$) ;
	        \drawdash  ($(A2)+(0,.2)$) -- (A2) -- ++ (0,-.2) arc (180:270:.2) -- ++ (.8,0.) arc (-90:-0:.2) -- ($(B2)+(0,.2)$);
	        \drawdash  ($(A4)+(0,-.2)$) -- (A4) -- ++ (0,.5) arc (180:90:.3) -- ++ (1.8,0) arc (90:0:.3) -- ($(B4)+(0,-.2)$);
	        \drawdash ($(A3)+(0,-.2)$) -- (A3) -- ++ (0,.2) arc (180:90:.2) -- ++ (.8,0) arc (90:0:.2) -- ($(B3)+(0,-.2)$);
  \fill[color=gray!30!white,opacity=1] ($(A2)+(.2,.1)$) rectangle ($(A4)+(-.2,-.1)$);  
  \draw ($(A1)+(.3,.5)$) node {$\scriptstyle{{\cal M}}$};
  \fill[color=gray!30!white,opacity=1] ($(B2)+(-.2,.1)$) rectangle ($(B4)+(.2,-.1)$);  
  \fill[color=white] ($(-.2,-1)$) rectangle ($(.2,2)$);  
  \draw[pattern=north west lines, pattern color=black] ($(-.2,-1)$) rectangle ($(.2,2)$);  
  \draw ($(B1)+(-.3,.5)$) node {$\scriptstyle{{\cal M}^*}$};
  \end{scope}
 }}

\newcommand{\Mtwototwo}[5] {
\tikeqbis{ \begin{scope}[scale=#5]
	        \coordinate (A2) at  (0,-0.3);  
	        \coordinate (A1) at  (0,0.3) ; 
	        \coordinate (A3) at  (1.,-0.3); 
	        \coordinate (A4) at  (1.,0.3);           
    \draw ($(A1)+(-0.6,.2)$) node {$#1$};
    \draw ($(A2)+(-0.6,-.2)$) node {$#2$};
    \draw ($(A3)+(0.6,-.2)$) node {$#4$};
    \draw ($(A3)+(1.1,0.)$) node {$\xi, \bm K$}; 
    \draw ($(A4)+(0.6,.2)$) node {$#3$}; 
    \draw ($(A4)+(1.2,0.)$) node {$\xibar, - \bm K$};      
	        \drawdash  (A1) -- ++ (-.7,0);
	        \drawdash (A2) -- ++ (-.7,0);
 		\drawdash  (A4) -- ++ (.7,0);
		\drawdash  (A3) -- ++ (.7,0);
  \fill[color=gray!30!white,opacity=1] ($(A1)+(0,.1)$) rectangle ($(A3)+(0,-.1)$);  
  \draw ($(A1)+(.5,-.3)$) node {${\cal M}$};
 \end{scope}}}

\def \IdentitySinglet (#1,#2) {
\tikeq{  \begin{scope}[scale=#1]
	        \coordinate (O) at (0,0); 
	        \coordinate (A) at  ($(O)+(0,-1.1)$) ;  
	        \coordinate (B) at  ($(O)+(0,-.7)$) ; 
	        \coordinate (C) at  ($(O)+(0,.7)$); 
	        \coordinate (D) at  ($(O)+(0,1.1)$);              
	        \drawdash  (A) -- ++ (#2,0);
	        \drawdash (B) -- ++ (#2,0);
	        \drawdash (C) -- ++ (#2,0);
	        \drawdash (D) -- ++ (#2,0);
  \end{scope}
 }}

\def \OrthoKet (#1,#2) {
\tikeq{  \begin{scope}[scale=#1]
	        \coordinate (O) at (0,0); 
	        \coordinate (A2) at (.5,-.3) ;  
	        \coordinate (A1) at (1.1,-.3) ; 
	        \coordinate (A3) at (.5,.3) ;  
	        \coordinate (A4) at  (1.1,.3) ;   
    \draw (-0.15,-1.1) node {$\mathsize{8}{\bar{1}}$};
    \draw (-0.15,-0.6) node {$\mathsize{8}{\bar{2}}$};
    \draw (-0.15,0.7) node {$\mathsize{8}{4}$};
    \draw (-0.15,1.1) node {$\mathsize{8}{3}$};
	        \drawdash ($(A1)+(0,.2)$) -- (A1) -- ++ (0,-.5) arc (360:270:.3) -- ++ (-0.8,0) ;
	        \drawdash  ($(A2)+(0,.2)$) -- (A2) -- ++ (0,-.2) arc (360:270:.2) -- ++ (-.3,0.) ;
	        \drawdash  ($(A4)+(0,-.2)$) -- (A4) -- ++ (0,.5) arc (0:90:.3) -- ++ (-0.8,0);
	        \drawdash ($(A3)+(0,-.2)$) -- (A3) -- ++ (0,.2) arc (0:90:.2) -- ++ (-.3,0);
   \fill[white,draw=black] ($(O)+(.8,0)$) circle (.42);  
  \draw ($(O)+(.8,0)$) node {$#2$};
  \end{scope}
 }}

\def \OrthoBra (#1,#2) {
\tikeq{  \begin{scope}[scale=#1]
 \coordinate (O) at (0,0); 
	        \coordinate (A2) at (-.5,-.3) ;  
	        \coordinate (A1) at (-1.1,-.3) ; 
	        \coordinate (A3) at (-.5,.3) ;  
	        \coordinate (A4) at  (-1.1,.3) ;  
    \draw (0.15,-1.1) node {$\mathsize{8}{\bar{1}}$};
    \draw (0.15,-0.6) node {$\mathsize{8}{\bar{2}}$};
    \draw (0.15,0.7) node {$\mathsize{8}{4}$};
    \draw (0.15,1.1) node {$\mathsize{8}{3}$}; 
	        \drawdash  ($(A1)+(0,.2)$) -- (A1) -- ++ (0,-.5) arc (180:270:.3) -- ++ (0.8,0) ;
	        \drawdash  ($(A2)+(0,.2)$) -- (A2) -- ++ (0,-.2) arc (180:270:.2) -- ++ (.3,0.) ;
	        \drawdash  ($(A4)+(0,-.2)$) -- (A4) -- ++ (0,.5) arc (180:90:.3) -- ++ (0.8,0);
	        \drawdash ($(A3)+(0,-.2)$) -- (A3) -- ++ (0,.2) arc (180:90:.2) -- ++ (.3,0);
   \fill[white,draw=black] ($(O)+(-.8,0)$) circle (.42);  
  \draw ($(O)+(-.8,0)$) node {$#2$};
  \end{scope}
 }}

\def \OrthoBrau (#1,#2) {
\tikeq{  \begin{scope}[scale=#1]
 \coordinate (O) at (0,0); 
	        \coordinate (A2) at (-.5,-.3) ;  
	        \coordinate (A1) at (-1.1,-.3) ; 
	        \coordinate (A3) at (-.5,.3) ;  
	        \coordinate (A4) at  (-1.1,.3) ;  
    \draw (1.45,-1.1) node {$\mathsize{8}{\bar{1}}$};
    \draw (1.45,-0.6) node {$\mathsize{8}{\bar{2}}$};
    \draw (1.45,0.7) node {$\mathsize{8}{4}$};
    \draw (1.45,1.1) node {$\mathsize{8}{3}$}; 
          \drawdash  ($(A1)+(0,.2)$) -- (A1) -- ++ (0,-.5) arc (180:270:.3) -- ++ (0.8,0) -- ++ (.7,0) -- ++ (.5,0);
          \drawdash  ($(A2)+(0,.2)$) -- (A2) -- ++ (0,-.2) arc (180:270:.2) -- ++ (.3,0.) 
          .. controls +(0.4,0) and +(-0.4,0) .. (0.8,0.7) 
          -- ++ (.4,0);
          \drawdash  ($(A4)+(0,-.2)$) -- (A4) -- ++ (0,.5) arc (180:90:.3) -- ++ (0.8,0) -- ++ (.7,0)-- ++ (.5,0);
          \drawdash ($(A3)+(0,-.2)$) -- (A3) -- ++ (0,.2) arc (180:90:.2) -- ++ (.3,0) 
          .. controls +(0.4,0) and +(-0.4,0) .. (0.8,-0.7) 
          -- ++ (.4,0);
   \fill[white,draw=black] ($(O)+(-.8,0)$) circle (.42);  
  \draw ($(O)+(-.8,0)$) node {$#2$};
  \end{scope}
 }}

\def \OrthoBrat (#1,#2) {
\tikeq{  \begin{scope}[scale=#1]
 \coordinate (O) at (0,0); 
	        \coordinate (A2) at (-.5,-.3) ;  
	        \coordinate (A1) at (-1.1,-.3) ; 
	        \coordinate (A3) at (-.5,.3) ;  
	        \coordinate (A4) at  (-1.1,.3) ;  
    \draw (1.45,-1.1) node {$\mathsize{8}{\bar{1}}$};
    \draw (1.45,-0.6) node {$\mathsize{8}{\bar{2}}$};
    \draw (1.45,0.7) node {$\mathsize{8}{4}$};
    \draw (1.45,1.1) node {$\mathsize{8}{3}$};   
          \drawdash  ($(A1)+(0,.2)$) -- (A1) -- ++ (0,-.5) arc (180:270:.3) -- ++ (0.8,0) -- ++ (.7,0) -- ++ (.5,0);
          \drawdash  ($(A2)+(0,.2)$) -- (A2) -- ++ (0,-.2) arc (180:270:.2) -- ++ (.1,0.) 
          .. controls +(0.4,0) and +(-0.4,0) .. (0.8,1.1)  
          -- ++ (.4,0);
          \drawdash  ($(A4)+(0,-.2)$) -- (A4) -- ++ (0,.5) arc (180:90:.3) -- ++ (0.6,0) 
          .. controls +(0.4,0) and +(-0.4,0) .. (0.8,-0.7) 
          -- ++ (.4,0);
          \drawdash ($(A3)+(0,-.2)$) -- (A3) -- ++ (0,.2) arc (180:90:.2) -- ++ (.3,0) -- ++ (.7,0) -- ++ (.5,0);
   \fill[white,draw=black] ($(O)+(-.8,0)$) circle (.42);  
  \draw ($(O)+(-.8,0)$) node {$#2$};
  \end{scope}
 }}

\def \Balphabeta(#1,#2) {
\tikeq{  \begin{scope}[scale=#1]
	        \coordinate (O) at (0,0); 
	        \coordinate (A2) at  ($(O)+(-.5,0)$) ;  
	        \coordinate (A1) at  ($(A2)+(-.6,0)$) ; 
	        \coordinate (A3) at  ($(A2)+(0,1)$); 
	        \coordinate (A4) at  ($(A1)+(0,1)$);  
	        \coordinate (B2) at  ($(O)-(-.5,0)$) ;  
	        \coordinate (B1) at  ($(B2)-(-.6,0)$) ;  
	        \coordinate (B3) at  ($(B2)+(0,1)$);  
	        \coordinate (B4) at  ($(B1)+(0,1)$);                
	        \drawdash  ($(A1)+(0,.2)$) -- (A1) -- ++ (0,-.5) arc (180:270:.3) -- ++ (1.6,0) arc (-90:0:.3) -- ($(B1)+(0,.2)$) ;
	        \drawdash  ($(A2)+(0,.2)$) -- (A2) -- ++ (0,-.2) arc (180:270:.2) -- ++ (.6,0.) arc (-90:-0:.2) -- ($(B2)+(0,.2)$);
	        \drawdash  ($(A4)+(0,-.2)$) -- (A4) -- ++ (0,.5) arc (180:90:.3) -- ++ (1.6,0) arc (90:0:.3) -- ($(B4)+(0,-.2)$);
	        \drawdash ($(A3)+(0,-.2)$) -- (A3) -- ++ (0,.2) arc (180:90:.2) -- ++ (.6,0) arc (90:0:.2) -- ($(B3)+(0,-.2)$);
   \fill[white,draw=black] ($(O)+(-.8,.5)$) circle (.42);  
  \draw ($(O)+(-.8,.5)$) node {$\al$};
    \fill[white,draw=black] ($(O)+(.8,.5)$) circle (.42);  
  \draw ($(O)+(.8,.5)$) node {$\beta$};
 \draw[glu] ($(O)+(0,-0.8)$) -- ($(O)+(0,#2)$) ; 
  \end{scope}
 }}
 
\def \Bbarone (#1) {
\tikeq{  \begin{scope}[scale=#1]
	        \coordinate (O) at (0,0); 
	        \coordinate (A2) at  ($(O)+(-.5,0)$) ;  
	        \coordinate (A1) at  ($(A2)+(-.6,0)$) ; 
	        \coordinate (A3) at  ($(A2)+(0,1)$); 
	        \coordinate (A4) at  ($(A1)+(0,1)$);  
	        \coordinate (B2) at  ($(O)-(-.5,0)$) ;  
	        \coordinate (B1) at  ($(B2)-(-.6,0)$) ;  
	        \coordinate (B3) at  ($(B2)+(0,1)$);  
	        \coordinate (B4) at  ($(B1)+(0,1)$);                
	        \draw[blue]  ($(A2)+(0,.2)$) -- (A2) -- ++ (0,-.2) arc (180:270:.2) -- ++ (.6,0.) arc (-90:-0:.2) -- ($(B2)+(0,.2)$);
	           \draw[brown] ($(A3)+(0,-.2)$) -- (A3) -- ++ (0,.2) arc (180:90:.2) -- ++ (.6,0) arc (90:0:.2) -- ($(B3)+(0,-.2)$);
	        \draw[qua]  ($(A4)+(0,-.2)$) -- (A4) -- ++ (0,.2) arc (0:90:.2) -- ++ (-.2,0) arc (90:180:.2) -- ++ (0,-1.4) arc (180:270:.2)  -- ++ (.2,0) arc (-90:0:.2) -- ($(A1)+(0,.2)$);
	         \draw[anti]  ($(B4)+(0,-.2)$) -- (B4) -- ++ (0,.2) arc (180:90:.2) -- ++ (.2,0) arc (90:0:.2) -- ++ (0,-1.4) arc (0:-90:.2)  -- ++ (-.2,0) arc (-90:-180:.2) -- ($(B1)+(0,.2)$);	        
   \fill[white,draw=black] ($(O)+(-.8,.5)$) circle (.42);  
  \draw ($(O)+(-.8,.5)$) node {$\al$};
    \fill[white,draw=black] ($(O)+(.8,.5)$) circle (.42);  
  \draw ($(O)+(.8,.5)$) node {$\beta$};
  \end{scope}
 }}
\def \Bbartwo (#1) {
\tikeq{  \begin{scope}[scale=#1]
	        \coordinate (O) at (0,0); 
	        \coordinate (A2) at  ($(O)+(-.5,0)$) ;  
	        \coordinate (A1) at  ($(A2)+(-.6,0)$) ; 
	        \coordinate (A3) at  ($(A2)+(0,1)$); 
	        \coordinate (A4) at  ($(A1)+(0,1)$);  
	        \coordinate (B2) at  ($(O)-(-.5,0)$) ;  
	        \coordinate (B1) at  ($(B2)-(-.6,0)$) ;  
	        \coordinate (B3) at  ($(B2)+(0,1)$);  
	        \coordinate (B4) at  ($(B1)+(0,1)$);               
	        \draw[anti]  ($(A1)+(0,.2)$) -- (A1) -- ++ (0,-.5) arc (180:270:.3) -- ++ (1.6,0) arc (-90:0:.3) -- ($(B1)+(0,.2)$) ;
	        \draw[blue]  ($(A2)+(0,.2)$) -- (A2) -- ++ (0,-.2) arc (180:270:.2) -- ++ (.6,0.) arc (-90:-0:.2) -- ($(B2)+(0,.2)$);
	        \draw[qua]  ($(A4)+(0,-.2)$) -- (A4) -- ++ (0,.5) arc (180:90:.3) -- ++ (1.6,0) arc (90:0:.3) -- ($(B4)+(0,-.2)$);
	        \draw[brown] ($(A3)+(0,-.2)$) -- (A3) -- ++ (0,.2) arc (180:90:.2) -- ++ (.6,0) arc (90:0:.2) -- ($(B3)+(0,-.2)$);
   \fill[white,draw=black] ($(O)+(-.8,.5)$) circle (.42);  
  \draw ($(O)+(-.8,.5)$) node {$\al$};
    \fill[white,draw=black] ($(O)+(.8,.5)$) circle (.42);  
  \draw ($(O)+(.8,.5)$) node {$\beta$};
  \end{scope}
 }}
\def \birdKtwo (#1) {
\tikeq{  \begin{scope}[scale=#1]
	        \coordinate (O) at (0,0); 
	        \coordinate (A2) at  ($(O)+(-.5,0)$) ;  
	        \coordinate (A1) at  ($(A2)+(-.6,0)$) ; 
	        \coordinate (A3) at  ($(A2)+(0,1)$); 
	        \coordinate (A4) at  ($(A1)+(0,1)$);               
	        \draw[blue]  ($(A2)+(0,.2)$) -- (A2) -- ++ (0,-.2) arc (180:270:.2) -- ++ (.3,0.);
	         \draw[brown] ($(A3)+(0,-.2)$) -- (A3) -- ++ (0,.2) arc (180:90:.2) -- ++ (.3,0) ;
	        \draw[qua]  ($(A4)+(0,-.2)$) -- (A4) -- ++ (0,.2) arc (0:90:.2) -- ++ (-.2,0) arc (90:180:.2) -- ++ (0,-1.4) arc (180:270:.2)  -- ++ (.2,0) arc (-90:0:.2) -- ($(A1)+(0,.2)$);              
   \fill[white,draw=black] ($(O)+(-.8,.5)$) circle (.42);  
  \draw ($(O)+(-.8,.5)$) node {$\al$};
  \end{scope}
 }
 = \frac{K_\al}{K_2} \ \ 
 \tikeq{  \begin{scope}[scale=#1]
	        \coordinate (O) at (0,0); 
	        \coordinate (A2) at  ($(O)+(-.5,0)$) ;  
	        \coordinate (A3) at  ($(A2)+(0,1)$);      
	        \draw[blue]  ($(A2)+(0,.5)$) -- (A2) -- ++ (0,-.2) arc (180:270:.2) -- ++ (.3,0.);
	         \draw[brown] ($(A3)+(0,-.5)$) -- (A3) -- ++ (0,.2) arc (180:90:.2) -- ++ (.3,0) ;
  \end{scope}
 }}

\def \Fierzthree {
2 \ \tikeq{
		\coordinate (O) at (0,0);
		\draw[qua] ($(O)+(0,0)$) -- ++(0.4,0) coordinate(A);
		\draw[qua] (A) -- ++(0.4,0);
		\draw[qua] ($(O)+(0,.5)$) -- ++(0.4,0)coordinate(B);
		\draw[qua] (B) -- ++(0.4,0);
		\draw[glu] (A) --(B);
} 
\ = \ - \ 
\tikeq{		
		\drawqua(0.25) (0,0) -- (0.7,0.5);  
		\drawqua(0.25) (0,0.5) -- (0.7,0);
}
\ + \ \frac{1}{\Nc}\ 
\tikeq{
		\drawqua(0.5) (0,0) -- (0.7,0);  
		\drawqua(0.5) (0,0.5) -- (0.7,0.5);	
}}

\newcommand{\vacuumamp}[2]{\tikeq{\begin{scope}[scale=#2]
 \draw[qua] (-0.5,0.4) -- (0,0.4); \draw[qua] (0,0.4) -- (0.5,0.4); 
 \draw (0.45,0.52) node {$\bm{\scriptstyle{q}}$}; 
\draw[glu][red] (0,-0.1) -- (0,0.4);
\draw[glu][red] (0,-0.1) -- (0.5,-0.1);
\draw (0.4,0.05) node {$\bm{\scriptstyle{-q}}$}; 
\draw (0.7,-0.2) node {$\scriptstyle{{{\scriptstyle p_4^+}} \simeq 0}$};
\draw[glu]  (-0.5,-0.1) -- (0,-0.1);
\draw (-0.2,0.2) node {$#1$}; 
\draw (-0.4,-0.25) node {$\scriptstyle{p_2}$}; 
\end{scope}
}}
 
\newcommand{\vacuumampblob}[2]{\tikeq{\begin{scope}[scale=#2]
\draw[qua] (-0.5,0.4) -- (0,0.4); \draw[qua] (0,0.4) -- (0.5,0.4); 
\draw[glu][red] (0,-0.1) -- (0,0.4);
\draw (-0.2,0.2) node {$#1$}; 
\draw[red,fill=red] (0,-0.1) circle (.1);
\end{scope}}}

\newcommand{\vacuum}[2]{\tikeqbis{\begin{scope}[scale=#2]
  \draw[qua] (-0.5,0.4) -- (0,0.4); \draw[qua] (0,0.4) -- (0.5,0.4); 
  \draw[anti] (-0.5,-0.4) -- (0,-0.4); \draw[anti] (0,-0.4) -- (0.5,-0.4); 
  \draw (-0.5,0.4) -- (-0.5,-0.4);  \draw (0.5,0.4) -- (0.5,-0.4);
 \draw[glu][red] (0,-0.4) -- (0,0.4) ;
 \draw (-0.2,0.2) node {$#1$}; 
 \draw[dashed] (-.8,0) -- (.8,0);
\draw[red,fill=red] (0,0) circle (.1);
 \end{scope}
 }}
 
\newcommand{\medium}[1]{\tikeqbis{\begin{scope}[scale=#1]
  \fill[color=blue!30!white,opacity=0.5] ($(-.4,-.6)$) rectangle (.4,.6);  
  \drawqua(0.3) (-0.8,0.4) -- (0,0.4); \drawqua(0.8) (0,0.4) -- (0.8,0.4); 
  \drawanti(0.3) (-0.8,-0.4) -- (0,-0.4); \drawanti(0.8) (0,-0.4) -- (0.8,-0.4); 
  \draw (-0.8,0.4) -- (-0.8,-0.4);  \draw (0.8,0.4) -- (0.8,-0.4);
 \draw[glu][red] (0,-0.4) -- (0,0.4) ;
  \draw (-0.2,0.2) node {$\bm{\scriptstyle{q}}$}; 
 \draw[dashed] (-1.2,0) -- (1.2,0);
 \draw[red,fill=red] (0,0) circle (.1);
 \end{scope}
 }}

\newcommand{\Cnkq}[2]{\tikeqbis{\begin{scope}[scale=#2]
  \fill[color=blue!30!white,opacity=0.5] (-1.3,-0.8) rectangle (1.4,1.2); 
  \drawqua(0.12) (-2,0.5) -- (0,0.5); \drawqua(0.9) (0,0.5) -- (2,0.5); 
  \drawanti(0.12) (-2,-0.5) -- (0,-0.5); \drawanti(0.9) (0,-0.5) -- (2,-0.5); 
  \draw (-2,0.5) -- (-2,-0.5);  \draw (2,0.5) -- (2,-0.5);
 \draw 
 (-1.,0) node{$\times$} 
 (-0.5,0) node {$\times$} 
 (0,0) node {$\times$} 
 (1.,0) node {$\times$} 
 (1.8,1.1) node {$\bm{\scriptstyle{k}}$}  (0.3,0.3) node {$\bm{\scriptstyle{q}}$} (-1.2,0.3) node {$\bm{\scriptstyle{\ell_1}}$} (-0.2,0.3) node {$\bm{\scriptstyle{\ell_i}}$} ; 
  \draw[glu] (-1,-0.5) -- (-1,0.5) ;
  \draw[glu] (-0.3,0.9) -- (-0.5,0); \draw[glu] (-0.5,0) -- (-0.7,0.5) ;
  \draw[glu] (0,-0.5) -- (0,0.9);
 \draw[glu][red] (0.5,-0.5) -- (0.5,0.5) ;
  \draw[glu] (0.85,-0.5) -- (1,0); \draw[glu] (1,0) -- (1.15,-0.5);
\draw[glu] (-1.6,.5) arc (180:90:.4) -- (2,0.9)  arc (90:0:.4) -- (2.4,-0.5)
arc (0:-180:.4); 
\draw[fill=white,opacity=#1] (1.6,-0.5) circle (0.1); 
\draw[dashed] (-2.2,0) -- (2.6,0);
\draw[red,fill=red] (0.5,0) circle (.1);
\end{scope}
}}

\newcommand{\Cnbisellipse}[1]{\tikeqbis{\begin{scope}[scale=#1]
  \fill[color=blue!30!white,opacity=0.5] (-1.2,-0.8) rectangle (1.3,1.2); 
  \drawqua(0.12) (-2,0.5) -- (0,0.5); \drawqua(0.92) (0,0.5) -- (2.1,0.5); 
  \drawanti(0.12) (-2,-0.5) -- (0,-0.5); \drawanti(0.92) (0,-0.5) -- (2.1,-0.5); 
  \draw (-2,0.5) -- (-2,-0.5);  \draw (2.1,0.5) -- (2.1,-0.5);
 \draw 
 (-1.,0) node{$\times$}  (-0.5,0) node {$\times$} (0,0) node {$\times$} 
 (1.,0) node {$\times$} 
 (1.8,1.1) node {$\bm{\scriptstyle{k}}$}  (1.6,0.3) node {$\bm{\scriptstyle{K}}$}
 (1.6,0.63) node {$\bm{\scriptstyle{-K}}$} 
 (-0.75,-0.25) node {$\bm{\scriptstyle{\ell_1}}$}
 (-0.2,-0.25) node {$\bm{\scriptstyle{\ell_i}}$} ; 
  \draw[glu] (-1,-0.5) -- (-1,0.5) ;
  \draw[glu] (-0.3,0.9) -- (-0.5,0); \draw[glu] (-0.5,0) -- (-0.7,0.3) ;
  \draw[glu] (0,-0.5) -- (0,0.9);
  \draw[glu][red] (1.5,-0.5) arc (180:90:.2) arc (-90:0:.2) 
-- (1.9,0.1) arc (0:90:.2) -- (-1,0.3) .. controls +(-0.3,0) and +(0.2,-0.2) .. (-1.5,0.5) ; 
 \draw[glu] (0.5,-0.5) -- (0.5,0.3) ;
 \draw[glu] (0.85,-0.5) -- (1,0); \draw[glu] (1,0) -- (1.15,-0.5);
\draw[glu] (-1.6,.5) arc (180:90:.4) -- (2,0.9)  arc (90:0:.4) -- (2.4,-0.5)
arc (0:-180:.35); 
\draw[fill=white,opacity=1] (1.7,-0.15) .. controls +(-0.1,0) and +(-0.1,0) .. (1.7,-0.65); 
\draw[fill=white,opacity=1] (1.7,-0.15) .. controls +(0.1,0) and +(0.1,0) .. (1.7,-0.65);
 \draw[dashed] (-2.2,0) -- (2.7,0);
\end{scope}
}}

\newcommand{\GammanKK}[1]{\tikeqbis{\begin{scope}[scale=#1]
  \fill[color=blue!30!white,opacity=0.5] (-1.2,-0.8) rectangle (1.3,1.2); 
  \drawqua(0.12) (-2,0.5) -- (0,0.5); \drawqua(0.92) (0,0.5) -- (2.1,0.5); 
  \drawanti(0.12) (-2,-0.5) -- (0,-0.5); \drawanti(0.92) (0,-0.5) -- (2.1,-0.5); 
  \draw (-2,0.5) -- (-2,-0.5);  \draw (2.1,0.5) -- (2.1,-0.5);
 \draw 
 (-1.,0) node{$\times$}  (-0.5,0) node {$\times$}  (0,0) node {$\times$} 
 (1.,0) node {$\times$} (2,1.1) node {$\bm{\scriptstyle{k}}$} (1.6,0.3) node {$\bm{\scriptstyle{K}}$}
 (1.6,0.63) node {$\bm{\scriptstyle{-K}}$} 
 (-0.75,-0.25) node {$\bm{\scriptstyle{\ell_1}}$}
 (-0.2,-0.25) node {$\bm{\scriptstyle{\ell_i}}$} ;
  \draw[glu] (-1,-0.5) -- (-1,0.5) ;
  \draw[glu] (-0.3,0.9) -- (-0.5,0); \draw[glu] (-0.5,0) -- (-0.7,0.3) ;
  \draw[glu] (0,-0.5) -- (0,0.9);
  \draw[glu][red] (1.5,-0.5) arc (180:90:.2) arc (-90:0:.2) 
-- (1.9,0.1) arc (0:90:.2) -- (-1,0.3) .. controls +(-0.3,0) and +(0.2,-0.2) .. (-1.5,0.5) ; 
 \draw[glu] (0.5,-0.5) -- (0.5,0.3) ;
 \draw[glu] (0.85,-0.5) -- (1,0); \draw[glu] (1,0) -- (1.15,-0.5);
\draw[glu] (-1.6,.5) arc (180:90:.4) -- (2,0.9)  arc (90:0:.4) -- (2.4,-0.5)
arc (0:-180:.35) -- (1.7,-0.3)  ; 
\draw[fill=white,opacity=1] (1.7,-0.3) circle (0.1); 
\draw[dashed] (-2.2,0) -- (2.7,0);
\end{scope}
}}

\newcommand{\GammazeroNorescatt}[2]{\tikeqbis{\begin{scope}[scale=#2]
 \drawqua(0.3) (-0.8,0.4) -- (0,0.4); \drawqua(0.8) (0,0.4) -- (0.8,0.4); 
  \drawanti(0.3) (-0.8,-0.4) -- (0,-0.4); \drawanti(0.8) (0,-0.4) -- (0.8,-0.4); 
  \draw (-0.8,0.4) -- (-0.8,-0.4);  \draw (0.8,0.4) -- (0.8,-0.4);
 \draw  (0.6,1) node {$\bm{\scriptstyle{k}}$};
 \draw[glu][red] (0,-0.4) -- (0,0.4) ;
  \draw[glu] (-0.4,0.4) arc (180:90:.4) -- (0.8,0.8) arc  (90:0:.4) -- (1.2,-0.4) 
  arc  (0:-180:.4) ; 
 \draw[fill=white,opacity=1] (0.4,-0.4) circle (0.1); 
 \draw[fill=white,opacity=#1] (-0.4,0.4) circle (0.1); 
 \draw[dashed] (-1,0) -- (1.4,0);
 \draw[red,fill=red] (0,0) circle (.1);
 \end{scope}
}}

\newcommand{\qtoqgtchannel}[1]{\tikeqbis{\begin{scope}[yshift=-2mm,scale=#1]
\drawqua(0.15) (-0.5,0.5) -- (1,0.5);
\draw (1,0.65) node {$\bm{\scriptstyle{p_3}}$} ; 
\draw[glu][red] (-0.1,0.5) .. controls +(0.2,-0.2) and +(-0.3,0) .. (0.5,0.3) -- (1,0.3);
\draw (1,0.15) node {$\bm{\scriptstyle{p_4}}$} ; 
\draw[glu](0.5,0) -- (0.5,0.3) ;
\end{scope}
}}

\newcommand{\qtoqguchannel}[1]{\tikeqbis{\begin{scope}[yshift=-2mm,scale=#1]
\drawqua(0.15) (-0.5,0.5) -- (1,0.5); 
\draw[glu][red] (-0.1,0.5) .. controls +(0.2,-0.2) and +(-0.3,0) .. (0.5,0.3) -- (1,0.3); 
\draw[glu](0.5,0) -- (0.5,0.5) ;
\end{scope}
}}

\newcommand{\qtoqgschannel}[1]{\tikeqbis{\begin{scope}[yshift=-2mm,scale=#1]
\drawqua(0.15) (-0.5,0.5) -- (1,0.5); 
\draw[glu][red] (0.4,0.5) .. controls +(0.2,-0.2) and +(-0.3,0) .. (0.9,0.3) -- (1,0.3) ; 
\draw[glu](0,0) -- (0,0.5) ;
\end{scope}
}}

\newcommand{\blueblob}[1]{\tikeqbis{\begin{scope}[scale=#1,yshift=-5]
  \coordinate (O) at (0,0);
  \coordinate (A) at ($(O)+(0,-0.3)$);
  \coordinate (B) at ($(O)+(0,0.3)$); 
  \coordinate (C) at ($(O)+(0,0.6)$); 
  \draw[anti] ($(A)-(0.5,0)$) -- ($(A)-(.2,0)$); 
  \draw[anti] ($(A)+(.2,0)$) -- ($(A)+(0.5,0)$);
  \draw[qua] ($(B)-(0.5,0)$) -- ($(B)-(.2,0)$); 
  \draw[qua] ($(B)+(.2,0)$) -- ($(B)+(0.5,0)$);
  \draw[glu] ($(C)-(0.5,0)$) -- ($(C)+(0.5,0)$) ;
  \draw ($(C)+(0.4,0.2)$) node {$\bm{\scriptstyle{k}}$} ;
  \fill[color=blue!15!white,opacity=1] ($(A)-(.2,.15)$) rectangle ($(C)+(.2,.2)$);
 \end{scope}
}}

\newcommand{\rescatta}[3]{\tikeq{\begin{scope}[scale=#3,yshift=-5]
  \coordinate (O) at (0,0);
  \coordinate (A) at ($(O)+(0,-0.3)$);
  \coordinate (B) at ($(O)+(0,0.3)$); 
  \coordinate (C) at ($(O)+(0,0.6)$); 
  \drawanti(0.35) ($(A)-(0.5,0)$) -- (A); \draw (A) -- ($(A)+(0.5,0)$);
  \drawqua(0.35) ($(B)-(0.5,0)$) -- (B); \draw (B) -- ($(B)+(0.5,0)$);
  \draw[glu] ($(C)-(0.5,0)$) -- ($(C)+(0.5,0)$);
  \draw ($(C)+(-0.5,0.17)$) node {$#1$} ;
  \draw ($(C)+(0.4,0.2)$) node {$#2$} ;
  \draw[glu]  (A) -- (C);
  \draw[color=blue] (O) node {$\times$};
 \end{scope}
}}

\newcommand{\rescattb}[3]{\tikeq{\begin{scope}[scale=#3,yshift=-5]
  \coordinate (O) at (0,0);
  \coordinate (A) at ($(O)+(0,-0.3)$);
  \coordinate (B) at ($(O)+(0,0.3)$); 
  \coordinate (C) at ($(O)+(0,0.6)$); 
 \drawanti(0.35) ($(A)-(0.5,0)$) -- (A); \draw (A) -- ($(A)+(0.5,0)$);
  \drawqua(0.35) ($(B)-(0.5,0)$) -- (B); \draw (B) -- ($(B)+(0.5,0)$);
  \draw[glu] ($(C)-(0.5,0)$) -- ($(C)+(0.5,0)$);
  \draw ($(C)+(-0.5,0.17)$) node {$#1$} ;
  \draw ($(C)+(0.4,0.2)$) node {$#2$} ;
  \draw[glu]  ($(C)+(-0.2,0)$) -- (O) ; \draw[glu] (O) --  ($(B)+(0.2,0)$); 
  \draw[color=blue] (O) node {$\times$};
 \end{scope}
}}
  
  \newcommand{\rescattc}[3]{\tikeq{\begin{scope}[scale=#3,yshift=-5]
  \coordinate (O) at (0,0);
  \coordinate (A) at ($(O)+(0,-0.3)$);
  \coordinate (B) at ($(O)+(0,0.3)$); 
  \coordinate (C) at ($(O)+(0,0.6)$); 
  \drawanti(0.35) ($(A)-(0.5,0)$) -- (A); \draw (A) -- ($(A)+(0.5,0)$);
  \drawqua(0.35) ($(B)-(0.5,0)$) -- (B); \draw (B) -- ($(B)+(0.5,0)$);
  \draw[glu] ($(C)-(0.5,0)$) -- ($(C)+(0.5,0)$);
  \draw ($(C)+(-0.5,0.2)$) node {$#1$} ;
  \draw ($(C)+(0.4,0.2)$) node {$#2$} ;
  \draw[glu]  ($(C)+(-0.2,0)$) -- (O) ;  \draw[glu]  (O) --  ($(C)+(0.2,0)$); 
  \draw[color=blue] (O) node {$\times$};
 \end{scope}
}}

\newcommand{\rescattd}[3]{\tikeq{\begin{scope}[scale=#3,yshift=-5]
  \coordinate (O) at (0,0);
  \coordinate (A) at ($(O)+(0,-0.3)$);
  \coordinate (B) at ($(O)+(0,0.3)$); 
  \coordinate (C) at ($(O)+(0,0.6)$); 
  \drawanti(0.35) ($(A)-(0.5,0)$) -- (A); \draw (A) -- ($(A)+(0.5,0)$);
  \drawqua(0.35) ($(B)-(0.5,0)$) -- (B); \draw (B) -- ($(B)+(0.5,0)$);
  \draw[glu] ($(C)-(0.5,0)$) -- ($(C)+(0.5,0)$);
  \draw ($(C)+(-0.5,0.2)$) node {$#1$} ;
  \draw ($(C)+(0.4,0.2)$) node {$#2$} ;
  \draw[glu]  ($(A)+(-0.2,0)$) -- (O) ; \draw[glu] (O) --  ($(A)+(0.2,0)$); 
  \draw[color=blue] (O) node {$\times$};
 \end{scope}
}}

\newcommand{\rescatte}[3]{\tikeq{\begin{scope}[scale=#3,yshift=-5]
  \coordinate (O) at (0,0);
  \coordinate (A) at ($(O)+(0,-0.3)$);
  \coordinate (B) at ($(O)+(0,0.3)$); 
  \coordinate (C) at ($(O)+(0,0.6)$); 
  \drawanti(0.35) ($(A)-(0.5,0)$) -- (A); \draw (A) -- ($(A)+(0.5,0)$);
  \drawqua(0.35) ($(B)-(0.5,0)$) -- (B); \draw (B) -- ($(B)+(0.5,0)$);
  \draw[glu] ($(C)-(0.5,0)$) -- ($(C)+(0.5,0)$);
  \draw ($(C)+(-0.5,0.2)$) node {$#1$} ;
  \draw ($(C)+(0.4,0.2)$) node {$#2$} ;
  \draw[glu]  ($(B)+(-0.2,0)$) -- (O);  \draw[glu]  (O) --  ($(B)+(0.2,0)$); 
  \draw[color=blue] (O) node {$\times$};
 \end{scope}
}}

\newcommand{\rescattf}[3]{\tikeq{\begin{scope}[scale=#3,yshift=-5]
  \coordinate (O) at (0,0);
  \coordinate (A) at ($(O)+(0,-0.3)$);
  \coordinate (B) at ($(O)+(0,0.3)$); 
  \coordinate (C) at ($(O)+(0,0.6)$); 
  \drawanti(0.35) ($(A)-(0.5,0)$) -- (A); \draw (A) -- ($(A)+(0.5,0)$);
  \drawqua(0.35) ($(B)-(0.5,0)$) -- (B); \draw (B) -- ($(B)+(0.5,0)$);
  \draw[glu] ($(C)-(0.5,0)$) -- ($(C)+(0.5,0)$);
  \draw ($(C)+(-0.5,0.2)$) node {$#1$} ;
  \draw ($(C)+(0.4,0.2)$) node {$#2$} ;
  \draw[glu]  ($(A)$) -- (O) --  ($(B)$); 
  \draw[color=blue] (O) node {$\times$};
 \end{scope}
}}

\newcommand{\HardRescatta}[1]{\tikeq{\begin{scope}[scale=#1,yshift=-5]
\drawqua(0.12) (-0.5,0.5) -- (1.5,0.5); 
\draw[glu](0.5,0.3) -- (0.5,0) ; \draw[glu](0.5,0) -- (0.5,-0.3) ;
\drawanti(0.1) (-0.5,-0.5) -- (1.5,-0.5); 
\draw[glu][red] (-0.1,-0.5) .. controls +(0.2,0.2) and +(-0.3,0) .. (0.5,-0.3) -- (1.1,-0.3) arc (-90:0:.2) 
-- (1.3,0.1) arc (0:90:.2) -- (1.1,0.3) -- (0.5,0.3) .. controls +(-0.3,0) and +(0.2,-0.2) .. (-0.1,0.5) ; 
\draw  (-0.5,0.5) --  (-0.5,-0.5);  \draw  (1.5,0.5) --  (1.5,-0.5); 
\draw[dashed] (-0.7,0) -- (1.7,0);
 \end{scope}
}}

\newcommand{\HardRescattb}[1]{\tikeq{\begin{scope}[scale=#1,yshift=-5]
\drawqua(0.12) (-0.5,0.5) -- (1.5,0.5); 
\draw[glu](0.5,0.5) -- (0.5,0) ; \draw[glu](0.5,0) -- (0.5,-0.5) ;
\drawanti(0.1) (-0.5,-0.5) -- (1.5,-0.5); 
\draw[glu][red] (-0.1,-0.5) .. controls +(0.2,0.2) and +(-0.3,0) .. (0.5,-0.3) -- (1.1,-0.3) arc (-90:0:.2) 
-- (1.3,0.1) arc (0:90:.2) -- (1.1,0.3) -- (0.5,0.3) .. controls +(-0.3,0) and +(0.2,-0.2) .. (-0.1,0.5) ; 
\draw  (-0.5,0.5) --  (-0.5,-0.5);  \draw  (1.5,0.5) --  (1.5,-0.5); 
\draw[dashed] (-0.7,0) -- (1.7,0);
 \end{scope}
}}

\newcommand{\HardRescattc}[1]{\tikeq{\begin{scope}[scale=#1,yshift=-5]
\drawqua(0.12) (-0.5,0.5) -- (1.5,0.5); 
\draw[glu] (0.3,0.5) -- (0.3,0) ; \draw[glu](0.3,0) -- (0.3,-0.5) ;
\drawanti(0.1) (-0.5,-0.5) -- (1.5,-0.5); 
\draw[glu][red] (0.6,-0.5) arc (180:90:.2) -- (1.1,-0.3) arc (-90:0:.2) 
-- (1.3,0.1) arc (0:90:.2) -- (1.1,0.3) -- (0.8,0.3) arc (-90:-180:.2) ; 
\draw  (-0.5,0.5) --  (-0.5,-0.5);  \draw  (1.5,0.5) -- (1.5,-0.5); 
\draw[dashed] (-0.7,0) -- (1.7,0);
 \end{scope}
}}

\newcommand{\HardRescattd}[1]{\tikeq{\begin{scope}[scale=#1,yshift=-5]
\drawqua(0.12) (-0.5,0.5) -- (1.5,0.5); 
\draw[glu](0.5,0.3) -- (0.5,0) ; \draw[glu](0.5,0) -- (0.5,-0.5) ;
\drawanti(0.1) (-0.5,-0.5) -- (1.5,-0.5); 
\draw[glu][red] (-0.1,-0.5) .. controls +(0.2,0.2) and +(-0.3,0) .. (0.5,-0.3) -- (1.1,-0.3) arc (-90:0:.2) 
-- (1.3,0.1) arc (0:90:.2) -- (1.1,0.3) -- (0.5,0.3) .. controls +(-0.3,0) and +(0.2,-0.2) .. (-0.1,0.5) ; 
\draw  (-0.5,0.5) --  (-0.5,-0.5);  \draw  (1.5,0.5) --  (1.5,-0.5); 
\draw[dashed] (-0.7,0) -- (1.7,0);
 \end{scope}
}}

\newcommand{\HardRescatte}[1]{\tikeq{\begin{scope}[scale=#1,yshift=-5]
\drawqua(0.12) (-0.5,0.5) -- (1.5,0.5); 
\draw[glu](0.5,0.5) -- (0.5,0) ; \draw[glu](0.5,0) -- (0.5,-0.5) ;
\drawanti(0.1) (-0.5,-0.5) -- (1.5,-0.5); 
\draw[glu][red] (0.8,-0.5) arc (180:90:.2) -- (1.1,-0.3) arc (-90:0:.2) 
-- (1.3,0.1) arc (0:90:.2) -- (1.1,0.3) -- (0.5,0.3) .. controls +(-0.3,0) and +(0.2,-0.2) .. (-0.1,0.5) ; 
\draw  (-0.5,0.5) --  (-0.5,-0.5);  \draw  (1.5,0.5) --  (1.5,-0.5); 
\draw[dashed] (-0.7,0) -- (1.7,0);
 \end{scope}
}}

\newcommand{\HardRescattf}[1]{\tikeq{\begin{scope}[scale=#1,yshift=-5]
\drawqua(0.12) (-0.5,0.5) -- (1.5,0.5); 
\draw[glu](0.5,0.3) -- (0.5,0) ; \draw[glu](0.5,0) -- (0.5,-0.5) ;
\drawanti(0.1) (-0.5,-0.5) -- (1.5,-0.5); 
\draw[glu][red] (0.8,-0.5) arc (180:90:.2) -- (1.1,-0.3) arc (-90:0:.2) 
-- (1.3,0.1) arc (0:90:.2) -- (1.1,0.3) -- (0.5,0.3) .. controls +(-0.3,0) and +(0.2,-0.2) .. (-0.1,0.5) ; 
\draw  (-0.5,0.5) --  (-0.5,-0.5);  \draw  (1.5,0.5) --  (1.5,-0.5); 
\draw[dashed] (-0.7,0) -- (1.7,0); 
 \end{scope}
}}

\newcommand{\Gammazerobis}[1]{\tikeqbis{\begin{scope}[scale=#1]
  \drawqua(0.2) (-1,0.5) -- (0,0.5); \drawqua(0.9) (0,0.5) -- (2.1,0.5); 
  \drawanti(0.2) (-1,-0.5) -- (0,-0.5); \drawanti(0.9) (0,-0.5) -- (2.1,-0.5); 
  \draw (-1,0.5) -- (-1,-0.5);  \draw (2.1,0.5) -- (2.1,-0.5);
  \draw[glu][red] (1,-0.5) arc (180:90:.2) -- (1.7,-0.3) arc (-90:0:.2) 
-- (1.9,0.1) arc (0:90:.2) -- (0.5,0.3) .. controls +(-0.3,0) and +(0.2,-0.2) .. (0,0.5) ; 
\draw[glu] (0.5,-0.5) -- (0.5,0.3) ;
\draw[glu] (-0.6,.5) arc (180:90:.4) -- (2,0.9)  arc (90:0:.4) -- (2.4,-0.5)
arc (0:-180:.35) -- (1.7,-0.3) ; 
 \draw[fill=white,opacity=1] (-0.6,0.5) circle (0.1); 
 \draw[fill=white,opacity=1] (1.7,-0.3) circle (0.1); 
 \draw[dashed] (-1.2,0) -- (2.65,0);
 \end{scope}
}}

\newcommand{\MMgluon}[1]{\tikeq{\begin{scope}[scale=#1,yshift=-5]
  \drawqua(0.2) (-1,0.5) -- (0,0.5); \drawqua(0.9) (0,0.5) -- (2.1,0.5); 
  \drawanti(0.2) (-1,-0.5) -- (0,-0.5); \drawanti(0.9) (0,-0.5) -- (2.1,-0.5); 
  \draw (-1,0.5) -- (-1,-0.5);  \draw (2.1,0.5) -- (2.1,-0.5);
  \draw[glu][red] (0.7,-0.5) arc (180:90:.2) -- (1.7,-0.3) arc (-90:0:.2) 
-- (1.9,0.1) arc (0:90:.2) -- (0.5,0.3) .. controls +(-0.3,0) and +(0.2,-0.2) .. (0.2,0.5) ; 
\draw[glu] (0.5,-0.5) -- (0.5,0.3) ;
  \fill[color=gray!30!white,opacity=1] (0.1,-0.6) rectangle (0.9,-0.2); 
  \fill[color=gray!30!white,opacity=1] (0.1,0.2) rectangle (0.9,0.6);  
\draw (0.5,0.4) node {$\scriptstyle{{\cal M}}$}  ;
\draw (0.5,-0.4) node {$\scriptstyle{{\cal M}^*}$}  ;
\draw[glu] (-0.6,.5) arc (180:90:.4) -- (2,0.9)  arc (90:0:.4) -- (2.4,-0.5)
arc (0:-180:.35) -- (1.7,-0.3) ; 
 \draw[fill=white,opacity=1] (-0.6,0.5) circle (0.1); 
 \draw[fill=white,opacity=1] (1.7,-0.3) circle (0.1); 
 \draw[dashed] (-1.2,0) -- (2.6,0);
 \end{scope}
}}

\newcommand{\MMquark}[1]{\tikeq{\begin{scope}[scale=#1,yshift=-5]
  \drawqua(0.2) (-1,0.5) -- (0,0.5); \drawqua(0.9) (0,0.5) -- (2.1,0.5); 
  \drawanti(0.2) (-1,-0.5) -- (0,-0.5); \drawanti(0.9) (0,-0.5) -- (2.1,-0.5); 
  \draw (-1,0.5) -- (-1,-0.5);  \draw (2.1,0.5) -- (2.1,-0.5); 
  \draw[glu][red] (0.7,-0.5) arc (180:90:.2) -- (1.7,-0.3) arc (-90:0:.2) 
-- (1.9,0.1) arc (0:90:.2) -- (0.5,0.3) .. controls +(-0.3,0) and +(0.2,-0.2) .. (0.2,0.5) ; 
\draw[glu] (0.5,-0.5) -- (0.5,0.3) ;
  \fill[color=gray!30!white,opacity=1] (0.1,-0.6) rectangle (0.9,-0.2); 
  \fill[color=gray!30!white,opacity=1] (0.1,0.2) rectangle (0.9,0.6);  
\draw (0.5,0.4) node {$\scriptstyle{{\cal M}}$}  ;
\draw (0.5,-0.4) node {$\scriptstyle{{\cal M}^*}$}  ;
\draw[glu] (-0.6,.5) arc (180:90:.4) -- (2,0.9)  arc (90:0:.4) -- (2.4,-0.5)
arc (0:-180:.35) ; 
 \draw[fill=white,opacity=1] (-0.6,0.5) circle (0.1); 
 \draw[fill=white,opacity=1] (1.7,-0.5) circle (0.1); 
 \draw[dashed] (-1.2,0) -- (2.6,0);
 \end{scope}
}}

\newcommand{\ggtoggschannel}[1]{\tikeq{\begin{scope}[scale=#1]
        \coordinate (O) at (0,0);
        \draw[glu] ($(O)+(0,.5)$) -- ++(.3,-.25);
        \draw[glu] (O) -- ++(.3,.25);
        \draw[glu] ($(O)+(.3,.25)$) -- ++(.5,0);
        \draw[glu] ($(O)+(.8,.25)$) -- ++(.3,.25);
        \draw[glu] ($(O)+(.8,.25)$) -- ++(.3,-.25);
 \end{scope}
}}

\newcommand{\ggtoggtchannel}[1]{\tikeq{\begin{scope}[scale=#1]
		\coordinate (O) at (0,0);
		\draw[glu] ($(O)+(0,0)$) -- ++(0.5,.1) coordinate(A);
		\draw[glu] (A) -- ++(0.5,-.1);
		\draw[glu] ($(O)+(0,.7)$) -- ++(0.5,-.1)coordinate(B);
		\draw[glu] (B) -- ++(0.5,.1);
		\draw[glu] (A) --(B);
\end{scope}
}}

\usepackage{tikz}

\usetikzlibrary{patterns}
\usetikzlibrary{decorations.pathmorphing,decorations.markings,trees,calc,shapes.geometric}
\tikzset{
  qua/.style={postaction={decorate},decoration={markings,mark=at position .55 with {\large\arrow{>}}}},
  anti/.style={postaction={decorate},decoration={markings,mark=at position .55 with {\large\arrow{<}}}},
  glu/.style={draw = black, decorate,decoration={coil,amplitude=1.4pt, segment length=2.6pt}} ,
  glu2/.style={draw = blue} ,
  vertical align/.style={baseline=-.5*(height("$+$")-depth("$+$"))},
  every picture/.style={vertical align}
}

\title{Coherent gluon radiation: beyond leading-log accuracy}

\author[a,b]{Greg Jackson,}
\author[b]{St\'ephane Peign\'e,}
\author[b,c]{Kazuhiro Watanabe} 

\affiliation[a]{Institute for Nuclear Theory, Box 351550, University of Washington, Seattle, WA 98195-1550, United States}
\affiliation[b]{SUBATECH UMR 6457 (IMT Atlantique, Universit\'e de Nantes, IN2P3/CNRS), 4 rue Alfred Kastler, 44307 Nantes, France}
\affiliation[c]{Faculty of Science and Technology, Seikei University, Musashino, Tokyo 180-8633, Japan} 

\emailAdd{jackson@subatech.in2p3.fr}
\emailAdd{peigne@subatech.in2p3.fr}
\emailAdd{kazuhiro-watanabe@st.seikei.ac.jp} 

\abstract{
Results are presented for the medium-induced, soft coherent radiation spectrum for all $2\to 2$ partonic channels in QCD, at leading-order in $\al_s$ but beyond leading logarithmic accuracy. The general formula is valid in the full kinematic range of the underlying process, and reduces to previous results in special cases. The soft gluon radiation spectrum is expressed in terms of the {\it color density matrix} specific to each channel, quantifying the entanglement between the color components of the $2 \to 2$ production amplitude. Beyond the leading logarithm, the spectrum depends explicitly on the off-diagonal elements of this matrix, owing to the soft gluon's ability to probe the internal color structure of the parton pair. 
}

\keywords{perturbative QCD; parton energy loss; coherent radiation.}

\begin{document} 

\begin{flushright}
INT-PUB-23-050\\
April 2024
\end{flushright}

\maketitle
\setcounter{footnote}{0}
\renewcommand{\thefootnote}{\arabic{footnote}}
\interfootnotelinepenalty=10000

\newpage

\section{Introduction}
\label{sec:intro}

The average energy lost through gluon radiation by a high-energy parton as it traverses a nuclear medium, and suffers small angle deflection 
(in the nuclear medium rest frame) due to multiple elastic scatterings,
is proportional to the incoming parton's energy, ${\Delta E \propto E}$~\cite{Arleo:2010rb}. This behaviour is a telltale sign of {\it fully coherent energy loss} (FCEL), the medium-induced energy loss mechanism expected in all perturbative QCD (pQCD) processes with a fast incoming parton and a colorful final state, which final state can be  a single parton~\cite{Arleo:2010rb,Peigne:2014uha,Munier:2016oih} or a composite parton system~\cite{Liou:2014rha,Peigne:2014rka}. FCEL stems from an interference between the gluon emission amplitudes off initial and final states, and the fully coherent regime is characterized by large induced gluon formation times $\tf$ relative to the medium (target) size $L\,$, $\tf \sim \omega/ k_\perp^2 \gg L\,$ (with $\omega$ and $k_\perp$ the energy and transverse momentum of the induced gluon, respectively). 
Although the coherent radiation thus views the target nucleus as a single effective scatterer, the ($k_\perp$-integrated) radiation spectrum ${\rm d}I/{\rm d}\omega \equiv {\rm d} \sigma_{\rm rad} / (\sigma_{\rm el} \, {\rm d}\omega)$ does depend on $L$ via the accumulated transverse momentum (or $p_\perp$-broadening) suffered by the fast parton system when crossing the target. 
At large $E$, FCEL surpasses parton energy loss in the Landau--Pomeranchuk--Migdal (LPM) regime which has a milder $E$-dependence \cite{Baier:1996sk,Baier:1996kr,Zakharov:1996fv,Zakharov:1997uu,Baier:1998kq}. 
FCEL has been shown to be crucial in explaining quarkonium~\cite{Arleo:2012hn,Arleo:2012rs,Arleo:2013zua}, light hadron~\cite{Arleo:2020eia,Arleo:2020hat} and heavy meson~\cite{Arleo:2021bpv,Arleo:2021krm} nuclear suppression in proton-nucleus (pA) collisions. 
It has also been shown to be an important cold nuclear matter effect for quarkonium production in nucleus-nucleus (AA) collisions~\cite{Arleo:2014oha}. 

In these phenomenological studies, the FCEL effect was estimated using leading-order pQCD, 
considering some $2\to 1$~\cite{Arleo:2012hn,Arleo:2012rs,Arleo:2013zua} 
or $2\to2$~\cite{Arleo:2020eia,Arleo:2020hat,Arleo:2021bpv,Arleo:2021krm} processes,\footnote{\label{numbers}
Previous studies referred to $2 \to n$ processes as `$1 \to n$ forward processes', emphasizing the identity of the incoming parton from the projectile. Presently, we adopt the $2\to n$ terminology, with the convention that the first mentioned incoming parton is from the projectile and the second is from the target. This distinguishes,\ e.g. $qg \to qg$ from $gq \to qg\,$ (which suffer FCEL differently).
}
and in the latter case by assuming that the induced radiation does not resolve individual color charges but sees the final parton pair effectively as a pointlike color charge. In this limit, referred to as leading-logarithmic (LL), the FCEL spectrum ${\rm d}I/{\rm d}\omega$ exhibits a large kinematic logarithm (see first term of \eq{Lxi-approx}), and a color dependence fully encoded in the `global' Casimir charge $C_\al$, where $\al$ labels an allowed irreducible representation (irrep) of the final parton pair (or more generally parton system)~\cite{Peigne:2014rka}. 
Thus, it was possible in refs.~\cite{Arleo:2020eia,Arleo:2020hat,Arleo:2021bpv,Arleo:2021krm} 
to implement FCEL as for $2 \to 1$ processes~\cite{Arleo:2012hn,Arleo:2012rs,Arleo:2013zua}, 
by relating the differential cross section $\dd \sigma/\dd E$ in pA collisions to that in pp collisions by means of a certain probability distribution, or `quenching weight', accounting for the incurred energy loss $\omega$. (The quenching weight can be constructed from the process-dependent radiation
spectrum $\dd I / \dd x\,$, where $x \simeq \omega / E$ is the fractional energy loss.) 
In refs.~\cite{Arleo:2020eia,Arleo:2020hat,Arleo:2021bpv,Arleo:2021krm}, the logarithm of the FCEL spectrum was checked to be quite large ($\sim 3 ... 4$) for the typical kinematic configuration of parton pair production, namely, for $\xi \sim \xibar \sim \frac12$ (with $\xi$ and $\xibar \equiv 1-\xi$ the light-cone longitudinal momentum fractions of the final partons w.r.t.~the incoming projectile parton), justifying {\em a posteriori} the LL approximation.  

However, formalizing the role of FCEL in the phenomenology of nuclear collisions will require {\it systematically} implementing FCEL in pQCD calculations of hadron production cross sections. 
To that end, extending the accuracy of the FCEL spectrum beyond LL is imperative. 
This is the first step before determining the quenching weight associated to this improved spectrum, 
and eventually convoluting the quenching weight with the pp cross section in the full phase space of the final pair, $0 \leq \xi \leq 1$.\footnote{We will consider $2 \to 2$ parton processes with {\it massless} initial partons, as motivated in section~\ref{sec:setup}. With this condition, $0 \leq \xi \leq 1$ is equivalent to $0 \leq m^2 - t \leq s$, with $s$ and $t$ Mandelstam variables of the $2 \to 2$ process and $m$ the mass of the equally massive final partons. } 
This would make a number of model assumptions in previous LL studies unnecessary, hence improving the accuracy of FCEL estimations. More importantly, and as argued previously~\cite{Arleo:2020hat,Arleo:2021bpv}, {\it first} implementing FCEL in pQCD calculations and {\it then} performing nuclear PDF (nPDF) global fit analyses (using the Hessian method~\cite{Paukkunen:2014zia} or reweighting techniques~\cite{Giele:1998gw,Ball:2011gg}), would allow to justify the use of hadron pA production data (which are sensitive to FCEL) in global fits, and thus to extract more reliable (and possibly more precise) nPDF sets. 
The main purpose of our study is to carry out the first stage of the aforementioned programme, namely, to derive the FCEL spectrum beyond LL for all $2\to 2$ parton processes (with massless initial particles).  

Independently of its interest for phenomenology, calculating the FCEL spectrum beyond LL for $2 \to 2$ processes is an interesting task in its own right, which highlights several theoretical aspects of gluon radiation in the coherent regime. Firstly, the soft induced radiation may probe individual color charges of the final parton pair (decomposed over available irreps as $a \otimes b = \sum_\al  {\rm R}_\al$) and thus allow {\it color transitions}\footnote{%
The exact meaning of color transitions will come later, see the paragraphs following eqs.~\eq{TuTt}.}
between different irreps, $\al \leftrightarrow \beta$, giving the spectrum a non-diagonal matrix structure in the space of irreps. The physics is thus richer than the LL limit, whereby each $\al$ is associated to a partial spectrum ${\rm d}I_\al / {\rm d} x$ contributing with a certain probability $\rho_\al$ to the full spectrum ${\rm d}I/{\rm d} x\,$~\cite{Peigne:2014rka}. 
Secondly, the FCEL spectrum can be {\it negative}, as previously found for the $qg \to q$ process~\cite{Peigne:2014uha,Munier:2016oih} and similarly for $qg \to qg$ (in the LL approximation) with a {\it color triplet} final $qg$ pair~\cite{Peigne:2014rka} (which has non-negligible effects in phenomenology~\cite{Arleo:2020eia,Arleo:2020hat}). Our study shows that this is not an exceptional case: ${\rm d}I/{\rm d}x < 0$ is possible for other $2\to 2$ processes, in a given range of the kinematical parameters $x$ and $\xi$. A negative FCEL, or fully coherent {\it energy gain} (FCEG),\footnote{%
Although the acronym FCEL is then a slight misnomer in general, 
we shall continue to use it as an umbrella term for both energy loss/gain situations.} 
is an interesting feature of coherent induced radiation, allowed by first principles (see~\cite{Peigne:2014uha} for a physical interpretation of FCEG). 
Lastly, we also stress that our beyond-LL calculation is done for the $\sun$ gauge group, 
with $\Nc \geq 3\,$. In particular, in the LL limit our results reproduce those of ref.~\cite{Peigne:2014rka}, where $\Nc \gg 1$ was assumed and the results for finite $\Nc$ {\it conjectured} on physical grounds. Our study thus proves this conjecture. 

The paper is organized as follows. 
In section~\ref{sec:2to2}, after a preliminary discussion of the kinematic setup and of the color structure of $2\to 2$ parton processes, we derive the FCEL spectrum beyond LL for each process. 
Of particular importance is the density matrix $\Phi_{\al\beta}$ defined in eq.~\eq{Phi-def},
which is constructed by decomposing the tree-level amplitude ${\cal M}$ into its available color states. This matrix is a gauge invariant, dimensionless function of $\xi\,$, and is given explicitly for each process in appendix~\ref{app:A}. Our main result for the FCEL spectrum ${\rm d}I/{\rm d}x$ is stated formally in the equivalent expressions \eq{master-spectrum} and~\eq{master-spectrum-2}, with details on the derivation of \eq{master-spectrum} relegated to appendix~\ref{app:master-spec} (where the spectrum is in fact derived more generally for $2\to \np$ processes). This is followed by an elaboration of the color matrices $\B_{\al\beta}$ and $\Bbar_{\al\beta}\,$, which are listed explicitly in appendix~\ref{app:BBbar} for each process.
We stress that the result~\eq{master-spectrum-2} is valid in the full kinematic range of the underlying reaction, and for any finite number of colors $\Nc \geq 3$. In section~\ref{sec:matching}, we explain how the previously obtained spectra for $2 \to 1$ processes~\cite{Arleo:2010rb,Peigne:2014uha,Munier:2016oih} and $2 \to 2$ processes in the logarithmic approximation~\cite{Liou:2014rha,Peigne:2014rka}, which turn out to be diagonal in a particular color basis, can be obtained from \eq{master-spectrum-2}. As a simple illustration, the case of quark-quark scattering is detailed in section~\ref{sec:illustration}. Finally, in section~\ref{sec:conclusion}, we briefly summarize our work, and present some phenomenological perspective while emphasizing the broad validity domain of our results in terms of kinematical variables in the c.m.~frame.

\section{FCEL spectrum beyond leading logarithm in $2\to 2$ processes}
\label{sec:2to2}

\subsection{Setup for underlying hard partonic process} 
\label{sec:setup}

Let us consider a general $2 \to 2$ partonic process arising when an incoming fast parton crosses a nuclear target,\footnote{%
    There are various experimental situations in which this may occur, like pA or AA collisions. In the AA case, one can view the $2\to 2$ process with either nucleus as the target and the fast incoming parton originating from the other nucleus. The equivalent pictures are related by a Lorentz transformation~\cite{Arleo:2014oha}.
  } which process will be treated within a leading-order (LO) pQCD picture, in a setup similar to that used in refs.~\cite{Peigne:2014rka,Arleo:2014oha,Arleo:2020hat,Arleo:2021bpv} (where however only processes involving a target gluon were addressed). In order to derive the medium-induced radiation associated with a given $2 \to 2$ hard process, it is convenient (though not mandatory) to regard the physics from the rest frame of the target. As illustrated in figure~\ref{fig:2to2}, an incident energetic parton with light-cone momentum $p_1 \equiv (p^+_1, p^-_1, \bm p_{1}) = (2E, 0, \bm 0)$\footnote{We define light-cone variables by $p^\pm = p^0 \pm p^z$, and denote transverse momenta by bold characters.} scatters off a parton from the nucleus with $p_2 = (0, p_2^-, \bm 0)$, to produce a pair of final partons with momenta $p_3$ and $p_4$. 

We will consider all tree-level $2 \to 2$ partonic channels, except those involving {\it initial} heavy quarks $Q=c, b$. Indeed, as previously argued~\cite{Arleo:2021bpv}, for our purpose processes such as $Q g \to Q g$ should preferably be interpreted as higher order flavor-excitation processes~\cite{Mangano:1998oia}, where heavy quarks typically arise from $g \to Q \bar{Q}$ perturbative splitting and actually belong to the final state. This `fixed flavor number scheme' allows one to keep track of the actual hard parton system, whose properties (in particular its color structure) determine the FCEL spectrum. In our study initial partons are thus massless,\footnote{\label{foot:vfns}However, note that the FCEL spectrum associated to partonic processes involving initial heavy quarks, \eg\ $Q g \to Q g$, can be easily inferred from our study. But the initial heavy quark should be interpreted as some non-perturbative, intrinsic heavy quark component in the projectile proton. (Otherwise, the accompanying $\bar{Q}$ arising from perturbative splitting would alter the structure of the hard process, resulting in a different FCEL spectrum.) The effect of FCEL on such processes could be studied within the `variable flavor number scheme'~\cite{Buza:1996wv}, which uses input heavy quark PDFs.}
and final partons have the same mass $m \equiv m_3 = m_4$, which is non-zero only in the case of heavy quark production (arising from $gg \to Q \bar{Q}$ and $q \bar{q} \to Q \bar{Q}$ channels). 

\begin{figure}[t]
\centerline{ \Mtwototwo{p_1}{p_2}{p_3}{p_4}{1.5} }  
\caption{Scattering amplitude for a generic $2 \to 2$ partonic process. The final partons of momenta $p_3$ and $p_4$ carry light-cone momentum fractions $\xibar$ and $\xi$, and have transverse momenta $-\bm{K}$ and $\bm{K}$, respectively. A dashed line stands for a  quark, antiquark or gluon. 
\label{fig:2to2} 
}
\end{figure}

Using the Mandelstam variables 
\be
\label{stu}
s=(p_1+p_2)^2 \, , \ \  t=(p_1-p_3)^2 \, , \ \ u=(p_1-p_4)^2 \, ,
\ee
we introduce the 
Lorentz-invariant quantities 
\be
\label{eq:xi}
\xi \equiv \frac{m^2-t}{s} \, , \ \ \xibar \equiv \frac{m^2-u}{s} = 1 - \xi \, .
\ee
Note that these variables coincide with the light-cone longitudinal momentum fractions of the final partons, namely, $\xi = p_4^+/p_1^+$ and $\bar \xi = p_3^+/p_1^+$, in all frames obtained from the nucleus rest frame by a boost along the $z$-axis. The kinematical dependence of any $2 \to 2$ (and thus dimensionless) scattering amplitude ${\cal M}$ considered in the following can be expressed 
in terms of either the single variable $\xi$ (for all processes with massless final partons), or $\xi$ and $m^2/s$ (for $q \bar{q} \to Q \bar{Q}$ and $gg \to Q \bar{Q}$ processes). 

In the setup of figure~\ref{fig:2to2}, the final partons are back-to-back in the transverse plane, 
$\bm{p}_4 = -\bm{p}_3 \equiv \bm{K}$. The final pair invariant mass $s$ and rapidity difference 
$\Delta y \equiv y_4- y_3$ are related to $\xi$ and $\xibar$ through
\be
\label{pair-kinematics}
s \; =\;  p_1^+ p_2^- = \frac{m_{\perp}^2}{\xi \, \xibar} \ \ ; \ \ \ \ \Delta y  = \ln{\left(\frac{\xi}{\xibar} \right)} \, ,
\ee
where we introduced the transverse mass $m_{\perp}^2 \equiv K_{\perp}^2 + m^2$. The typical scale of the hard partonic process is given by $m_{\perp} \sim \morder{K_{\perp}}$.

An essential assumption we make, as in previous FCEL studies, will be the kinematical limit $p^+_1 = 2 E \to \infty$ at fixed $\xi$ and $K_\perp$, corresponding to small angle, `forward scattering' in the nucleus rest frame. This limit is suitable for studying hadron production at moderate $p_\perp$ in high-energy pA collisions.
Note that the assumption $E, \, \xi E, \, \xibar E \gg K_\perp$ is not very demanding in practice (the incoming parton energy $E$ in the nucleus rest frame being huge at modern collider energies), and holds from large positive down to large negative rapidities in the c.m.~frame of the collision. (The assumption $E \gg K_\perp$ would start to be invalid only at {\it very large} negative rapidities, close to the nucleus fragmentation region.) 

The second relation from \eq{pair-kinematics} can be rewritten as 
\be
\label{eq-xixibar}
\xi = \frac{1}{1+e^{-\Delta y}} \ \ ;  \ \ \ \ \xibar = \frac{1}{1+e^{\Delta y}} \ ,
\ee
relating explicitly $\xi$ (and $\xibar = 1-\xi$) to $\Delta y$. In previous studies of $2\to 2$ processes~\cite{Peigne:2014rka,Arleo:2020eia,Arleo:2020hat,Arleo:2021bpv,Arleo:2021krm} 
we considered the FCEL spectrum for a typical configuration of the parton subprocess, 
$|\Delta y| \lsim \morder{1} \Leftrightarrow \xi \sim \xibar \sim 1/2\,$. As we shall see, this latter approximation is closely connected with the LL limit. In the present study, we generalize the calculation to any value of $|\Delta y|$ (including $|\Delta y| \gg 1$), \ie\ in the whole range $0 \leq \xi \leq 1$. 

\subsection{Color density matrix} 
\label{sec:hard-process-color}

In order to derive the soft induced radiation associated with a given hard partonic process, we need to decompose the $2 \to 2$ parton scattering amplitude ${\cal M}$ in terms of the available color structures. At tree-level, ${\cal M}$ can be expressed as a linear combination of either Hermitian color projectors $\proj_{\al}$ (for those channels where the initial and final parton pairs are of the same type, $gg \to gg$, $q g \to q g$, $q \bar{q} \to q \bar{q}$, \ldots, irrespective of quark flavors), or transition operators $\trans_{\al}$ (for the remaining channels, namely, $gg \to q\bar{q}$ and $q\bar{q} \to gg$).  

The projectors $\proj_\al$ are represented graphically as
\bea
{\proj}^{q{\bar q}}_\alpha \ \equiv\ \Pqqbar (\al,0.7) \ \ , \qquad
& &
{\proj}^{gg}_\alpha \ \equiv\  \Pgg (\al,0.7) \ \ , 
\eea
and similarly for other channels, where $\al$ denotes some available $\sun$ irrep for the parton pair under consideration.
For convenience, we adopt color pictorial rules, cf.~refs.~\cite{Cvitanovic:2008zz,Dokshitzer:1995fv,Keppeler:2017kwt,Peigne:2023iwm}, and thus draw color graphs (referred to as `birdtracks') instead of writing explicit color indices.\footnote{\label{foot:birdtrack-rules} These rules slightly vary in the literature, depending on sign and notational conventions. The rules we will use are as follows. Edges are used to identify color indices:
\bea
\qEdge (1) = \delta^{j}_{\ i} \ ; \quad \gEdge (1) = \delta_{ab} \, , \nn 
\eea
where quark and antiquark indices are denoted by $i,j \in \{ 1,...\,, \Nc\}$ and gluon indices by $a,b \in \{ 1, ...\,, \Nc^2-1 \}$\,. (The index of an outgoing quark or incoming antiquark is written as a raised index by convention.) The vertices in a graph represent color tensors, associating a color index to each connecting edge. For any representation $\al\,$, let $T^a_{\al}$ be the (Hermitian) generators of the associated Lie algebra. For the fundamental and adjoint representations, let us denote their matrix elements  by $(t^a)^{i}_{\ j} \equiv (T_{\rm F}^a)^{i}_{\ j}$ and $-i f_{abc} \equiv (T_{\rm A}^a)_{bc}$, respectively. We thus introduce
\bea
  \begin{tabular}{ccc}
$\gqqVert (1) = (t^a)^{j}_{\ i}\  ;$
 &
$\gqbarqbarVert (1)  = (-t^a)^{i}_{\ j}\  ;$
 &
$\gggVert (1) = -if_{abc}\  .$
\end{tabular}
\nn 
\eea
In addition, we need the 3-gluon `star' vertex,
\bea
\starVert (1)  \! \! &=& d_{abc}
  \ \equiv \ 2 \sum_{ijk} \big\{ 
  (t^a)^j_{\ i} (t^c)^{i}_{\ k} (t^b)^{k}_{\ j} + 
  (t^a)^j_{\ i} (t^b)^{i}_{\ k} (t^c)^{k}_{\ j} 
  \big\}
= \ 2 \left( \gggqloopCCW \, - \, \gggqloopCW \right) ,  \nn 
\eea
which is fully symmetric in the permutation of two gluon lines. Note that for $\Nc=2$, quarks and antiquarks are equivalent objects in terms of color, and the star vertex identically vanishes (as is pictorially clear from the r.h.s.~of the above equation). In this paper we consider $\Nc\geq 3$.}
This leads to the expressions of the projectors $\proj_\al$ displayed in table~\ref{tab:irreps} below, for (flavor-blind) $qq$, $q \bar{q}$, $qg$, and $gg$ pairs. Results for $\bar{q} \bar{q}$ and $\bar{q} g$ pairs follow trivially from complex conjugation. In this table we also recall the dimensions $K_\al$ and Casimirs $C_\al$ of the corresponding irreps $\al$. 

\begin{table}[h]
\renewcommand{\arraystretch}{.8}
\setlength\tabcolsep{8pt}
\hskip 0mm
\begin{tabular}{c|c|c|c|c}
  system & irrep $\al$ & projector $\proj_{\al}$ & dimension $K_{\al}$ & Casimir $C_{\al}$ 
  \\[2mm] \hline \hline &&&& \\[-2mm]
  $\bm 3 \otimes \bm 3$ & 
  $\ba{c} \bm{\bar 3} \\[4mm] \bm 6 \ea$ & 
  $\ba{c} \Pthreebarqq \\[4mm] \Psixqq \ea$ & 
  $\ba{c} \frac12\Nc(\Nc-1) \\[4mm] \frac12\Nc(\Nc+1) \ea$ & 
  $\ba{c} 2 \CF -\frac{\Nc+1}{\Nc} \\[4mm] 2 \CF + \frac{\Nc-1}{\Nc} \ea$ 
  \\[8mm]
  \hline &&&&
  \\[-.2cm]
  $\bm 3 \otimes \bm{\bar 3}$ & 
  $\ba{c} \bm{1} \\[4.5mm] \bm{8} \ea$ &
  $\ba{c} \mathsize{10}{\Poneqqbar} \\[4mm] \mathsize{10}{\Peightqqbar} \ea$ &
  $\ba{c} 1 \\[4mm] \Nc^2-1 \ea$ & 
  $\ba{c} 0 \\[4mm] 
  \Nc \ea$
  \\[8mm]
  \hline &&&&
  \\[-.2cm]
   $\ba{c} \bm 3 \otimes \bm 8 \\[2.5mm] \ea$ & 
   $\ba{c} \bm{3} \\[5.5mm] \bm{\bar 6} \\[5.5mm] \bm{15} \\[2mm] \ea$ &
   $\ba{c} \mathsize{8}{\Pthreeqg} \\[4mm] \mathsize{8}{\Psixqg} \\[4mm] \mathsize{8}{\Pfifteenqg} \\[4mm] \ea$ &
   $\ba{c} \Nc \\[5mm] \frac12 \Nc(\Nc+1)(\Nc-2) \\[5mm] \frac12 \Nc(\Nc-1)(\Nc+2) \\[2mm] \ea$ &
   $\ba{c} \CF \\[5mm] \CF+\Nc-1 \\[5mm]  \CF+\Nc+1 \\[2mm] \ea$ 
  \\[8mm]
  \hline &&&&
  \\[-.2cm]
   & $\bm{8}_{\rm a}$ & $\mathsize{8}{\PeightAgg}$ & $\Nc^2-1$ & $\Nc$
  \\[4mm]
  & $\bm{10}\oplus\bm{\overline{10}}$ & $\mathsize{8}{\Ptengg}$ & $\frac12(\Nc^2-1)(\Nc^2-4)$ & $2\Nc$
  \\[4mm]
  & $\bm 1$ & $\mathsize{8}{\Ponegg}$ & $1$ & $0$
  \\[1mm]
  $\ba{c} \bm 8 \otimes \bm 8 \\[8mm] \ea$ & $\bm{8}_{\rm s}$ & $\mathsize{8}{\PeightSgg}$ & $\Nc^2-1$ & $\Nc$ 
  \\[4mm]
  & $\bm{27}$ & $\mathsize{8}{\Ptwentysevenggcut} \, \mathds{Q} $  & $\frac14\Nc^2(\Nc-1)(\Nc+3)$ & $2(\Nc+1)$
  \\[4mm]
  & $\bm 0$ & $\mathsize{8}{\Pzeroggcut} \, \mathds{Q} $  & $\frac14 \Nc^2 (\Nc+1)(\Nc-3)$ & $2(\Nc-1)$
   \\[4mm]
\hline
\end{tabular}
\caption{\label{tab:irreps}
  Projectors, dimensions and quadratic Casimirs of the $\sun$ irreps associated to the $qq$, $q \bar{q}$, $qg$ and $gg$ systems. An $\sun$ irrep is labelled according to its dimension for $\Nc=3$. We express the arising color factors via $\Nc$ and $\CF = (\Nc^2-1)/(2\Nc)\,$. 
  In the last two rows of the table, for the projectors $\Proj{\bm{27}}$ and $\Proj{\bm 0}$ of 
  a $gg$ pair we use the shorthand notation $\mathds{Q} \equiv \mathsize{8}{\protect\Q}$.}  
\end{table}

The transition operators $\trans_{\al}$ for $gg \to q\bar{q}$, 
labelled by the gluon pair irreps which are available in this channel, 
namely, $\al= \{ \bm{1}, \bm{8}_{\rm a}, \bm{8}_{\rm s} \}$, read
\be \label{trans-op}
\Trans{\bm 1} = \frac{1}{\sqrt{\Nc(\Nc^2-1)}} \  \Tonegg  \ \, , \ \  \Trans{{\bm 8_{\rm a}}} = \frac{\sqrt{2}}{\sqrt{\Nc}} \  \TeightAgg  \ \, , \ \ \Trans{{\bm 8_{\rm s}}} =  \frac{\sqrt{2\Nc}}{\sqrt{\Nc^2-4}}  \ \TeightSgg  \ . 
\ee
The transition operators for $q\bar{q} \to gg$ are given by $\trans_{\al}^\dagger$, and the $\trans_{\al}$'s are normalized as
\be \label{trans-ortho-rules}
\trans_{\al}^\dagger \cdot \trans_{\beta} = \delta_{\al\beta} \, \proj_{\al}^{gg} \ \ ; \ \ \ \ 
\trans_{\al} \cdot \trans_{\beta}^\dagger = \delta_{\al\beta} \, \proj_{\tilde{\al}}^{q \bar{q}} \ ,
\ee
where
in the rightmost identity, 
$\tilde{\al}\in \{ \bm{1}, \bm{8} \}$ denotes the $q \bar{q}$ irrep which is 
equivalent to the $gg$ irrep $\al \in \{ \bm{1}, \bm{8}_{\rm a}, \bm{8}_{\rm s} \}$. 
(For example, $\tilde{\al}=\bm{8}$ if $\al=\bm{8}_{\rm a}$ or $\al=\bm{8}_{\rm s}$.) 

It will be convenient to trade $\proj_{\al}$, $\trans_{\al}$ or $\trans_{\al}^\dagger$ (depending on the partonic process), viewed as {\it operators} mapping the $12$ and $34$ parton pair color spaces, 
for the {\it vectors} 
\be
\label{s-basis}
\bra{\alpha} = \frac{1}{\sqrt{K_\alpha}} \ \OrthoBra(0.6,\al)  \ , 
\ee
which are elements of the space of color singlet four-parton states (denoted as $\bar{1}\bar{2}34$). 
In \eq{s-basis} the blob $\alpha$ denotes $\proj_{\al}$, $\trans_{\al}$ or $\trans_{\al}^\dagger$, corresponding to an $s$-channel irrep $\al$ of the $12 \to 34$ partonic process, and the factor $1/\sqrt{K_\alpha}$ ensures that the vectors $\bra{\alpha}$ form an orthonormal basis, $\langle \alpha \vert \beta \rangle = \delta_{\alpha \beta}$. In the following, the basis \eq{s-basis} will be referred to as the `$s$-basis' of the $12 \to 34$ process. Note that {\it kets} $\ket{\alpha}$ are obtained from {\it bras} $\bra{\alpha}$ by complex conjugation, which corresponds pictorially to taking the mirror image and reversing quark arrows,\footnote{Under complex conjugation, an outgoing parton in the bra \eq{s-basis} thus becomes an {\it incoming} parton of the same type in the ket \eq{s-basis-ket}.}  
\be
\label{s-basis-ket}
 \ket{\alpha} \equiv  \bra{\alpha}^* = \frac{1}{\sqrt{K_\alpha}} \ \OrthoKet(0.6,\al^*) \ . 
\ee

The completeness relation
\be
\label{eq:completeness}
\sum_{\al} \ket{\al} \bra{\al} \ = \  \sum_{\al}  \frac{1}{K_\al} \OrthoKet(0.6,\al^*) \ \OrthoBra(0.6,\al) \ = \ \IdentitySinglet (0.6,1) \ \ ,
\ee 
simply follows from the fact that $\{p_\al \equiv \ket{\al} \bra{\al} \}$ is a set of orthogonal projectors of rank 1 acting in the space spanned by the vectors $\bra{\alpha}$, whose dimension is the number of possible values of $\al$.\footnote{Strictly speaking, the identity operator on the r.h.s.~of \eq{eq:completeness} is the identity of the space ${\cal V} \equiv \{ \bra{\alpha} \}$ spanned by the vectors $\bra{\alpha}$, rather than the identity of the space ${\cal W}$ of {\it all} color singlet four-parton states. Indeed, at tree-level there is one process, namely, $gg \to gg$, for which ${\cal V} \varsubsetneq {\cal W}$. 
In a study of $gg \to gg$ at NLO, the label $\al$ in the blob of \eq{s-basis} should list, in addition to the projectors ${\proj}^{gg}_\alpha$, 
two transition operators $\bm{8}_{\rm a} \leftrightarrow \bm{8}_{\rm s}$. This would exhaust all linearly independent operators mapping $gg \to gg$, implying ${\cal V} = {\cal W}$ for all $2 \to 2$ parton processes at NLO.}

The color decomposition of the amplitude ${\cal M}$ in the $s$-basis \eq{s-basis} reads 
\be
\label{M-deco}
{\cal M}_{12 \to 34} = \sum_{\al} \, \nu_\al \, \bra{\al} \ ,
\ee
where the dependence of ${\cal M}$ on external color indices is contained in $\bra{\al}$, and the kinematical, spin and flavor dependence in the coefficients $\nu_{\al}$. Those coefficients depend on the specific partonic channel and are given in table~\ref{tab:channels} of appendix~\ref{app:A}. 
(Note that $\nu_\alpha$ may also depend on the parameter $\Nc$ for some channels.) 
We expect $\nu_\alpha$ to satisfy the Ward identity, for physical polarizations, and therefore to inherit the same gauge invariance as ${\cal M}$ at tree-level~\cite{Combridge:1977dm,Cutler:1977qm}. 

Squaring the amplitude of figure~\ref{fig:2to2}, performing the sums over external color indices 
(by connecting birdtrack lines) for fixed spins/polarizations, and using the decomposition \eq{M-deco} we can write
\be
\Msquaredbird(0.6) \ \equiv \ \trc |{\cal M}|^2 
= \sum_{\al\beta} \nu_{\al} \, \nu_{\beta}^{*} \, \langle \alpha \vert \beta \rangle 
= \ \sum_{\al} |\nu_{\al}|^2  \ , 
\label{eq:Msquared}
\ee
where the $\trc$ symbol indicates the sum over external color indices. In our study we focus on the FCEL spectrum associated to {\it unpolarized} $2 \to 2$ parton processes. Denoting the sum over 
external Dirac indices by the $\trd$ symbol, the total unpolarized squared amplitude is given by
$\trd \trc |{\cal M}|^2$ which is the quantity listed in the third column of table~\ref{tab:channels} (appendix~\ref{app:A}), for each parton process. 

In what follows, we will need to evaluate `expectation values' of some color operator $\mathds{O}\,$ (represented below by a hatched rectangle),
\be
\label{expec-value}
\ave{ \mathds{O} }_{12\to34} \equiv 
\frac1{\trd \trc |{\cal M}|^2} \ \trd \ \ 
\Operatorbird(0.6) =
\sum_{\al\beta} \ \Phi_{\al\beta} \, \mathds{O}_{\al\beta} = \Tr \! \left\{ \Phi \cdot \mathds{O} \right\} \, ,  
\ee
where we used \eq{M-deco}, and $\Phi_{\al\beta}$ and $\mathds{O}_{\al\beta}$ are matrices in the space of irrep labels $\al$ for the partonic channel under consideration (the symbol $\Tr$ denoting the color trace in this space), defined by\footnote{\label{trd convention}Note that in \eq{Phi-def}, we display $\nu_{\al} \nu_{\beta}^*$, even though the $\nu_{\al}$'s are real-valued for tree-level amplitudes.}
\be
\label{Phi-def}
\Phi_{\al\beta} 
= \frac{\trd  ( \nu_{\al} \, \nu_{\beta}^{*} )}{\trd \trc |{\cal M}|^2} 
=  \frac{\trd  ( \nu_{\al} \, \nu_{\beta}^{*} )}{\sum_\gamma \trd |\nu_\gamma|^2} 
\, , 
\ee
and 
\be
\label{O-def}
\mathds{O}_{\al\beta} = \bra{\al} \mathds{O} \ket{\beta} \ .
\ee
For the color graph in \eq{expec-value}, $\mathds{O}$ should be viewed as an operator in the space of color indices of the four-parton state $\bar{1}\bar{2}34$, whereas on the r.h.s.~of \eq{expec-value} [and in \eq{O-def}], it is an 
operator 
in the space of irrep labels $\al$. Pictorially, this distinction is not so important since the two `representations' of the operator $\mathds{O}$ are related with the help of the birdtrack completeness relation \eq{eq:completeness}, namely, 
\be
\label{indices1}
\mathds{O}_{i_{\bar{1}} i_{\bar{2}} i_3 i_4}^{j_{\bar{1}} j_{\bar{2}} j_3 j_4} = \sum_{\al\beta} \ \ket{\al}_{i_{\bar{1}} i_{\bar{2}} i_3 i_4} \ \bra{\al} \mathds{O} \ket{\beta} \ \bra{\beta}^{j_{\bar{1}} j_{\bar{2}} j_3 j_4} \, , 
\ee
where the $i$'s and $j$'s are the color indices of the four-parton state $\bar{1}\bar{2}34$
which are respectively incoming and outgoing in the hatched rectangle of~\eq{expec-value}.
Alternatively we have (with repeated indices being summed over): 
\be
\label{indices2}
\bra{\al}_{i_{\bar{1}} i_{\bar{2}} i_3 i_4} \mathds{O}_{i_{\bar{1}} i_{\bar{2}} i_3 i_4}^{j_{\bar{1}} j_{\bar{2}} j_3 j_4} \ket{\beta}^{j_{\bar{1}} j_{\bar{2}} j_3 j_4} =  \ \bra{\al} \mathds{O} \ket{\beta} \  \, .
\ee

The matrix $\Phi$ defined in \eq{Phi-def} is a crucial quantity in our study. It is a characteristic of the hard partonic process, and may be computed directly from the coefficients $\nu_\al\,$.
These coefficients are easily found in light-cone gauge
(see appendix~\ref{app:A}), 
but we stress that $\Phi$ is gauge invariant, as shown by a calculation in all covariant $R_\xi$-gauges (namely, explicitly verifying that there is no dependence on the gauge 
parameter). 
We supply a {\sc form}~\cite{Kuipers:2012rf} code implementation as an ancillary file to 
this paper, which uses tree-level graphs generated by {\sc qgraf}~\cite{Nogueira:1991ex}, 
to compute $\Phi$ in a general covariant gauge with Faddeev-Popov ghosts. Listed in appendix~\ref{app:Phi-matrix}, are the expressions of the matrix $\Phi$ for all tree-level $2\to2$ QCD processes. 

Before proceeding, let us remark that the matrix $\Phi$ exhibits the main features needed to construct a quantum mechanical density operator $\sum_{\al\beta} \Phi_{\al\beta} \ket{\al} \bra{\beta}$~\cite{LL}, and we shall dub $\Phi$ the {\em color density matrix}. In particular, $\Phi$ is Hermitian (being real and symmetric) and satisfies $\Tr \{ \Phi \} = \sum_{\al} \Phi_{\al\al} = 1\,$, following directly from \eq{Phi-def} and ensuring the normalization in \eq{expec-value} is such that $\ave{\mathds{1}}_{12\to34} = 1$. The diagonal element $\Phi_{\al\al}$ is the probability $\rho_\al$ for the final parton pair to be produced through the color component $\al$ of the amplitude ${\cal M}$, 
\be \label{rho-def}
\rho_{\al} \equiv \Phi_{\al\al} =  \frac{\trd |\nu_{\al}|^2}{ \sum_{\gamma} \trd |\nu_{\gamma}|^2 } \ .
\ee
It also follows from the 
definition \eq{Phi-def} that 
$\Phi$ is a positive semi-definite matrix. 
(Let us remark that some off-diagonal elements of $\Phi$ can be negative, implying that $\Phi$ is not a Markov matrix.) 
We see from table~\ref{tab:channels} 
of appendix~\ref{app:A} 
(last column), that for any $2 \to 2$ process except $qq\to qq$ and $q \bar{q} \to q \bar{q}$, the spin dependence of the coefficient $\nu_\al$ factorizes (in the same overall factor for each $\al$). 
For those processes, the $\trd$ symbol appearing in \eq{Phi-def} can be formally dropped, and $\Phi$ acquires the exact form of an outer product of vectors. This implies $\Phi^2 = \Phi$, a criterion 
that characterizes a `pure' (color) state~\cite{LL}. For $qq\to qq$ and $q \bar q\to q \bar q$ processes, the above-mentioned factorization does not hold, the sum over Dirac indices in the numerator of \eq{Phi-def} depends on the given values of $\al$ and $\beta$, and $\Phi$ is not an outer product. In these two cases, $\Phi$ describes a `mixed' state, of purity $\Tr\{\Phi^2\} < 1\,$.
One may evaluate, for example, the von Neumann entropy ${\cal S} = - \Tr\{ \Phi \log \Phi \}$ 
to assess the degree of color entanglement achieved by 
these particular processes. 

\subsection{Induced gluon radiation spectrum}
\label{sec:spectrum} 

We now discuss the fully coherent induced radiation associated to $2 \to 2$ processes beyond the leading-logarithmic approximation.

The successive parton rescatterings responsible for nuclear $p_\perp$-broadening are modelled as rescatterings off 
screened 
Coulomb potentials~\cite{Baier:1996sk,Baier:1996kr,Baier:1998kq,Gyulassy:2000er,Gyulassy:2000fs,Gyulassy:1999zd}. It is assumed that the 
potential screening length $1/\mu$ satisfies $1/\mu \ll \lambda$, where $\lambda$ is the `color stripped' elastic mean free path,\footnote{\label{lambda_alpha}%
  Let us recall that $\mu$ is on the order of the typical transverse momentum exchange in a single scattering, and that the elastic mean free path of a parton (or pointlike parton system) of Casimir charge $C_\al$ is related to $\lambda$ as $\lambda_\al = \lambda / C_\al$.} allowing the successive rescatterings to be considered independent. 
The total broadening $\Delta p_\perp$ delivered to the final parton pair, $\Delta p_\perp^2 \sim \mu^2 (L/\lambda)$, is on the order of the saturation scale $Q_s$ in the nucleus, and will be simply denoted by $Q_s$ in the following. Nuclear $p_\perp$-broadening induces radiation, and as a result the FCEL spectrum will depend on $Q_s$.\footnote{To avoid any confusion that might result from the use of the word `saturation' in our study, let us stress that FCEL is fully determined by $\Delta p_\perp$ viewed as a theoretical input, independently of the details of this input. Saturation or small-$x$ evolution effects~\cite{Gelis:2010nm} are responsible for the $\xtwo$-dependence of $\Delta p_\perp \sim Q_s$ at small $\xtwo < 10^{-2}$, modifying the magnitude of $\Delta p_\perp$ in this domain of $\xtwo$, but are not responsible for the presence of FCEL. In particular, FCEL is also present at $\xtwo \gsim 10^{-2}$, where saturation effects are absent (and $\Delta p_\perp$ thus independent of $\xtwo$), emphasizing that FCEL and saturation are different physical effects.} 
We will assume the hard scale $K_\perp$ of the $2 \to 2$ process to satisfy $K_\perp \gg Q_s$. 
In this limit, nuclear $p_\perp$-broadening affects negligibly the hard process kinematics, 
which remains well approximated by `back-to-back' production in pA collisions. 

Another approximation concerns the induced gluon radiation, which is considered soft compared to the partons it is emitted from. Denoting by $k \equiv (k^+,k^-,\bm k)$ the radiated gluon light-cone momentum, we have $k^+ \simeq 2 \omega \ll p^+_1, p^+_3, p^+_4$ (implying $x \equiv k^+/p^+_1 \ll \xibar, \xi$), and $k_{_\perp} \equiv |\bm k| \ll K_{_\perp}$. Like nuclear broadening, the induced soft radiation does not affect the hard process kinematics. Moreover, the induced radiation formation time $\tf$ is kept larger than the hard process production time $t_{\rm hard}$~\cite{Arleo:2010rb}, 
\be
\label{x-range}
\tf \sim \frac{\omega}{k_\perp^2} \  \gg \ t_{\rm hard} \sim \frac{\xi \xibar E}{K_\perp^2} \ \  \Longleftrightarrow \ \ \xi \xibar \frac{k_\perp^2}{K_\perp^2} \  \ll \ x  \ . 
\ee

The detailed derivation of the induced radiation spectrum for $2 \to 2$ processes, within the above approximations and beyond leading-logarithm, can be found in appendix~\ref{app:master-spec}, with the final result given by \eq{eq:final-spec2to2}. Note that in this derivation, the spectrum is defined w.r.t.~an ideal target of zero size (in which no soft rescattering occurs), and thus vanishes when $L \to 0$. In order to compare pA and pp collisions, we define the spectrum in a nucleus of size $L$ w.r.t.~a proton target of size $L_{\rm p}$, which is simply obtained from \eq{eq:final-spec2to2} by subtracting the same expression evaluated at $L = L_{\rm p}$. As explained in section~\ref{sec:setup}, we will also restrict to $2 \to 2$ processes with massless incoming partons. In \eq{eq:final-spec2to2} 
we thus set $m_1=0$ [implying $\tilde{m}_1 \to 0$ in  \eq{sig-fun}] and $m_3=m_4=m$ (with $m \neq 0$ when a heavy $Q \bar{Q}$ pair is produced, in either $gg \to Q \bar{Q}$ or $q \bar{q} \to Q \bar{Q}$, and $m = 0$ in all other $2 \to 2$ processes). 

The induced spectrum associated to a $2 \to 2$ process of amplitude ${\cal M}$ is thus 
\be \label{master-spectrum}
\frac{\dd I}{\dd x} = \frac{\alpha_s}{\pi\,x} 
\Big( {\cal L}_{\xi} \, \T_4 + {\cal L}_{\xibar} \, \T_3 \Big) \, , 
\ee
the effect of multiple soft scatterings by the medium being encapsulated in the function
\be 
\label{LuLt}
{\cal L}_{\xi} \equiv F\left( \frac{x K_{\perp}}{\xi \mu}, \frac{x m}{\xi \mu}; \frac{L}{\lambda_g} \right) \, - \, \big(\, L \to L_{\rm p}\, \big) \, , 
\ee
where $F(x,y;r) = 2 y \int_0^\infty \dd B  \, {\rm J}_0(x B) \, {\rm K}_1 (yB) \big[ 1 - \exp\{-r\bm(1-B\, {\rm K}_1(B)\bm)\} \big]$ (with ${\rm J}_\nu$ and ${\rm K}_\nu$ Bessel functions), and 
$\lambda_g = \lambda/\Nc$ is the gluon mean free path. 

The coefficients $\T_3$ and $\T_4$ originate from the two possible soft color connections to the final parton pair: 
\be
\label{TuTt} 
\T_4 = \frac{2}{\trd \trc |{\cal M}|^2} \, \trd \ \xibird(0.6,-0.8) \ \ ; \ \ \ \ 
\T_3 = \frac{2}{\trd \trc |{\cal M}|^2} \, \trd \ \xibird(0.6,-0.2) \ \ ,
\ee
where the radiated gluon line carries only color indices. 
We stress that the spectrum \eq{master-spectrum} is valid at finite $\Nc$ and beyond LL 
(a limit to be properly defined below), 
thus generalizing the result of ref.~\cite{Peigne:2014rka} obtained in the LL and large $\Nc$ limits. 
Note that the function $F$ can be approximated by a logarithm 
(which can be shown to arise from the integration over the soft radiated gluon $k_\perp$)~\cite{Peigne:2014uha}, 
allowing to rewrite \eq{LuLt} as 
\be 
\label{Lxi-approx}
{\cal L}_{\xi} \ \simeq \ 
\log \Big(1+\xi^2 \frac{Q_{s}^2 }{x^2 m_\perp^2} \Big)
- \log \Big(1+\xi^2 \frac{Q_{s,{\rm p}}^2 }{x^2 m_\perp^2} \Big) \ , 
\ee
where $m_\perp$ was defined
below 
\eq{pair-kinematics}, 
$Q_s^2 = \mu^2 (L/\lambda_g) \log{(L/\lambda_g)}$, 
and $Q_{s, {\rm p}}$ denotes the saturation scale in a proton.

As should be clear from appendix~\ref{app:master-spec}, the factors ${\cal L}_{\xi}$ and ${\cal L}_{\xibar}$ in \eq{master-spectrum} are `soft factors' which arise by dressing the squared radiation amplitude by any number of soft rescatterings. They also include the Lorentz part of the initial and final soft gluon emission vertices (leading to the above-mentioned logarithm when integrating over $k_\perp$). 
As for the associated factors $\T_4$ and $\T_3$, they are entirely determined by the $2 \to 2$ amplitude ${\cal M}$, although their precise color structure depends on how the soft gluon line is connected to the final state. We readily see that the induced radiation may probe the color structure of the hard process, and the spectrum can be expected to depend on the irrep of the produced parton pair. 

By rotating the graphs in the expression \eq{TuTt} of $\T_4$ and $\T_3$ (by $90^\circ$), and stretching the soft gluon to a straight line, we see that $\T_4$ and $\T_3$ are of the form \eq{expec-value}, namely, $\T_4 = \Tr \! \left\{ \Phi \cdot \B \right\}$ and $\T_3 = \Tr \! \left\{ \Phi \cdot \Bbar \right\}$, where $\Phi$ is defined by \eq{Phi-def} and the color matrices $\B$ and $\Bbar$ read 
[see \eq{O-def}] 
\bea 
\B_{\al\beta} &\equiv& \bra{\al} \B \ket{\beta} = \bra{\al} 2T_1T_4 \ket{\beta}  = \frac{2}{\sqrt{K_\al K_\beta}} \ \Balphabeta(0.6,1.4)  \label{eq:B} \ , \\[1mm]
\Bbar_{\al\beta} &\equiv&  \bra{\al} \Bbar \ket{\beta} = \bra{\al} 2T_1T_3 \ket{\beta} = \frac{2}{\sqrt{K_\al K_\beta}} \ \Balphabeta(0.6,1.8) \ , \label{eq:Bbar}
\label{eq:Bbirds-3} 
\eea
with $T_i \equiv T_i^a$ the $\sun$ generators of parton $i$. For each partonic process, the explicit form of the matrices $\B_\pm \equiv \B \pm \Bbar = 2T_1(T_4 \pm T_3)$ can be found in appendix~\ref{app:BBbar}.\footnote{We also include the calculation of $\B_{\pm}$ in the ancillary {\sc form} code supplied 
with this paper.} 
The physical relevance of $\B_{\pm}$ will be made clear shortly.
 
The spectrum \eq{master-spectrum} can thus be expressed as
\be
\label{master-spectrum-2}
\frac{\dd I}{\dd x} = \Tr \! \left\{ \Phi \cdot S(x) \right\} \ \ ; \ \ \ \ 
S(x) \equiv \frac{\alpha_s}{\pi\,x} \left( {\cal L}_{\xi} \, \B + {\cal L}_{\xibar} \, \Bbar \, \right) \ ,
\ee
in terms of the density matrix $\Phi$ and of the `soft color matrix' $S(x)$, for each $2 \to 2$ process. 
For the sake of the following discussion, the two bracketed terms in 
\eq{master-spectrum-2} are 
rewritten as 
\bea
\label{eq:introduce-Bpm}
{\cal L}_{\xi} \, \B + {\cal L}_{\xibar} \, \Bbar &=&
  \frac{{\cal L}_{\xi} + {\cal L}_{\xibar}}{2}  \, \B_{+} 
+ \frac{{\cal L}_{\xi} - {\cal L}_{\xibar}}{2} \, \B_{-} \ . 
\eea

It is evident that $\B$ and $\Bbar$ add coherently if ${\cal L}_{\xi} = {\cal L}_{\xibar}$, implying that the soft gluon sees the final parton pair as a pointlike object and cannot change its color state. In this case, the spectrum proportional to $\B_+$ must arise only from contributions with $\al = \beta$, and depend on the final parton pair only through its global color charge $C_\al$. Indeed, adding \eq{eq:B} and \eq{eq:Bbar} and using color conservation $T_1 +T_2 = T_3 +T_4 = T_\al$ (with $T_\al \equiv T_\al^a$ the $\sun$ generators in the irrep $\al$), we directly find that $\B_+$ is a diagonal matrix, 
\be
\label{Bplus}
\left( \B_{+} \, \right)_{\al\beta} =  \bra{\al} 2 \, T_1 \, T_\al  \ket{\beta} =  \bra{\al} T_1^2 + T_\al^2 - T_2^2  \ket{\beta}= (C_1+ C_\al - C_2) \, \delta_{\al\beta} \, ,
\ee
where $C_i$ is the Casimir charge of parton $i$.  

If, on the other hand, $\B$ and $\Bbar$ do not add coherently, \ie\ ${\cal L}_{\xi} \neq {\cal L}_{\xibar}$, the soft gluon can probe the individual colors of the parton pair, and thus allow {\it color transitions} between different irreps of the pair. From \eq{eq:introduce-Bpm} we see that color transitions are encoded in $\B_{-}\,$.
The matrices $\B_{-}\,$ for all partonic processes are listed in appendix~\ref{app:BBbar}, see eqs.~\eq{BmBbar matrix: qq}--\eq{BmBbar matrix: ggqqbar}. For each process, some (if not all) non-diagonal elements of $\B_{-}$ are non-zero, confirming the presence of color transitions contributing to the spectrum when ${\cal L}_{\xi} \neq {\cal L}_{\xibar}$. 
Let us point out that some color transition may be allowed by the structure of $\B_{-}$, but not contribute to the spectrum \eq{master-spectrum-2} simply because the amplitude ${\cal M}$ vetoes some particular irrep. 
For instance, for $gg \to gg$, $\left( \B_{-} \, \right)_{\al\beta}$ with $\al$ or $\beta = \bm{10} \oplus \bm{\overline{10}}$ can be non-zero [see \eq{BmBbar matrix: gg}], but the corresponding amplitude $\nu_\al$ (or $\nu_\beta$) vanishes (see table~\ref{tab:channels}), leading to vanishing rows and columns of the matrix $\Phi$ for $\al, \beta = \bm{10} \oplus \bm{\overline{10}}$ [see \eq{phi 7}].

Note that color transitions are absent if $\xi=\xibar=\frac{1}{2}$ (implying ${\cal L}_{\xi} = {\cal L}_{\xibar}$), but may also be neglected for asymmetric final parton pairs ($\xi \neq \frac{1}{2}$) 
in the logarithmic approximation ${\cal L}_{\xi} \simeq {\cal L}_{\xibar} \gg 1$, 
where $|{\cal L}_{\xi} - {\cal L}_{\xibar}| \ll {\cal L}_{\xi} + {\cal L}_{\xibar}$ 
and the $\B_{+}$ part of the spectrum dominates. Previous calculations of the FCEL spectrum for $2 \to 2$ processes~\cite{Liou:2014rha,Peigne:2014rka} were performed using such approximations, 
and were thus insensitive to color transitions. 
The spectrum \eq{master-spectrum-2} is valid beyond LL accuracy and can be used for any $\xi$. In particular, this will allow us in section~\ref{sec:matching} to show that in the limit $\xi \to 0$ (where parton 3 carries most of the energy), the spectrum \eq{master-spectrum-2} matches with the results obtained previously for $2 \to 1$ processes~\cite{Arleo:2010rb,Peigne:2014uha,Munier:2016oih}. 
Interestingly, having 
a proper matching requires keeping the $\B_{-}$ part of the spectrum, and thus considering color transitions. 

Let us mention that the matrices $\B$ and $\Bbar$ defined by \eq{eq:B}--\eq{eq:Bbar} also appear in the study of parton pair transverse momentum broadening~\cite{Cougoulic:2017ust} and, more generally, are related to the soft anomalous dimension matrix $\ADM$ of $2\to 2$ processes~\cite{Botts:1989kf,Sotiropoulos:1993rd,Contopanagos:1996nh,Kidonakis:1998nf, Oderda:1999kr,Bonciani:2003nt,Appleby:2003hp,Banfi:2004yd,Kyrieleis:2005dt,Dokshitzer:2005ek,Dokshitzer:2005ig,Sjodahl:2008fz,Forshaw:2008cq}. The relation between $\ADM$ and $\B_{-}$ is given in section~\ref{app:BBbar-softADM} of appendix~\ref{app:BBbar}, see eq.~\eq{Q-expr}.

\subsection{Validity domain\label{sec:validity}}

We now specify the conditions under which the spectrum \eq{master-spectrum-2} is valid. Firstly, 
we mentioned before that the radiated gluon should be soft compared to both final partons, $x \ll \xi$ and $x \ll \xibar$. However, the condition $x \ll \xi$ (resp.~$x \ll \xibar$) is only required to derive the first term $\sim {\cal L}_{\xi}$ (resp.~second term $\sim {\cal L}_{\xibar}$) of the spectrum, arising from a final emission off parton 4 (resp.~off parton 3), see appendix~\ref{app:master-spectrum}. 
When $\xi \to 0$ at fixed $x$, the condition $x \ll \xi$ is violated, but this is irrelevant since the term $\sim {\cal L}_{\xi}$ formally vanishes and the spectrum is dominated by the term $\sim {\cal L}_{\xibar}$, with $\xibar \to 1$. Thus, the condition for the gluon to be soft enough for \eq{master-spectrum-2} 
to be valid is simply $x \ll {\rm max}(\xi,\xibar)$, \ie\ $x \ll 1$, for any $\xi$. Secondly, the result \eq{master-spectrum-2} has been obtained under the condition $\tf \gg t_{\rm hard}$, see \eq{x-range}. When $\xi$ is finite (typically $\xi  \sim \frac12$), \ie\ when the condition \eq{x-range} is the most restrictive, it can be shown that \eq{x-range} holds 
provided $Q_s^2/K_\perp^2 \ll x$.\footnote{%
  See section C.2 of ref.~\cite{Peigne:2014uha}. 
  The typical $k_\perp$ (contributing to \eq{LuLt} or equivalently \eq{Lxi-approx}) 
  satisfies $k_\perp \lsim {\rm max}(Q_s, x K_\perp)$. 
  Thus, when ${Q_s}/{K_\perp} \ll x \ll 1$, 
  we have $x \gg x^2 \gsim {k_\perp^2}/{K_\perp^2}$, 
  and \eq{x-range} is automatically satisfied. 
  When $x \lsim {Q_s}/{K_\perp}$ however, \eq{x-range} holds only if ${Q_s^2}/{K_\perp^2} \ll x$. 
} 
Finally, the calculation of the spectrum \eq{master-spectrum} 
performed in appendix~\ref{app:master-spec} uses the fact that in the fully coherent regime, there is a cancellation of purely initial-state and purely final-state radiation, as demonstrated in ref.~\cite{Peigne:2014uha}. 
Strictly speaking, this cancellation is not exact~\cite{Munier:2016oih}, leading to corrections to the spectrum 
for $x \gsim 1/\log{(K_\perp/\mu)}$. 

Ultimately, the validity range of the spectrum \eq{master-spectrum-2} reads, for any $\xi$: 
\be
\label{validity-domain}
\frac{Q_s^2}{K_\perp^2} \ll x \ll \frac{1}{\log{(K_\perp/\mu)}} \ .
\ee

In the limit $K_\perp \gg Q_s$ we are considering, we have $Q_s^2/K_\perp^2 \ll Q_s/K_\perp \ll 1/\log{(K_\perp/\mu)}$, and the domain \eq{validity-domain} has an overlap with the regions $x \ll {Q_s}/{K_\perp}$ (where at least one of the two terms of \eq{master-spectrum-2} is a large logarithm) and $x \gg {Q_s}/{K_\perp}$ [where both ${\cal L}_{\xi}$ and ${\cal L}_{\xibar}$ are smaller than unity, see \eq{Lxi-approx}]. 
In practice, due to the rapid decrease of \eq{Lxi-approx} when $x \gg Q_s/K_\perp$, the typical values of $x$ may be on the order of the average fractional energy loss, $x \sim \ave{x} \sim Q_s/K_\perp$, in which domain ${\rm max}({\cal L}_{\xi},{\cal L}_{\xibar}) \sim \morder{1}$. 
Being conservative, the above discussion shows that the spectrum \eq{master-spectrum-2} is valid 
{\it beyond leading-logarithmic accuracy}, which does not imply (nor exclude) the presence of a large logarithm, and is formally defined by: 
\be
{\rm max}({\cal L}_{\xi},{\cal L}_{\xibar}) \gsim \morder{1} \ . 
\ee

\section{Matching with FCEL spectrum for $2 \to 1$ processes}
\label{sec:matching}

Here we show that the spectrum \eq{master-spectrum-2} reduces to the FCEL spectra obtained previously for $2 \to 1$ processes~\cite{Arleo:2010rb,Peigne:2014uha,Munier:2016oih} and for $2 \to 2$ processes in the logarithmic approximation~\cite{Liou:2014rha,Peigne:2014rka}. The former are obtained from \eq{master-spectrum-2} by taking the limit $\xi \to 0$ (or $\xi \to 1$), and the latter by taking $\xi \to \frac12$. As mentioned in the Introduction, the spectrum \eq{master-spectrum-2} should thus allow implementing FCEL in the full phase space of the produced parton pair ($0 \leq \xi \leq 1$), improving the accuracy of phenomenological studies of the FCEL effect.\footnote{In particular, previous studies of the FCEL effect on $2\to 2$ processes~\cite{Arleo:2020hat,Arleo:2021bpv} focussed on the typical kinematic configuration $\xi \simeq \frac12$, thus using the LL form (proportional to $\B_+$) of the FCEL spectrum. To estimate the theoretical uncertainty associated with this assumption, $\xi$ was varied as $\xi = 0.50 \pm 0.25$~\cite{Arleo:2020hat,Arleo:2021bpv}, however by still assuming the LL form of the spectrum, with the recipe of using the pair invariant mass [given in \eq{pair-kinematics}] in the logarithm. In the $\xi \to 0$ limit, this recipe amounts to neglect FCEL, and thus cannot match with the FCEL spectrum associated with $2 \to 1$ processes [correctly captured by \eq{master-spectrum-2}].} 
As we shall see, the above correspondence 
with previous results is not entirely trivial, and relies on the precise color structure of \eq{master-spectrum-2}. In order to discuss the matching, it will be useful to first introduce general orthonormal color bases. 

\subsection{Orthonormal color bases, and rotations between them} 
\label{sec:ortho-bases}

Let us define a rotated orthonormal basis [w.r.t.~to the $s$-channel basis \eq{s-basis}] by
\be
\label{ortho-basis}
\bra{\wv{\al}} = \sum_\beta \, A_{\wv{\al}\beta} \, \bra{\beta} \ , 
\ee
where $A$ is an orthogonal matrix, \ie\ $A^{-1} = A^\intercal \,$. We use the `tilde' notation to distinguish the sets of states $\{ \bra{\wv{\al}} \}$ and $\{ \bra{\beta} \}$ which may be different. (This happens for instance when the available irreps in the $s$ and $t$-channels are different and 
$A$ is the matrix rotating between the $s$ and $t$-bases, see \eq{A matrix} below.)

The amplitude ${\cal M}$ being independent of the color basis, using \eq{M-deco} we have
\be
\sum_{\wv{\al}} \, \wv{\nu}_{\wv{\al}} \, \bra{\wv{\al}} =  \sum_{\al} \, \nu_\al \, \bra{\al} \ .
\ee
This shows that the coefficients $\nu_\al$ transform as the vectors $\bra{\al}$, namely, 
\be
\wv{\nu} = A \cdot \nu \ .
\ee
The matrices $\Phi$ and $S(x)\,$, defined in \eq{Phi-def} and \eq{master-spectrum-2} respectively,
transform as
\be
\wv{S} = A \cdot S \cdot A^{-1} \ \ ; \ \ \ \ \wv{\Phi} = A \cdot \Phi \cdot A^{-1} \ . 
\label{S and A trans}
\ee
Obviously, the spectrum \eq{master-spectrum-2}, $\dd I/\dd x =  \Tr \! \left\{ \Phi \cdot S(x) \right\}$, 
being a trace, is independent of the color basis.

In addition to the $s$-basis \eq{s-basis}, we introduce the $t$-basis and the $u$-basis by exchanging certain lines but keeping fixed the external color indices of partons $1,\,2,\,3$ and $4$. To be concrete, define
\be
\label{ortho-basis-tu}
\bra{ \al^t } \ \equiv \ \frac1{\sqrt{K_{\al^t}}} \ \OrthoBrat(0.6,\al^t)   \ \ \ \ , \ \ \ \
\bra{ \al^u } \ \equiv \ \frac1{\sqrt{K_{\al^u}}} \ \OrthoBrau(0.6,\al^u) \ \ , 
\ee
where the blob $\alpha^t$ (resp.~$\alpha^u$) now indicates a projector or transition operator\footnote{In some cases, the directions of quark lines and/or the position of gluons may appear incompatible with the explicit projectors $\proj_\al$ listed in table~\ref{tab:irreps} [or transition operators $\trans_\al$ defined in \eq{trans-op}]. For such cases, it is implicit that $\proj_\al$ (or $\trans_\al$) should be supplemented by a permutation of lines, \eg
\bea
\Pqqbaranti (\al,1.) \ \ = \ \ \Pqqbaranticross (\al,1.) \ \  , \nn 
\eea
which arises in the $t$-channel basis for $qq\to qq\,$.
}
relevant to the $t$-channel $1\bar{3} \to \bar{2}4$ (resp.~$u$-channel $1\bar{4} \to \bar{2}3$) 
partonic process, with $K_{\alpha^t}$ (resp.~$K_{\alpha^u}$) the dimension of the corresponding $t$-channel (resp.~$u$-channel) irrep. With this convention, the orthonormality condition $\langle \alpha^t \vert \beta^t \rangle = \delta_{\alpha^t \beta^t}$ (resp.~$\langle \alpha^u \vert \beta^u \rangle = \delta_{\alpha^u \beta^u}$) directly follows from the pictorial expression \eq{ortho-basis-tu} of the basis vectors. 

The matrices that rotate between the $s$, $t$, $u$ color bases directly follow from \eq{ortho-basis}:
\be
\label{A matrix}
\big( A_{ts} \big)_{\al^t \beta}
\ = \ \bra{ \al^t} \, \beta  \rangle
\quad ; \quad
\big( A_{us} \big)_{\al^u \beta}
\ = \ \bra{ \al^u} \, \beta  \rangle
\quad ; \quad
\big( A_{tu} \big)_{\al^t \beta^u}
\ = \ \bra{ \al^t} \, \beta^u  \rangle \, .
\ee
These matrices are orthogonal and, denoting $A_{st}  \equiv A_{ts}^{-1} = A_{ts}^\intercal\,$,
satisfy the `composition relation' $A_{tu} = A_{ts}\cdot A_{su}\,$, which implies that only two of the matrices \eq{A matrix} are independent. Some of these color matrices have already been computed in the literature (\eg\ appendix~A of ref.~\cite{Dokshitzer:2005ig}). 
We don't list them here, but our attached {\sc form} code may be adjusted to compute them all, using \eq{A matrix} and the birdtracks \eq{s-basis} and \eq{ortho-basis-tu}. 

\subsection{Compatibility with $2 \to 1$ processes}
\label{subsec:matching}

There are three special values of $\xi$, namely, $\xi= \{ 0, \frac12, 1 \}$, for which the spectrum \eq{master-spectrum-2} is effectively equivalent to the FCEL spectrum associated with $2 \to 1$ processes. For these values of $\xi$, $S(x)$ is diagonal in the {\it same} basis for any $x$. The basis $\{ \bra{\wv{\al}} \}$ which diagonalizes $S(x)$ is the $t$-basis, $s$-basis and $u$-basis for $\xi=0$, $\xi=\frac12$  and $\xi=1$, respectively. In each of these cases, the spectrum \eq{master-spectrum-2} takes the form $\sum_{\wv{\al}} \Phi_{\wv{\al}\wv{\al}} S(x)_{\wv{\al}\wv{\al}}\,$, where $\rho_{\wv{\al}} (\xi) \equiv \Phi_{\wv{\al}\wv{\al}}$ can be interpreted as the probability associated with the irrep $\wv{\al}$, and $S(x)_{\wv{\al}\wv{\al}}$ provides a color factor $\sim C_a+C_c-C_b$ characteristic of $2 \to 1$ processes. (For other values of $\xi$, the basis diagonalizing $S(x)$ depends on $x$, the above simple interpretation does not hold, and there is no basis where the spectrum is effectively the same as for $2 \to 1$ processes.)

\subsubsection{Matching for $\xi = \frac{1}{2}$}
\label{sec:xi-onehalf}

As we have seen in section~\ref{sec:spectrum}, for a final parton pair in a symmetric configuration, $\xi = \frac{1}{2}$, we have ${\cal L}_{\xi} = {\cal L}_{\xibar} = {\cal L}_{\mathsize{7}{1/2}}$, and the matrix $S(x)$ defined in \eq{master-spectrum-2} is proportional to $\B_{+}$ [given in \eq{Bplus}] and thus diagonal in the $s$-basis, 
\be
S(x)_{\al\beta} =  \delta_{\al\beta} \, (C_1 + C_\al - C_2) \, \frac{\alpha_s}{\pi \, x} \, {\cal L}_{\mathsize{7}{1/2}} \ .
\ee
The spectrum \eq{master-spectrum-2} thus becomes 
\be
\label{spectrum-symmetric}
\left. \frac{\dd I}{\dd x} \right|_{\xi = \frac{1}{2}}
=  \Tr \! \left\{ \Phi \cdot S(x) \right\} = \sum_{\al} \, \rho_{\al} \, (C_1 + C_\al  - C_2) 
\, \frac{\alpha_s}{\pi \, x} \, {\cal L}_{\mathsize{7}{1/2}} \,  \ ,
\ee
where $\rho_{\al}$ is the probability of the $s$-channel irrep $\al$ (evaluated at $\xi=\frac12$).

The expression \eq{spectrum-symmetric} corresponds to the spectrum associated to $2 \to 2$ processes obtained in ref.~\cite{Peigne:2014rka} to LL accuracy.\footnote{Note that in this limit, 
the precise value of $\xi$ in the argument of the logarithm of ref.~\cite{Peigne:2014rka} is not important, but setting $\xi =  \frac{1}{2}$ is a natural choice in phenomenological studies~\cite{Arleo:2020eia,Arleo:2020hat,Arleo:2021bpv,Arleo:2021krm}.
} 
This spectrum depends only on the global charge $C_{\al}$ of the final parton pair, 
and this situation is thus analogous to $2 \to 1$ scattering (with a color charge $C_2$ exchanged in the $t$-channel), up to the average over $\al$ (with weights $\rho_{\al}$) to be performed in \eq{spectrum-symmetric}. 
Let us recall that the LL result of ref.~\cite{Peigne:2014rka} was rigorously obtained in the large $\Nc$ limit and conjectured, on physical grounds, to be valid for any $\Nc$. The present study holds for any $\Nc$ and provides a proof of this conjecture.

\subsubsection{Matching in $\xi \to 0$ and $\xi \to 1$ limits}

When $\xi \to 0$, we set ${\cal L}_{\xi} \to 0$ and ${\cal L}_{\xibar} \to {\cal L}_{1}$ in the spectrum \eq{master-spectrum-2}, which reads 
\be
\label{xitozero-spectrum}
\left. \frac{\dd I}{\dd x} \right|_{\xi \to 0}  =  \frac{\alpha_s}{\pi \, x} \, {\cal L}_{1} \, \Tr \! \left\{ \Phi \cdot \Bbar \right\} \ .
\ee
Using the basis-independent form of $\Bbar$ [see eq.~\eq{eq:Bbar}], 
\be
\Bbar = 2 T_1 T_3 = T_1^2 + T_3^2 - T_t^2 = C_1 + C_3 - T_t^2 \ , 
\ee
with $T_t \equiv T_1-T_3$ the $t$-channel color exchange, we readily see that $\Bbar$ is diagonal in the $t$-basis $\bra{\al^{t}}$ introduced in section~\ref{sec:ortho-bases}. Using \eq{ortho-basis-tu} we obtain the matrix elements of $\Bbar$ in the $t$-basis: 
\be
\Bbar^{\,t}_{\al^t \beta^t} \equiv \bra{\al^t} \, \Bbar \, \ket{\beta^t} 
= (C_1+C_3 -C_{\al^t}) \, \delta_{\al^t \beta^t} \ , 
\ee
where $\al^t$ labels the various $t$-channel irreps, and $C_{\al^t}$ their Casimir charges. 

The basis-independent spectrum \eq{xitozero-spectrum} is conveniently expressed in the $t$-basis as  
\be
\label{xitozero-spectrum-t}
\left. \frac{\dd I}{\dd x} \right|_{\xi \to 0} 
= \frac{\alpha_s}{\pi \, x} \, {\cal L}_{1} \, \sum_{\al^t} \, \Phi_{\al^t \al^t}^{\,t} \, \Bbar^{\,t} _{\al^t \al^t} 
= \sum_{\al^t} \,  \rho_{\al^t}^t \, (C_1+C_3 -C_{\al^t}) \, \frac{\alpha_s}{\pi \, x} \, {\cal L}_{1} \  , 
\ee
where $\rho_{\al^t}^t \equiv \Phi_{\al^t \al^t}^{\,t}$ is the probability for the $t$-channel parton pair to be produced through the irrep ${\al}^t$. This probability is provided by the density matrix $\Phi^{\,t}$ in the $t$-basis,  which is obtained from eq.~\eq{S and A trans} by using the matrix $\Phi$ in the $s$-basis (defined in \eq{Phi-def}, and listed in appendix~\ref{app:Phi-matrix} for each process) and the rotation matrix $A_{ts}$ relating the $s$ and $t$-bases. (See section~\ref{sec:illustration} for some practical examples.) 

The expression \eq{xitozero-spectrum-t} encompasses the FCEL spectra associated with
$2 \to 1$ processes previously derived in~\cite{Arleo:2010rb,Peigne:2014uha,Munier:2016oih} 
and used in phenomenology in~\cite{Arleo:2012hn,Arleo:2012rs,Arleo:2013zua}. In these studies, the spectrum was derived assuming a single color state in the $t$-channel, \ie, $\rho_{\al^t}^t = 1$ for some $\al^t$. Equation~\eq{xitozero-spectrum-t} is a generalization to cases where several 
$t$-channel irreps contribute to the production amplitude ${\cal M}$.  

When $\xi \to 1$, we have ${\cal L}_{\xi} \to {\cal L}_{1}$ and ${\cal L}_{\xibar} \to 0$ in 
\eq{master-spectrum-2}, which thus reads as for $\xi \to 0$, up to the replacement $ \Bbar \to \B$. 
The matrix $\B$ is diagonal in the $u$-basis $\bra{\al^{u}}$ ($\al^u$ labelling a $u$-channel irrep of Casimir $C_{\al^u}$),
\be
\B^{\,u}_{\al^u \beta^u} \equiv 
\bra{\al^u} \, \B \, \ket{\beta^u} = (C_1+C_4 -C_{\al^u}) \, \delta_{\al^u \beta^u} \ . 
\ee
Expressing the spectrum in the $u$-basis we obtain
\be
\label{xitoone-spectrum-u}
\left. \frac{\dd I}{\dd x} \right|_{\xi \to 1} 
= \frac{\alpha_s}{\pi \, x} \, {\cal L}_{1} \, \sum_{\al^u} \, \Phi_{\al^u \al^u}^{\,u} \, \B^{\,u} _{\al^u \al^u} 
= \sum_{\al^u} \,  \rho_{\al^u}^u \, (C_1+C_4 -C_{\al^u}) \, \frac{\alpha_s}{\pi \, x} \, {\cal L}_{1} \ .
\ee

Let us remark that the spectra \eq{spectrum-symmetric}, \eq{xitozero-spectrum-t}, \eq{xitoone-spectrum-u} corresponding to the cases $\xi = \frac{1}{2}, 0, 1$ all contain a color factor of the form $C_a+C_c-C_b$ expected in $a+b \to c$ processes~\cite{Peigne:2014uha}. Indeed, the effective $2 \to 1$ process at play for $\xi = \frac{1}{2}, 0, 1$ is, respectively, $1 + 2 \to [34]_{\al}$, $1 + [2\bar{4}]_{\al^t} \to 3$, and $1 + [2\bar{3}]_{\al^u} \to 4$ (see figure~\ref{fig:2to2}), where $ [ {\rm ab}]_{\wv{\al}}$ refers to an effectively pointlike ab parton pair in color state ${\wv{\al}}$.

\subsection{Example: quark-quark scattering}
\label{sec:illustration}

Let us study two specific partonic channels, namely, $qq^\prime \to qq^\prime$ and $qq \to qq$, to illustrate the general expressions above. For both these channels, in the preferred basis for each of the limiting cases $\xi=\frac{1}{2}$, $\xi = 0$, and $\xi=1$, the spectrum takes the form specified by \eq{spectrum-symmetric}, \eq{xitozero-spectrum-t} and \eq{xitoone-spectrum-u}, where the corresponding available irreps are $\al = \{ \bm{\bar 3}, \bm{6} \}$, $\alpha^t= \{ \bm{{1}}, \bm 8 \}$, and $\alpha^u= \{ \bm{1}, \bm{8} \}$ in the $s$, $t$ or $u$-channel, respectively. To be specific, we have
\bea
\label{spectrum-symmetric-qqprime-0}
&& \left. \frac{\dd I}{\dd x} \right|_{\xi = \frac{1}{2}}
\ = \ \frac{\alpha_s}{\pi \, x} \, {\cal L}_{\mathsize{7}{1/2}}  \, \big[ \, \Rho{\bm{\bar{3}}} \Cas{\bm{\bar{3}}} + \Rho{\bm{{6}}} \Cas{\bm{{6}}} \, \big] 
\  , \\ 
\label{xitozero-spectrum-t-qqprime-0}
&& 
\left. \frac{\dd I}{\dd x} \right|_{\xi \to 0} \ = \ 
\frac{\alpha_s}{\pi \, x} \, {\cal L}_{1} \, \big[ \, \Rho{\bm{1}}^t (2\CF) + \Rho{\bm{8}}^t  (2\CF-\Nc ) \, \big] 
\  , \\ 
\label{xitoone-spectrum-u-qqprime-0}
&& \left. \frac{\dd I}{\dd x} \right|_{\xi \to 1} \ = \ 
\frac{\alpha_s}{\pi \, x} \, {\cal L}_{1} \,  \big[ \, \Rho{\bm{1}}^u (2\CF) + \Rho{\bm{8}}^u  (2\CF-\Nc ) \, \big]  
\  .
\eea
The expressions above
  highlight the characteristic structure of spectra associated to $2 \to 1$ processes, 
  showing explicitly, for the specific cases of 
  the $qq^\prime \to qq^\prime$ 
  and  $qq \to qq$ 
  processes, 
  the matching demonstrated in general in section~\ref{subsec:matching}. 
Moreover, they allow one to observe an interesting effect. 
For $\xi \to 0$ and $\xi \to 1$, the spectrum expressed in the $t$- and $u$-basis contains some component proportional to the color factor $2\CF-\Nc = -\frac{1}{\Nc} <0$. This can be attributed to a {\it negative} FCEL, or fully coherent energy gain (FCEG), 
as previously observed for the $qg \to q$ process~\cite{Peigne:2014uha,Peigne:2014rka,Munier:2016oih}. For $\xi = \frac{1}{2}$, each contribution to the FCEL spectrum (for any of the $s$-channel irreps, $\al =\bm{\bar 3}$ or $\al = \bm{6}$) is positive. 

Let us first focus on the scattering of quarks of different flavors, $qq^\prime \to qq^\prime\,$, for which the ingredients needed to evaluate eq.~\eq{master-spectrum-2} are the matrix $\Phi$ in \eq{phi 1}, the matrix $\B_+$ in \eq{BpBbar matrix: qq} and the matrix $\B_-$ in \eq{BmBbar matrix: qq}. Direct evaluation gives
\be
\label{qqprime-spec}
 \frac{\dd I}{\dd x} = 
 \frac{\alpha_s}{\pi\, x} \bigg[ \,
    \Big( \Nc - \frac2{\Nc} \Big) {\cal L}_{\xi}
    \, - \,  \frac{{\cal L}_{\xibar}}{\Nc} 
    \, \bigg] \ , 
\ee
from which one can readily obtain the spectra for $\xi=\frac{1}{2}$, $\xi = 0$, and $\xi=1$: 
\bea
\label{spectrum-symmetric-qqprime}
&& \left. \frac{\dd I}{\dd x} \right|_{\xi = \frac{1}{2}} \ 
= \ \frac{\alpha_s}{\pi \, x} \, {\cal L}_{\mathsize{7}{1/2}} \, \bigg( \Nc - \frac{3}{\Nc} \bigg) \  , \\ 
\label{xitozero-spectrum-t-qqprime}
&& 
\left. \frac{\dd I}{\dd x} \right|_{\xi \to 0} \ = 
\ \frac{\alpha_s}{\pi \, x} \, {\cal L}_{1} \,  \bigg( -\frac{1}{\Nc} \bigg) \  , \\ 
\label{xitoone-spectrum-u-qqprime}
&& \left. \frac{\dd I}{\dd x} \right|_{\xi \to 1} \ = \ 
\frac{\alpha_s}{\pi \, x} \, {\cal L}_{1} \,  \bigg( \Nc - \frac{2}{\Nc} \bigg) \  .
\eea

In the $s$-basis, the diagonal entries of the matrix $\Phi$ [see \eq{phi 1}]  yield the $s$-channel probabilities $\Rho{\bm{\bar 3}} = \frac{\Nc+1}{2\Nc}$ and $\Rho{\bm{6}} = \frac{\Nc-1}{2\Nc}$,
which corroborates eq.~\eq{spectrum-symmetric-qqprime}. To obtain the probabilities in the other bases, and thus verify that eqs.~\eq{xitozero-spectrum-t-qqprime}--\eq{xitoone-spectrum-u-qqprime} 
are indeed of the form \eq{xitozero-spectrum-t-qqprime-0} or \eq{xitoone-spectrum-u-qqprime-0},
one needs the (flavor-blind) rotation matrices $A_{ts}$ and $A_{us}$ obtainable from \eq{A matrix}. 
Explicitly, they read\footnote{%
  The birdtrack expressions of \eq{A matrix} make it clear that $A_{ts}$ and $A_{us}$ 
  are related by permuting two incoming quark lines in the blob defining the ket $\ket{\beta}$. 
  Therefore, since the irrep $\bm{\bar{3}}$ is antisymmetric, 
  the first columns of the two matrices in \eq{A-qq2qq} are related by a minus sign: 
  $\bra{\al^t} \bm{\bar{3}} \rangle = - \bra{\al^u} \bm{\bar{3}} \rangle\,$. 
  The second columns are identical because $\bm{6}$ is symmetric.}
\be
\label{A-qq2qq}
A_{ts} \ =\ \frac{1}{\sqrt{2\Nc}}  \left( 
\begin{matrix}  D_1  &  \  U_1   \\  - U_1  &\ D_1 \end{matrix} \right)  \ \quad ; \quad \ 
A_{us} \ = \ \frac{1}{\sqrt{2\Nc}}  \left( \begin{matrix}  -D_1  &  \  U_1   \\  U_1  &\ D_1 \end{matrix}  \right) \, , 
\ee
where we use the shorthand notation $U_k \equiv \sqrt{\Nc +k}$ and $D_k \equiv \sqrt{\Nc -k}$~\cite{Dokshitzer:2005ig}. Using these matrices, one can rotate $\Phi$ to the $t$ and $u$-bases:
\be
  \Phi^{(t)} \ = \ A_{ts} \cdot \Phi \cdot A_{ts}^{-1}
\ =\  \left( \begin{matrix}  0  &  0   \\  0  & 1 \end{matrix} \right) ; \ \ 
\Phi^{(u)} \ = \ A_{us} \cdot \Phi \cdot A_{us}^{-1} \ = \ \frac{1}{\Nc^2}  \left( \begin{matrix}  K_A  &  -\sqrt{K_A}   \\  -\sqrt{K_A}  & 1 \end{matrix} \right)  \, , 
\ee
where $K_A \equiv \Nc^2 -1\,$. From the above, we infer the probabilities $\Rho{\bm{1}}^t = 0$, $\Rho{\bm{8}}^t =1$, and $\Rho{\bm{1}}^u = K_A/\Nc^2$, $\Rho{\bm{8}}^u =1/\Nc^2$. (The fact that the color exchange in the $t$-channel is purely octet is pictorially obvious from the expression of ${\cal M}$ for $qq^\prime \to qq^\prime$, see table~\ref{tab:channels} of appendix~\ref{app:A}.) 

\begin{figure}[t]
\begin{center}
\includegraphics[scale=.72]{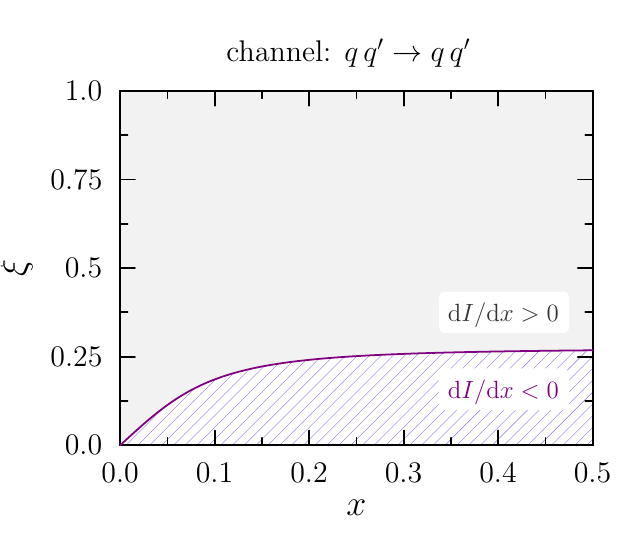}  
\hskip -8mm 
\includegraphics[scale=.72]{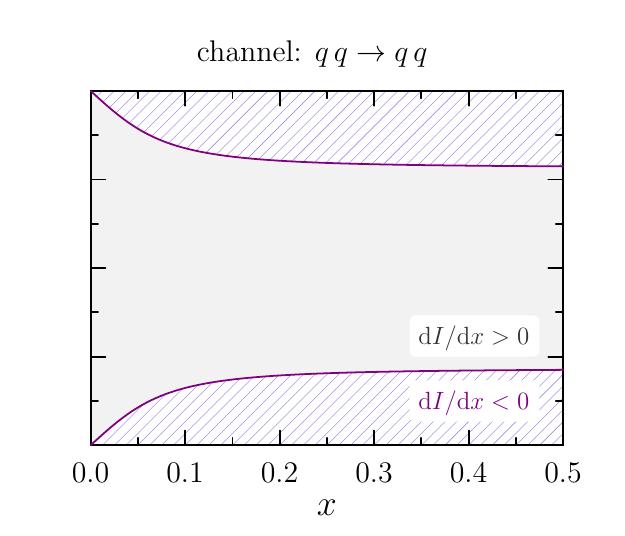}  
\end{center}
\vskip -5mm 
\caption{
  Regions in the $(x,\xi)$-plane, corresponding 
  to energy-loss (solid gray) or energy-gain (hatched purple).
  In this figure, $\Nc=3$, $Q_{s,{\rm A}} =  \frac14 m_\perp$ and 
  $Q_{s,{\rm p}} =  \frac1{10} m_\perp\,$. 
  }
\label{fig:xi_vs_x}
\end{figure}

For $\xi \to 0$, the spectrum \eq{xitozero-spectrum-t-qqprime} expressed in the $t$-basis 
[and thus of the form \eq{xitozero-spectrum-t-qqprime-0}] 
depends only on the $t$-channel irrep $\al= \bm{8}$, and is proportional to the negative color factor $-\frac{1}{\Nc}$ leading to FCEG. A similar effect is visible in the $\xi \to 1$ limit [see \eq{xitoone-spectrum-u-qqprime-0}], when the {\it $u$-channel} $q \bar{q}$ pair is color 
octet (with probability $\Rho{\bm{8}}^u =1/\Nc^2$). However, this negative contribution to the spectrum is in that case overcome by a positive contribution when the $q \bar{q}$ pair is color singlet (with probability $\Rho{\bm{1}}^u = K_A/\Nc^2$), leading to the positive spectrum \eq{xitoone-spectrum-u-qqprime}. In figure~\ref{fig:xi_vs_x} (left) we give a more detailed chart of the sign of ${\rm d}I/{\rm d}x$ in the $(x,\xi)$-plane, showing the domains where medium-induced radiation can be interpreted as actual (positive) FCEL, or FCEG, with a boundary between them where the induced radiation spectrum vanishes. 

We now consider the scattering of quarks of identical flavor, $qq \to qq\,$, and repeat the above discussion. The matrices $\B_+$, $\B_-$, $A_{ts}$, $A_{us}$ are flavor-blind and thus the same as for $qq^\prime \to qq^\prime$, but the matrix $\Phi$ is now given by \eq{phi 2}, yielding the $s$-channel probabilities (which now depend on $\xi$),
\be \Rho{\bm{\bar 3}}(\xi) =
\frac{(\Nc+1)\left[(\xi-\xibar)^2+\xi^4+\bar{\xi}^4 \right]}{2\Nc  \big(\xi^4+ \bar{\xi}^4+\xi^2+\xibar^2 \big) -4\xi\bar{\xi} \, }  \ ; \ \ \ \ 
\Rho{\bm{6}}(\xi) =  
\frac{(\Nc-1)\left[1+\xi^4+\bar{\xi}^4 \right]}{ 2\Nc \big(\xi^4+ \bar{\xi}^4+\xi^2+\xibar^2 \big) -4\xi\bar{\xi} \, } \  \, .
\ee
For $\xi = \frac{1}{2}$, this gives: $\Rho{\bm{\bar 3}}(\frac{1}{2}) =\frac{\Nc+1}{10\Nc-8}$, $\Rho{\bm{6}}(\frac{1}{2}) =\frac{9(\Nc-1)}{10\Nc-8}$. For $\xi \to 0$, we find the same $t$-channel probabilities as for $qq^\prime \to qq^\prime$: $\Rho{\bm{1}}^t = 0$, $\Rho{\bm{8}}^t =1$. 
(This follows from the fact that the contribution to ${\cal M}_{qq \to qq}$ of the exchange diagram proportional to $\mathcal{B}_u$, see  table~\ref{tab:channels}, vanishes when $\xi \to 0$.)  
Due to the symmetry in $\xi \leftrightarrow \xibar$ of the $qq \to qq$ process, for $\xi \to 1$ we find the same $u$-channel probabilities: $\Rho{\bm{1}}^u = 0$, $\Rho{\bm{8}}^u =1$. Using \eq{spectrum-symmetric}, \eq{xitozero-spectrum-t}, \eq{xitoone-spectrum-u}, the spectra associated to $qq \to qq$ and corresponding to $\xi=\{ \frac{1}{2}, 0, 1\}$ read: 
\bea
\label{spectrum-symmetric-qq}
&& \left. \frac{\dd I}{\dd x} \right|_{\xi = \frac{1}{2}} \ 
= \ \frac{\alpha_s}{\pi \, x} \, {\cal L}_{\mathsize{7}{1/2}} \, \bigg( \frac{8 - 15 \Nc +5 \Nc^3}{\Nc (5\Nc-4)} \bigg) \, ,\\
\label{xitozero-spectrum-t-qq}
&& \left. \frac{\dd I}{\dd x} \right|_{\xi \to 0} \ = 
\ \frac{\alpha_s}{\pi \, x} \, {\cal L}_{1} \,  \bigg( -\frac{1}{\Nc} \bigg) \, ,\\
\label{xitoone-spectrum-u-qq}
&& \left. \frac{\dd I}{\dd x} \right|_{\xi \to 1} \ = \ 
\frac{\alpha_s}{\pi \, x} \, {\cal L}_{1} \,  \bigg( -\frac{1}{\Nc} \bigg) \, .
\eea

For $\xi \to 0\,$, we get energy gain with probability $\Rho{\bm{8}}^t = 1\,$ (as for $qq^\prime \to qq^\prime$). Due to the aforementioned symmetry under $\xi \leftrightarrow \xibar$, when $\xi \to 1$ the spectrum has the same limit, and thus corresponds to FCEG (with probability $\Rho{\bm{8}}^u = 1$). Again, all those features are contained in the spectrum~\eq{master-spectrum-2}, which is fully determined for any $\xi$ once the matrices $\Phi$, $\B_+$, $\B_-$ are given. We show in figure~\ref{fig:xi_vs_x} (right) the chart of the sign of  ${\rm d}I/{\rm d}x$ in the $(x,\xi)$-plane, 
for the $qq \to qq$ process. Due to the symmetry in $\xi \leftrightarrow \xibar$, 
we now have two regions corresponding to FCEG. 

The above discussion of $qq^\prime \to qq^\prime$ and $qq \to qq$ processes can be done for any $2 \to 2$ partonic process, highlighting the structure of the FCEL spectra for $\xi = \{ \frac{1}{2}, 0, 1 \}$, and providing---in those three limits---a physical interpretation of the various contributions to ${\rm d}I/{\rm d}x$ (including FCEG-type contributions which occur for some partonic channels and some values of $\xi$) in terms of the available irreps in the $s$, $t$, and $u$-bases.

\section{Discussion}
\label{sec:conclusion}

We have derived the medium-induced gluon radiation spectrum in the fully coherent regime for all $2\to2$ partonic scatterings (with massless initial partons). Using ~\eq{master-spectrum-2}, the matrix $\Phi$ given in appendix~\ref{app:Phi-matrix}, and the matrices $\B_+$ and $\B_-$ given in appendix~\ref{app:BBbar}, the spectrum can be obtained for any $2\to2$ channel. The spectrum \eq{master-spectrum-2} is independent of the color basis used to decompose the hard subprocess $2 \to 2$ amplitude, is valid beyond leading-logarithmic accuracy,\footnote{The precise domain of validity, and the formal accuracy of our result, is explained in section~\ref{sec:validity}.} for any number of colors $\Nc$, and in the full kinematic range of the underlying process. We have noted that for $\xi \neq \frac{1}{2}$, soft gluon radiation can probe the individual color charges of the pair constituents (provided the two logarithms in \eq{master-spectrum-2} are not simultaneously large, otherwise the first term of \eq{eq:introduce-Bpm} would dominate), and the spectrum~\eq{master-spectrum-2} thus involves {\it color transitions} between different irreps of the parton pair. Another feature of the spectrum~\eq{master-spectrum-2} is the exact matching with known limits for $\xi = \frac{1}{2}$, $\xi \to 0$ and $\xi \to 1$, for which the spectrum is effectively the same as for $2 \to 1$ processes. 

We thus expect the spectrum~\eq{master-spectrum-2} to be applicable in phenomenology across the entire $\xi$ range, $0 < \xi < 1$. Strictly speaking, the formal limit $\xi \to 0$ (resp.~$\xibar \equiv 1-\xi \to 0$) should be understood as $\xi \ll 1$ (resp.~$\xibar \ll 1$) at {\it finite} $\xi$ (resp.~$\xibar$), due to the kinematic limit we considered, namely, $E \to \infty$ at fixed $\xi$ and $K_\perp$, ensuring $E,\, \xi E, \,\bar \xi E \gg K_\perp\,$, \ie\ small angle scattering {\it in the nucleus rest frame}. (As mentioned earlier, this setup encompasses the situation of production at mid-rapidity in the c.m.~frame.) In the limit where $\xi$ or $\xibar$ becomes very small, some final parton has a very little longitudinal momentum even in the nucleus frame (corresponding to a very large negative rapidity in the c.m.~frame), invalidating the assumption $\xi E \gg K_\perp\,$. In practice, however, the drastic $\xi \to 0$ limit (or $\xibar \to 0$) is irrelevant. 

By way of illustration, let's consider one of the simplest situations where FCEL is at work, namely, single inclusive hadron production in pA collisions. We view this process within collinear factorization, assuming the tagged hadron arises in our setup from the fragmentation of parton 4 of the $2\to 2$ partonic subprocess (see figure~\ref{fig:2to2}), 
with given transverse momentum $p_\perp = z K_\perp$ and rapidity $y=y_4$ (in the proton-nucleon c.m.~frame). 
The hadron cross section $\dd \sigma /(\dd p_\perp \dd y)$ involves an integral over the fragmentation variable $z$, as well as an integral over the rapidity $y_3 = y_4 - \Delta y$ of the unobserved parton 3, which using \eq{pair-kinematics} can be traded for an integral over $\xi\,$. In the LO kinematics, the longitudinal momentum fractions $x_1$ and $x_2$ of the incoming partons from the projectile and target are not independent variables: they are related to $z$ and $\xi$ as $x_1 = m_\perp\, e^{+y}/(\xi \sqrt{s_{_{\rm NN}}})$ and $x_2 = m_\perp\, e^{-y}/\bm((1-\xi)\sqrt{s_{_{\rm NN}}}\bm)\,$, where $m_\perp = \sqrt{m^2 + (p_\perp/z)^2}$.

It follows that in the LO kinematics, $\xi_{\rm min} \leq \xi \leq \xi_{\rm max}$ where $\xi_{\rm min} \equiv  m_\perp e^{y}/\sqrt{s_{_{\rm NN}}}$ and $\xi_{\rm max} \equiv  1- m_\perp e^{-y}/\sqrt{s_{_{\rm NN}}}\,$. 
The exact matching limits for the spectrum at $\xi=0$ and $\xi=1$ therefore cannot be reached,
so long as we are considering single hadron production with a 
non-zero $m_\perp$. 
In other words, the limit where the spectrum~\eq{master-spectrum-2} exactly matches with $2 \to 1$ processes is only accessible asymptotically at large $|y|\,$. As an example, let us consider $\sqrt{s_{_{\rm NN}}}={\cal O}(10\,{\rm TeV})$, taking $m_\perp = 3$~GeV and $y = -5$, to represent values attainable at the LHC. This gives $\xi_{\rm min} \simeq 2 \times 10^{-6}$, showing that the asymptotic $2 \to 1$ case is not merely academic for the observable under consideration, and emphasizing the relevance of the spectrum \eq{master-spectrum-2} (which encompasses this limiting case). Furthermore, the longitudinal momentum of parton 4 (in the nuclear rest frame) is then $p_4^z \simeq \xi x_1 E_{\rm p} = \xi_{\rm min} E_{\rm p} \sim 100\,{\rm GeV}$ [using $E_{\rm p} = s_{_{\rm NN}}/(2 m_{\rm p}) \simeq 5\times10^{7}\,{\rm GeV}$]. Thus, parton 4 is fast in the nuclear target down to quite large negative rapidities. This validates the use of the spectrum, which was calculated assuming both final partons to be fast in the nuclear target, in a broad rapidity range. 
We stress that this rapidity range encompasses mid-rapidities as well as negative rapidities  (defined in the c.m.~frame), where the hard process looks like $90^\circ$ scattering and backward scattering, respectively. However, as mentioned in section~\ref{sec:setup}, both these situations correspond to small angle, forward scattering in the nucleus rest frame, which is the main condition for the validity of our results. 

As mentioned in the introduction, the induced spectrum ${\rm d}I/{\rm d}x$ should ultimately be used to determine the specific energy loss probability distribution (or {\em quenching weight}) valid beyond LL, in order to predict the effect of FCEL on hadron production pA (or AA) cross sections without restricting to a typical $\xi$ value as in previous LL studies. The new features of the spectrum that arise beyond LL, in particular the presence of color transitions, will need to be incorporated. The explicit construction of the quenching weight and its consequences for phenomenology will be addressed in later studies. 

\acknowledgments
S.~P. and K.~W. were funded by the Agence Nationale de la Recherche (ANR)
under grant No.~ANR-18-CE31-0024 (COLDLOSS). G.~J. was funded by the U.S. Department of Energy (DOE), under grant No.~DE-FG02-00ER41132, and now by the ANR under grant No.~ANR-22-CE31-0018 (AUTOTHERM).
We thank Paul Caucal for fruitful discussions on the different nature of FCEL and small-$x$ evolution effects.

\appendix

\section{Color structure of $2 \to 2$ partonic channels}
\label{app:A}

In this appendix, we discuss the ${\rm SU}(\Nc)$ structure of the hard, leading-order $2 \to 2$ scattering amplitudes, and how to construct the matrix $\Phi$ defined in eq.~\eq{Phi-def}. The calculation of $\Phi$ and of the required coefficients $\nu_\al$ in~\eq{M-deco} is a self-contained task, unrelated to the physics of soft gluon radiation discussed in the main text.

\subsection{Color decomposition of amplitudes} 
\label{app:color-dec}

In table~\ref{tab:channels} we give the color decomposition of the amplitude ${\cal M}$ for each partonic channel. In contrast with table~\ref{tab:irreps}, here flavor matters, and we denote by $q$ and $q'$ different quark flavors. Observe that for $qg\to qg\,$, $gg\to gg$ and $gg \to q \bar{q}$ channels, for the {\it pictorial form} of ${\cal M}$ given in the second column of table~\ref{tab:irreps} we chose the $s$-channel and $t$-channel color graphs as the independent color structures to express it. (Of course, using color conservation another choice could be made to obtain equivalent expressions of ${\cal M}$.)

As an illustration of how the coefficients $\nu_{\al}$ of the last column of table~\ref{tab:channels} are obtained, we consider below the case of a few specific channels. 
 
\paragraph{Example 1: $\bm{q q \to q q}$ channel}

The amplitude for the scattering of two quarks of same flavor reads
\be \label{Muuuu}
{\cal M} =  \ \mathcal{B}_t \ \qqtchannel{1} \ + \ \mathcal{B}_u \ \qquchannel{1}  \ \ ,  
\ee
where the graphs stand for color factors only, and $\mathcal{B}_t$ and $\mathcal{B}_u$ are given by:
\be \label{Bt and Bu}
\mathcal{B}_t = \frac{g^2}{t}\, \left[ \bar{u}(p_3) \gamma^\mu u(p_1) \right] \,  \left[ \bar{u}(p_4) \gamma_\mu u(p_2) \right] \ ; \ \ \mathcal{B}_u = - \frac{g^2}{u}\, \left[ \bar{u}(p_4) \gamma^\mu u(p_1) \right] \,  \left[ \bar{u}(p_3) \gamma_\mu u(p_2) \right] \ , 
\ee

with $u(p_i)$ and $\bar u(p_i)$ the usual Dirac spinors for particle $i$. Using the Fierz identity 
\be \label{Fierz} 
\Fierzthree \ \ , 
\ee
we can express the color graphs of \eq{Muuuu} as linear combinations of the projectors $\Proj{\bm {\bar 3}}$ and $\Proj{\bm 6}$ of a $qq$ pair (given in table~\ref{tab:irreps} in the $\bm 3 \otimes \bm 3$ entry). One obtains
\be \label{MqqqqBtBu}
{\cal M} = - \frac{1}{2} \left\{  \frac{\Nc+1}{2\Nc} (\mathcal{B}_u -\mathcal{B}_t) \, \Proj{\bm {\bar 3}} + \frac{\Nc-1}{2\Nc} (\mathcal{B}_t +\mathcal{B}_u) \, \Proj{\bm 6}  \right\}  \, ,
\ee 
which yields the coefficients $\nu_{\al}$ (last column of table~\ref{tab:channels}) for this channel. 

\begin{table}[h]
\renewcommand{\arraystretch}{.92}
\setlength\tabcolsep{5pt}
\hskip -4mm
\vspace{3mm}
\begin{tabular}{c|c|c|c|c}
 channel & 
 ${\cal M}$ &  
 $\displaystyle \frac{\trd \trc |{\cal M}|^2 }{4 g^4 (N^2-1)}$ &
 $\al$ & 
  $\displaystyle \frac{\nu_{\al}}{\sqrt{K_{\al}}}$  
 \\[.4cm] \hline \hline &&&& \\[-.2cm]
 $q q' \to q q'$ & 
 $\mathcal{A} \ \Mqoneqtwo$ & 
 $\displaystyle \frac{1+\xibar^2}{2\xi^2}$ & 
 $\ba{c} \bm{\bar 3} \\[2mm] \bm 6 \ea$ & 
 $\ba{c} \mathcal{A} \, \frac{\Nc+1}{2\Nc} \\[2mm] - \mathcal{A} \, \frac{\Nc-1}{2\Nc} \ea$    
  \\[5mm]
  \hline &&&&
  \\[-.2cm]
  $q q \to q q$ & 
  $\Mqoneqone$ &  
  $\displaystyle \frac{1+\xi^2}{2\xibar^2} + \frac{1+\xibar^2}{2\xi^2} - \frac1{\Nc \xi \xibar}$ & 
  $\ba{c} \bm{\bar 3} \\[2mm] \bm 6 \ea$ & 
  $\ba{c} \frac{\Nc + 1}{4\Nc}\,(\mathcal{B}_t -\mathcal{B}_u) \\[2mm]  -\frac{\Nc - 1}{4\Nc}\,(\mathcal{B}_t +\mathcal{B}_u) \ea$ 
  \\[5mm]
  \hline &&&&
\\[-.2cm]
  $q \bar{q}' \to q \bar{q}'$ & 
  $\mathcal{C} \ \Mqoneqtwobar$ & 
  $\displaystyle \frac{1+\xibar^2}{2\xi^2}$ &
  $\ba{c} \bm{1} \\[2mm] \bm{8} \ea$ &
  $\ba{c} \CF \, \mathcal{C} \\[2mm] -\frac{1}{2\Nc} \,\mathcal{C} \ea$ 
  \\[5mm]
  \hline &&&&
\\[-.2cm]
  $q \bar{q} \to q' \bar{q}'$ & 
  $\mathcal{D} \ \Mqoneqonebar$ & 
  $\displaystyle \frac{\xi^2+\xibar^2}{2}$ &
  $\ba{c} \bm{1} \\[2mm] \bm{8} \ea$ &
  $\ba{c} 0 \\[2mm] \frac{1}{2} \,\mathcal{D} \ea$ 
  \\[5mm]
  \hline &&&&
\\[-.2cm]
  $q \bar{q} \to q \bar{q}$ &  
  $\Mqoneqonebarbis$ & 
  $\displaystyle \frac{\xi^2+\xibar^2}2 + \frac{1+\xibar^2}{2\xi^2} + \frac{\xibar^2}{\Nc \xi}$ & 
  $\ba{c} \bm{1} \\[2mm] \bm{8} \ea$ &
  $\ba{c} \CF \, \mathcal{E}_t \\[2mm] \frac12 \big( \mathcal{E}_s - \frac1{\Nc} \mathcal{E}_t \big) \ea$ 
  \\[5mm]
  \hline &&&&
\\[-.2cm]
  $qg \to qg $ & 
  $\mathcal{F} \, \left[ \ \Mqg \ \right]$ & 
  $\displaystyle (1+\xibar^2) \Big( \frac{\Nc}{\xi^2} + \frac{\CF}{\xibar} \Big)$ & 
  $\ba{c} \bm{3} \\[2.3mm] \bm{\bar 6} \\[2.3mm] \bm{15} \ea$ &
  $\ba{c} \left(\frac{1}{2\Nc} + \xibar \CF\right) \mathcal{F} \\[2mm] \frac12  \,\mathcal{F} \\[2mm] -\frac12 \,\mathcal{F}\ea$ 
  \\[8mm]
  \hline &&&&
\\[-.2cm]
  $gg \to gg$ & 
  $\mathcal{G} \, \left[ \ \ggtoggtchannel{1} - \xi \ \ggtoggschannel{1.2} \ \right]$ &
  $\displaystyle 4  \Nc^2 \frac{(1 - \xi \xibar)^3}{\xi^2 \xibar^2} $ & 
  $\ba{c} \bm{8}_{\rm a} \\[2.5mm] \bm{10}\oplus\bm{\overline{10}} \\[2.3mm] \bm{1} \\[2mm] \bm{8}_{\rm s} \\[2mm] \bm{27}  \\[2mm] \bm{0}  \\[2mm] \ea$ & 
  $\ba{c} \frac{\Nc}{2} (\xibar -\xi) \, \mathcal{G}  \\[1.8mm] 0  \\[2.2mm] \Nc  \, \mathcal{G}  \\[2mm] \frac{\Nc}{2} \, \mathcal{G} \\[2mm]  - \mathcal{G} \\[2mm]  \mathcal{G}  \\[1mm] \ea$ 
  \\[8mm] 
  \hline &&&&
\\[-2mm]
  $gg \to q \bar{q} $ & 
  $\mathcal{H} \,\left[ \ \Mggqq \ \right]$ & 
  $\displaystyle (\xi^2+\xibar^2) \Big( \frac{\CF}{\xi \xibar} - \Nc \Big)$ & 
  $\ba{c} \bm{1}  \\[2.7mm] \bm{8}_{\rm a} \\[2.7mm]  \bm{8}_{\rm s} \ea$ & 
  $\ba{c} \frac{\sqrt{N^2-1}}{2\sqrt{\Nc}} \, \mathcal{H}  \\[2mm] \frac12 \left( \xibar - \xi \right) \frac{\sqrt{\Nc}}{\sqrt{2}} \,\mathcal{H} \\[2mm] \frac{\sqrt{\Nc^2-4}}{2\sqrt{2\Nc}}\,\mathcal{H}  \ea$
  \\[10mm]
  \hline 
\end{tabular}
  \caption{\label{tab:channels} 
For each partonic channel, we give the production amplitude ${\cal M}$, where the graphs stand for color factors only, and the coefficients $\mathcal{A}$, $\mathcal{B}_t$, $\mathcal{B}_u$, $\mathcal{C}$, $\ldots$ for the Lorentz structure. In the next column, we give the square of the amplitude, $\trd \trc |{\cal M}|^2$, summed over spins, helicities and colors of all partons. We then list the coefficients $\nu_{\al}$ (normalized by $\sqrt{K_{\al}}$) of the color decomposition \eq{M-deco} of ${\cal M}$, for each irrep $\al$ of the channel under consideration.} 
\end{table}

\paragraph{Example 2: $\bm{qg \to qg}$, $\bm{gg \to gg}$ and $\bm{gg \to q \bar{q}}$ channels}

These three processes, for which the incoming parton from the target (parton 2) is a gluon, can be treated on the same footing. The gauge invariant amplitude ${\cal M}$ is most easily derived in $A^+=0$ gauge, and for each process, there are a priori three possible different color graphs in ${\cal M}$. As is readily obtained from color conservation (viewing the target gluon as the generator of an infinitesimal color rotation), only two of the latter are independent, leading to the entries of the second column of table~\ref{tab:channels} for those processes. In addition, the spin/helicity dependence factors out and is encoded in the factors $\mathcal{F}, \mathcal{G}, \mathcal{H}$ given by 
\be \label{FGH}
\mathcal{F} = \frac{g^2}{\sqrt{\xi \xibar}} \  {\bm \varepsilon}  \cdot {\bm U}_{q}^{g} \ \ ; \ \ \ \
\mathcal{G} = \frac{g^2}{\sqrt{\xi \xibar}} \  {\bm \varepsilon}  \cdot {\bm U}_{g}^{g} \ \ ; \ \ \ \ 
\mathcal{H} = \frac{g^2}{\sqrt{\xi \xibar}} \  {\bm \varepsilon}  \cdot {\bm U}_{g}^{\bar{q}} \, .
\ee
Here ${\bm \varepsilon} \equiv {\bm \varepsilon}_\lambda$ ($\lambda = \pm 1$) is the target gluon  
transverse polarization vector (taken to be orthogonal to the $z$-axis in $A^+=0$ gauge), and ${\bm U}_{a_1}^{a_3}$ is also orthogonal to the $z$-axis.

The vector ${\bm U}\,$ (which depends on $\xi$ and on the helicities of partons $a_1$, $a_3$ and $a_4$)
is different for the three processes, but it satisfies the general property
\be \label{Usum}
\sum_{\rm hel.}  U^i U^{j \, *} = 2 V_{a_1}^{a_3}(\xi) \, \delta^{ij} \ , 
\ee
where $V_{a_1}^{a_3}(\xi)$ is the `color-stripped' splitting function of parton $a_1$ into parton $a_3$, namely,
\be \label{splitting}
V_{q}^{g}(\xi) = V_{q}^{q}(\xibar) = 2\,  \frac{1+\xibar^2}{\xi} \ ; \ \ 
V_{g}^{g}(\xi) = 4 \, \frac{(1 - \xi \xibar)^2}{\xi \xibar} \ ; \ \ 
V_{g}^{\bar{q}}(\xi) = V_{g}^{q}(\xibar) = 2\, (\xi^2+ \xibar^2) \ .  
\ee
Using~\eq{FGH}, \eq{Usum} and \eq{splitting}, one easily recovers the textbook expressions of $\tr |{\cal M}|^2$ for the three processes, given for convenience in the third column of table~\ref{tab:channels}. 

In order to obtain the coefficients $\nu_{\al}$ of the color decomposition \eq{M-deco} of ${\cal M}$, we proceed as in the $q q \to q q$ case described previously. For each channel we express the color graphs appearing in ${\cal M}$ (second column of table~\ref{tab:channels}) as a linear combination of the projectors $\proj_{\al}$ (given in table~\ref{tab:irreps}) or transition operators $\trans_{\al}$ [given in \eq{trans-op}] corresponding to the accessible color states in that channel. For the $qg \to qg$ channel, this is straightforward. 
For $gg \to gg$, we use 
\be
\ggtoggschannel{1.2} \ = \ \Nc \, \Proj{\bm{8}_a}  \ \ \ {\rm and} \ \ \ 
\ggtoggtchannel{1} \ = \ \sum_{\al} b_{\al} \, \proj_{\al} \, , 
\ee
where by projecting the latter equation on $\proj_{\al}$, we readily interpret $-2 b_\al$ as the `color interaction potential' of a gluon pair in color state $\al$~\cite{Dokshitzer:1995fv} and thus get  
\be
b_{\al}  = -\frac12 (C_\al - 2 \Nc)  \, . 
\ee
From the above equations we readily obtain the coefficients $\nu_{\al}$ of the $gg \to gg$ amplitude. Note that the coefficient of the antisymmetric octet ($\al = \bm{8}_{\rm a}$) is odd under the exchange $\xi \leftrightarrow \xibar$ of the two final state partons, and the coefficient of $\bm{10}\oplus\bm{\overline{10}}$ identically vanishes (for the $gg \to gg$ amplitude at leading-order). 
For the $gg \to q \bar{q}$ amplitude, the coefficients $\nu_{\al}$ of the decomposition 
\eq{M-deco} in terms of the transition operators \eq{trans-op} are obtained by using 
\bea
\tikeq{\begin{scope}[scale=1]
		\coordinate (O) at (0,0);
		\draw[glu] ($(O)+(0,0)$) -- ++(0.5,.1) coordinate(A);
		\draw[anti] (A) -- ++(0.5,-.1);
		\draw[glu] ($(O)+(0,.7)$) -- ++(0.5,-.1)coordinate(B);
		\draw[qua] (B) -- ++(0.5,.1);
		\draw[qua] (A) --(B);
\end{scope}} &=&  \frac{1}{2 \Nc} \ \Tonegg + \frac12 \ \TeightAgg  + \frac12 \ \TeightSgg  \nn \\[2mm] 
&=&  \mathsize{14}{\frac{\sqrt{N^2-1}}{2 \sqrt{\Nc}} \, \Trans{\bm 1}  + \frac{\sqrt{\Nc}}{2\sqrt{2}} \,  \Trans{\bm{8}_a}  +  \frac{\sqrt{N^2-4}}{2 \sqrt{2 \Nc}} \, \Trans{\bm{8}_s}} \ . 
\eea

The derivation of the coefficients $\nu_{\al}$ for all other channels of table~\ref{tab:channels} goes along the same lines. 

Let us mention that for the process $gg \to Q\bar{Q}$ with massive quarks (not listed in table~\ref{tab:channels}) the results are the same as for $gg \to q\bar{q}$ (last row of table~\ref{tab:channels}), up to a modification of the helicity dependent factor, $\mathcal{H} \to \mathcal{H}'$, leading to a modification of the squared amplitude:
\be
\mathsize{14}{\frac{1}{4 g^4 (N^2-1)}} \, \tr |{\cal M}(gg \to Q\bar{Q})|^2 = \left[ 1 +\frac{4 m^2 K_\perp^2}{\xi \xibar (\xi^2+\xibar^2) s^2} \right]  (\xi^2+\xibar^2) \Big( \frac{\CF}{\xi \xibar} - \Nc \Big) \ .
\ee
Here $\xi$ and $\xibar$ are still given by their Lorentz-invariant definition \eq{eq:xi} (with now $m \neq 0$), and $K_\perp^2 = \xi  \xibar s -m^2$ [see \eq{pair-kinematics}]. 
In the present study we only need the coefficients $\nu_{\al}$, to be used in the expression of the density matrix $\Phi$ defined in eq.~\eq{Phi-def}. Since for $gg \to Q\bar{Q}$, all $\nu_{\al}$'s are proportional to the {\it same} helicity dependent factor $\mathcal{H}'$, the latter drops out in the expression of $\Phi$, which is thus the same for $gg \to Q\bar{Q}$ and $gg \to q\bar{q}$. 
A similar remark applies to $q \bar{q} \to Q \bar{Q}$ as compared to $q \bar{q} \to q' \bar{q}'$, due to the triviality of this process: the amplitude being given by a single Feynman diagram, there is only one helicity dependent factor (say, $\mathcal{D}$ for $q \bar{q} \to q' \bar{q}'$ and $\mathcal{D}'$ for $q \bar{q} \to Q \bar{Q}$), leading to the same density matrix for the two processes. 

Finally, let us recall that in our setup (see section~\ref{sec:setup}), we do not need to consider those $2 \to 2$ processes where some of the initial parton is a heavy quark. For our purpose, the list of channels in table~\ref{tab:channels}, complemented by $q \bar{q} \to Q \bar{Q}$ and $gg \to Q\bar{Q}$, is thus complete. (The processes ${\bar q} {\bar q}^{\,\prime} \to {\bar q} {\bar q}^{\,\prime}$, ${\bar q} {\bar q} \to {\bar q} {\bar q}$, ${\bar q} g \to {\bar q} g$ are trivially obtained from $q q' \to q q'$, $q q \to q q$, $q g \to q g$ by complex conjugation, and the process $q \bar{q} \to gg$ from $gg \to q \bar{q}$ by time reversal.)

\subsection{Explicit form of matrix $\Phi$} 
\label{app:Phi-matrix}

Here we give the density matrix $\Phi$ defined in \eq{Phi-def}, for each of the partonic channels of table~\ref{tab:channels}. Note also that the diagonal elements of $\Phi$ are the probabilities $\rho_\al$ defined in \eq{rho-def}, which for some of the processes (namely, $qg \to qg$, $gg \to gg$ and $gg \to q \bar{q}$) were evaluated previously~\cite{Peigne:2014rka,Arleo:2020hat,Arleo:2021bpv}. 

For each partonic channel, we define the matrix $\Phi$ by ordering the irreps $\al$ as in the second to last column of table~\ref{tab:channels}. The hard production amplitudes ${\cal M}$ are given in lightcone gauge, but the gauge invariance of the expressions below may be demonstrated in a general covariant gauge. A {\sc form}~\cite{Kuipers:2012rf} code implementation is attached as an ancillary file to the arXiv record of this paper, using tree-level graphs generated by {\sc qgraf}~\cite{Nogueira:1991ex} and including Faddeev-Popov ghosts. With the notation $U_k \equiv \sqrt{\Nc +k}\,$, $D_k \equiv \sqrt{\Nc -k}$ and $K_A \equiv \Nc^2 -1\,$, the matrices are given by (recall that $q$ and $q'$ denote distinct quark flavors)
\be
\label{phi 1}
\Phi_{ q q^\prime \to q q^\prime} = \frac{1}{2\Nc}  \left( 
\begin{matrix}  \Nc+1  &  \ - \sqrt{K_A}   \\  - \sqrt{K_A}  &\ \Nc-1
\end{matrix} 
\right)  \, , \hskip 50.5mm 
\ee
\bea
\label{phi 2}
\Phi_{qq\to qq} &=& \frac1{\Nc \big(\xi^4+ \bar{\xi}^4+\xi^2+\xibar^2 \big) -2\xi\bar{\xi} \, } \nn \\
&\times& \left( 
\begin{matrix}  
    \frac12 (\Nc+1) \left[(\xi-\xibar)^2+\xi^4+\bar{\xi}^4 \right]
    & \,
  - \sqrt{K_A} (\xibar-\xi) (1 - \xi\xibar)\\[1mm]  - 
  \sqrt{K_A} (\xibar-\xi) (1 - \xi\xibar)
   &  \,
   \frac12 (\Nc-1) \left[1+\xi^4+\bar{\xi}^4 \right]
\end{matrix} 
\right)  \, , \hskip 4mm
\eea
\be
\label{phi 3}
\Phi_{q\bar{q}^\prime \to q\bar{q}^\prime} = \frac{1}{\Nc^2}  \left( 
\begin{matrix}  K_A  &  - \sqrt{K_A}   \\  - \sqrt{K_A}  & 1
\end{matrix} 
\right) \, , \hskip 52mm
\ee
\be
\label{phi 4}
\Phi_{q\bar{q} \to q^\prime \bar{q}^\prime} = \left( 
\begin{matrix} \, 0 \  & \ 0 \, \\  \, 0 \  & \ 1 \,  
\end{matrix} 
\right) \, , \hskip 76.5mm
\ee
\bea
\label{phi 5}
\Phi_{q\bar{q} \to q\bar{q}} &=& 
  \frac1{\Nc \big[1+\xibar^2+\xi^2(\xi^2+\xibar^2)\big] + 2 \xi \xibar^2}  \nn \\
&\times& \left( 
\begin{matrix} 
  2\CF\big(\xibar^2+1\big) & \ 
  - \frac{\sqrt{K_A}}{\Nc} \big[ 1 + (1+\Nc \xi) \xibar^2 \big]
  \\[1mm]
  - \frac{\sqrt{K_A}}{\Nc} \big[ 1 + (1+\Nc \xi) \xibar^2 \big]
  & \ \frac1{\Nc} \big[ 1 + \xibar^2 (1+\Nc \xi)^2 + \Nc^2 \xi^4 \big]
\end{matrix} 
\right) \, , \hskip 4mm
\eea
\bea
\label{phi 6}
\Phi_{qg \to qg} &=&  \frac{1}{\CF \xi^2 + \Nc \xibar} \nn \\
&\times& { \left( 
\begin{matrix} \CF \left( \xibar + \frac{1}{K_A} \right)^2 &  \frac{U_1 D_2}{2\sqrt{2}} \left( \xibar + \frac{1}{K_A} \right) & -  \frac{U_2 D_1}{2\sqrt{2}} \left( \xibar + \frac{1}{K_A} \right) \\[2mm]  \frac{U_1 D_2}{2\sqrt{2}}  \left( \xibar + \frac{1}{K_A} \right)  & \frac{\Nc(\Nc-2)}{4(\Nc-1)} & - \frac{\Nc}{4} \frac{U_2 D_2}{U_1 D_1} \\[2mm] 
-  \frac{U_2 D_1}{2\sqrt{2}} \left( \xibar + \frac{1}{K_A} \right) & - \frac{\Nc}{4} \frac{U_2 D_2}{U_1 D_1}  & \frac{\Nc(\Nc+2)}{4(\Nc+1)}  
\end{matrix} 
\right) } \, , \hskip 8.5mm
\eea
\bea
\label{phi 7}
\Phi_{gg \to gg}  &=&  
\frac{1}{2\left(1+ \xi^2 + \xibar^2 \right)} \nn \\
&\times& \left(
\begin{matrix}
   \left(  \xibar - \xi \right)^2 \ 
  & 
   0
  & 
   \frac{2(\xibar - \xi)}{\sqrt{K_A}} 
  &
   (\xibar - \xi)
  &
   \frac{U_3 (\xi - \xibar)}{U_1}
  &
   \frac{D_3 (\xibar - \xi)}{D_1}
  \\[2mm]
   0 \ 
  &
   0
  &
   0
  &
   0
  &
   0
  &
   0
  \\[2mm]
   \frac{2(\xibar - \xi)}{\sqrt{K_A}} \ 
  &
   0
  &
   \frac{4}{K_A}
  &
   \frac{2}{\sqrt{K_A}}
  &
   -\frac{2 U_3}{U_1 \sqrt{K_A}}
  &
   \frac{2 D_3}{D_1 \sqrt{K_A}}
  \\[2mm]
   (\xibar - \xi) \ 
  &
   0
  &
  \frac{2}{\sqrt{K_A}}
  &
   1
  &
   -\frac{U_3}{U_1}
  &
   \frac{D_3}{D_1}
  \\[2mm]
   \frac{U_3 (\xi - \xibar)}{U_1} \ 
  & 
   0
  &
   -\frac{2 U_3}{U_1 \sqrt{K_A}}
  &
   -\frac{U_3}{U_1}
  &
   \frac{\Nc+3}{\Nc+1}
  &
   -\frac{U_3 D_3}{U_1 D_1}
  \\[2mm]
   \frac{D_3 (\xibar - \xi)}{D_1} \ 
  &
   0
  &
    \frac{2 D_3}{D_1 \sqrt{K_A}}
  &
   \frac{D_3}{D_1}
  &
   -\frac{U_3 D_3}{U_1 D_1}
  &
  \frac{\Nc-3}{\Nc-1}
\end{matrix}
\right) \, , \hskip 18.5mm
\eea
\bea
\label{phi 8}
\Phi_{gg \to q\bar{q}} &=& \frac{1}{\Nc^2\left( \xi^2 + \xibar^2 \right) -1} \nn\\
&\times& \left( 
\begin{matrix} 1 & 
  \frac{\Nc(\xibar-\xi)}{\sqrt{2}} & \frac{U_2 D_2}{\sqrt{2}} \\[2mm] 
  \frac{\Nc(\xibar-\xi)}{\sqrt{2}} & \frac12 \Nc^2 (\xi - \xibar)^2   & \frac12 \Nc U_2 D_2 (\xibar - \xi) \\[2mm]
   \frac{U_2 D_2}{\sqrt{2}}  & \frac12 \Nc U_2 D_2 (\xibar - \xi) & \frac12(\Nc^2-4) 
\end{matrix} 
\right)  \, . \hskip 27mm
\eea
It is apparent from eqs.~\eq{phi 1}--\eq{phi 8} that all the matrices $\Phi$ satisfy $\Tr \, \Phi = 1\,$, as they should. Furthermore, for the channels which depend on a single helicity-dependent factor 
(\ie, all channels but $qq \to qq$ and $q\bar{q} \to q\bar{q}$) the matrix $\Phi$ has the property that every $2\times 2$ `sub-determinant' vanishes (implying ${\rm Det}\, \Phi = 0$). This directly follows from \eq{Phi-def} because, for those channels, the sum over quark helicities acts similarly for any matrix element and can thus be effectively removed. In those cases $\Phi$ can be rapidly constructed from the knowledge of its diagonal elements, and by noting that the sign of a non-diagonal element $\Phi_{\al\beta}$ is given by the sign of $\nu_{\al}^{*} \, \nu_{\beta}$, which is directly inferred from the last column of table~\ref{tab:channels}. 

The $2\to 2$ matrix elements obey crossing symmetries which enable certain matrices to be related to one another by interchanging external particles. Specifically, using $\bar i$ to denote the antiparticle of $i\,$, one has 
\bea
{\cal M}_{12 \to 34}(p_1,p_2;p_3,p_4) 
&=& {\cal M}_{1\bar3 \to \bar2 4}(p_1,-p_3;-p_2,p_4)  \label{cross t}\\
&=& {\cal M}_{1\bar4 \to 3\bar2}(p_1,-p_4;p_3,-p_2)  \label{cross u}\, .
\eea
The first equality \eq{cross t} amounts to exchange $s \leftrightarrow t$ (or $\xi \to \frac1{\xi}$) in the original scattering amplitude, and the second equality \eq{cross u} amounts to exchange $s \leftrightarrow u$ (or $\xi \to \frac{\xi}{\xi-1}$). However, when it comes to $\Phi$ we also need to change the orthonormal color basis in accordance with \eq{S and A trans}. Hence we obtain, for example, 
\bea
\label{cross-rel-1}
\Phi_{_{\!1\bar3\to \bar2 4}} (\xi) &=&
A_{ts} \cdot \Phi_{_{\!12\to34}}\bigg( \frac1{\xi} \bigg) \cdot A^{-1}_{ts} \, ,
\eea
where $A_{ts}$ is the rotation matrix relating the appropriate bases as discussed in section~\ref{sec:matching}. What plays the role of the $s$-channel basis for $12 \to 34\,$,
is the $t$-channel basis for $1\bar3 \to \bar2 4\,$. Similarly, the second crossing relation \eq{cross u} leads to
\bea
\label{cross-rel-2}
\Phi_{_{\!1\bar4\to 3\bar2}} (\xi) &=&
A_{us} \cdot \Phi_{_{\!12\to34}}\bigg( \frac{\xi}{\xi-1} \bigg) \cdot A^{-1}_{us} \, .
\eea
Another trivial crossing relation stems from exchanging particle 3 with particle 4, which does not
require any change of color basis:
\bea
\label{cross-rel-3}
\Phi_{_{\!12\to43}} (\xi) &=& \Phi_{_{\!12\to34}}(1-\xi) \, ,
\eea
corresponding to $t \leftrightarrow u\,$.

\section{FCEL spectrum beyond leading-log for $2 \to \np$ processes} 
\label{app:master-spec}

The purpose of this appendix is to derive the expression~\eq{master-spectrum} of the FCEL spectrum associated to hard $2 \to 2$ partonic processes (viewed as $1\to 2$ {\em forward} processes) studied in this paper. We first rederive (section~\ref{app:qtoq}) the medium-induced spectrum found in ref.~\cite{Peigne:2014uha} for hard $2\to 1$ (\ie, $1\to 1$ forward) processes in a more synthetic way, in particular by making use of color conservation at a certain stage of the calculation. We then show how this generalizes to $2\to 2$ processes (section~\ref{app:master-spectrum}), thus obtaining the associated induced spectrum beyond logarithmic accuracy~\eq{master-spectrum}, as well as to $2 \to \np$ processes with $\np >2$. In particular, a conjecture of ref.~\cite{Peigne:2014rka} made within logarithmic accuracy for the induced spectrum of the latter processes is validated.

\subsection{Reminder of $2 \to 1$ processes}
\label{app:qtoq} 

Here we consider a simple theoretical setup where the underlying hard partonic process is $qg \to q\,$, viewed as forward quark scattering mediated by a single $t$-channel gluon exchange~\cite{Peigne:2014uha}. This case will be sufficient to infer the FCEL spectrum associated to any $2 \to 1$ process, given by Eqs.~\eq{spectrum-1to1}--\eq{sig-fun}. The derivation below follows the lines of ref.~\cite{Peigne:2014uha}, but presents several simplications, which will make the generalization to $2 \to \np$ processes quite straightforward. 

First, we define the $qg \to q$ process as a special kinematic limit of the $qg \to qg$ process, namely, when the final gluon (corresponding to parton 4 of figure~\ref{fig:2to2}) has a negligible longitudinal momentum, $p_4^{+} \simeq 0\,$: 
\be
\label{2to1vs2to2}
{\cal M}_{qg\to q}^{\rm vac} \ \equiv\  \lim_{\xi \to 0} \Big( {\cal M}_{qg \to qg}^{\rm vac} \Big)
\ = \ \vacuumamp{\bm{\scriptstyle{q}}\!\uparrow}{1.3} \ . 
\ee
In the $\xi \to 0$ limit, the $qg \to qg$ amplitude is dominated by a single diagram corresponding to a gluon exchange in the $t$-channel. (This is consistent with the color structure of the $qg \to qg$ amplitude when $\xi \to 0$, see the first column of Table~\ref{tab:channels}.) The ``$qg \to q$ amplitude''~\eq{2to1vs2to2} is thus well-defined and, in particular, gauge-invariant. 
In~\eq{2to1vs2to2} and the following, a red gluon will always denote a gluon carrying the hard transverse momentum scale. For example, the final gluon in~\eq{2to1vs2to2} has $p_4^{+} \simeq 0$ but carries the hard transverse momentum $-\bm q$.

Due to the presence of only one energetic parton in the final state, the calculation of the induced spectrum associated to $qg \to q$ is simpler than in the general case of $2 \to 2$ processes (with no kinematic restriction) addressed in the next section. In particular, we may think of \eq{2to1vs2to2}, viewed in the target rest frame, as an incoming quark with light-cone momentum $p_1 = (2E, 0, \bm 0)$ undergoing a transfer of momentum $q = (0, q^-, \bm q)$,\footnote{Unlike $p_2$ and $p_4$, which are on-shell, the four-momentum $q$ is virtual.} drawn as
\be
\label{2to1amp-blob}
{\cal M}_{qg\to q}^{\rm vac}  
\ \equiv \ \vacuumampblob{\bm{\scriptstyle{q}}\!\uparrow}{1.3} \ .
\ee
Here the red gluon highlights the hard transverse exchange with the incoming quark, which is crucial to keep track of the color structure of the process, and the red blob hides partons 2 and 4 [parton 4 being the final {\it recoil} gluon of transverse momentum $-\bm q$ in~\eq{2to1vs2to2}], whose kinematic details are irrelevant for the calculation of the radiation spectrum in the present section.

The final quark of momentum $p_3= p_1+q$ is assumed to hadronize into a tagged hadron of sufficiently hard transverse momentum $\bm p$, implying that $\bm q = \bm p /z$ (with $z$ the fragmentation variable) is hard, $ |\bm q|  \equiv q_\perp \gg \lqcd$. However, throughout this study we focus on {\it moderately large} $q_\perp$, namely, $q_\perp \ll E$, corresponding to small angle ({\it forward}) scattering in the target rest frame. 

The squared matrix element for the hard $qg \to q$ process (referred to as the `vacuum' process in what follows) may be illustrated as
\be
\label{2to1:vac}
\big\vert {\cal M}_{qg\to q}^{\rm vac} \big\vert^2 \ =\ \vacuum{\bm{\scriptstyle{q}}}{1.2} \ \  ,
\ee
where the upper and lower halfs (separated by the dashed horizontal line) 
represent the amplitude and conjugate amplitude respectively.

In the case of a nuclear target of finite size $L$, the above picture for $qg \to q$ scattering is altered by multiple rescatterings, contributing to the final hadron transverse momentum $\bm p$. However, when $p_\perp$ is much larger than the average transverse momentum broadening, the parent quark with $\bm K = \bm p /z$ must undergo a {\it single hard} exchange $\bm q$ (since the probability of a second hard exchange is power suppressed), accompanied by  in-medium {\em soft} rescatterings $\bm \ell_i$ such that $\bm \ell \equiv \sum_i \bm \ell_i$ satisfies $|\bm \ell| \ll q_\perp$, implying $\bm K = \bm q + \bm \ell \simeq \bm q\,$. 
This is the regime entailed by our initial assumption 
$Q_s \ll K_\perp$ (see section~\ref{sec:spectrum}), where the saturation 
scale $Q_s$ is of the same order as $|\bm \ell|\,$. 
Although soft rescatterings modify negligibly the kinematics of the vacuum process, they affect the color field of the energetic parton, resulting in induced gluon radiation. 
Compared with \eq{2to1:vac}, in-medium parton propagation will be denoted by a blue rectangle, 
encoding the additional rescatterings $\bm \ell_i$ (which may occur in any ordering w.r.t.~to the hard exchange $\bm q$), 
\be
\label{2to1:med}
\big\vert {\cal M}_{qg\to q}^{\rm med} \big\vert^2 \ = \ \ \medium{1.2} \ \ .
\ee

For the next step, consider gluon radiation associated with this specific process and kinematics, carrying away a momentum $k=(k^+, \bm k^2/k^+, \bm k)$, and assumed to be soft ($x \equiv k^+/(2E) \ll 1$) and emitted at small angle ($|\bm k| \equiv k_\perp \ll k^+$).  The radiation spectrum can be obtained by resumming to all orders, as done in the BDMPS-Z~\cite{Baier:1996kr,Zakharov:1997uu} and AMY~\cite{Arnold:2002ja} formalisms, the number $\ns$ of scattering centers (located at longitudinal positions $z_i$) on which the energetic quark has rescattered. The zeroth order term corresponds to the spectrum associated with the vacuum process, and the expansion in terms of $\ns$ is called the {\em opacity expansion}~\cite{Gyulassy:2000er}. Here we are looking for the additional, {\it medium-induced} spectrum, obtained by summing over $\ns \geq 1$ the radiation spectrum for a given $\ns$,
which reads 
(see appendix A of ref.~\cite{Peigne:2014uha})
\bea
x \frac{\dd I^{(\ns)}}{\dd x} = 
\frac{\alpha_s}{\pi} 
\int \frac{\dd^2 \bm k}{\pi} \left[ \, \prod_{i=1}^{\ns} 
\int \frac{\dd z_i}{\Nc \lambda_g} 
\int \dd^2 \bm \ell_i \, V(\bm \ell_i) \, \right]  \, C_\ns(\bm k, \bm q) \ ,  && 
\label{spec-order-n} 
\\ 
C_\ns(\bm k, \bm q) = \, \frac{2}{ \big\vert {\cal M}_{qg\to q}^{\rm vac} \big\vert^2 } 
\ \Cnkq{0}{1} \ . \hskip 7mm && 
\label{eq:Cn}
\eea
In eq.~\eq{eq:Cn} and in the following, within any blue rectangle a sum is implicitly performed over all possible connections of rescattering gluons to the hard quark and antiquark lines and to the soft radiated gluon. Note that the color structure of graphs is evaluated using the pictorial rules recalled in footnote~\ref{foot:birdtrack-rules}. 

\bigskip 

A few comments on eqs.~\eq{spec-order-n}--\eq{eq:Cn} are in order~\cite{Peigne:2014uha}:
\begin{itemize}
  \item[(1)]{The rescatterings $\bm \ell_i$ are modelled as a sequence of individual scatterings off static centers [denoted by crosses in \eq{eq:Cn}], typically separated by the elastic mean free path 
$\lambda_1$ of the fast parton of Casimir $C_1$~\cite{Gyulassy:1993hr}. An average over $\bm \ell_i$ is performed using the screened Coulomb potential $V(\bm \ell_i) = \mu^2/[\pi (\bm \ell_i^2 + \mu^2)^2]$ (normalized as $\int \dd^2\bm \ell \, V(\bm \ell) = 1$). With the screening length $1/\mu$ of the medium satisfying $1/\mu \ll \lambda_1$, the potential is screened enough so that successive rescatterings may be treated independently. Note that $\mu$ is the typical magnitude of a single transfer, $|\bm \ell_i| \sim \mu$. Averages over the $\bm \ell_i$'s will be implicit in the following.}
\item[(2)]{\label{bullet2}We do not need to assume $L \gg \lambda_1$. Resumming all orders in opacity should encompass the typical cases of a large nuclear target ($L \gg \lambda_1$) and of a small proton target ($L \sim \lambda_1$). In the latter case, the spectrum will be simply dominated by the first term (linear in $L$) of the opacity expansion. Note that defining $\hat{q} \equiv \mu^2/\lambda_g$, with $\lambda_g$ the gluon mean free path, the estimate of the average broadening $\sqrt{\hat{q}L} = \mu \sqrt{L/\lambda_g}$ holds at both small and large $L$.}
\item[(3)]{The factor $\Nc \lambda_g$ in \eq{spec-order-n} arises independently of the type of parton which propagates in the 
medium~\cite{Peigne:2014uha}, and can be viewed as the `color stripped' elastic mean free path 
$\lambda \equiv \Nc \lambda_g = C_F \lambda_q$. This explains why the spectrum \eq{arb order opacity} below turns out to depend on the {\it gluon} mean free path (due to the fact that $C_\ns(\bm k, \bm q) \propto \Nc^\ns$, see eqs.~\eq{CnGamma} and \eq{Gamma-psi-n}), for any partonic process.}
\item[(4)]{The specific setup under consideration leads to important simplifications~\cite{Peigne:2014uha}:  
i) the induced radiation spectrum arises from large gluon formation times $t_f \gg L$, corresponding to the emission vertices in the amplitude and its conjugate being {\em outside} the nuclear medium, 
ii) diagrams where the hard $t$-channel gluon couples to the soft radiation are negligible (which follows from $k_\perp \ll q_\perp$), and 
iii)
purely initial and purely final state radiation cancels out in the induced spectrum, leaving only a contribution from interference terms [of the form drawn in \eq{eq:Cn}].}
\end{itemize}

For a given $\ns \geq 1$, we now derive the quantity $C_\ns$ defined by \eq{eq:Cn}. The derivation is similar to that of ref.~\cite{Peigne:2014uha}, but with important technical simplifications. First, there is no need to consider quantities depending on the medium size $L$, since the integrals over the longitudinal positions of the scattering centers in \eq{spec-order-n} are trivial, and just provide a factor $L^\ns/(\ns!)\,$. We simply draw crosses [in \eq{eq:Cn} and the diagrams below] to remind that the $\ns$ rescatterings are ordered. Second, the treatment of the color structure of \eq{eq:Cn} will be simplified, making the generalization to $2 \to \np$ processes easier (see section~\ref{app:master-spectrum}). 

Since the soft scatterings $\bm \ell_i$ affect negligibly the final quark momentum, $C_\ns(\bm k , \bm q)$ is always proportional to the factor $-\psi_0(\bm k - x \bm q)^*$ (arising from the final gluon emission vertex in the conjugate amplitude), where $\psi_0(\bm k - x \bm q)$ is the light-cone wavefunction of the final quark-gluon fluctuation (in a parent quark of momentum $\bm q$). 
For a quark of mass $m$ the function $\psi_0(\bm k)$ reads 
\be
\label{psi0-k}
\psi_0(\bm k) \equiv \frac{\bm k \cdot \bm \varepsilon}{\bm k^2 + x^2 m^2} \, , 
\ee
with $\bm \varepsilon$ the polarization of the radiated gluon. Importantly, the wavefunction of $g \to g+g(k)$ is given by \eq{psi0-k} with $m=0$ and the variable $x$ appearing in $\psi_0(\bm k - x \bm q)$ and in \eq{psi0-k} is the radiated gluon momentum fraction w.r.t.~the parent parton. 

We can thus rewrite $C_\ns$ as 
\bea
\label{CnGamma}
C_\ns(\bm k, \bm q) &=& -  \psi_0(\bm k - x \bm q)^*  \cdot \Gamma_\ns(\bm k) \, , \hskip 1.5cm   \\
\label{Gamma-n}
\Gamma_\ns(\bm k) &\equiv& \, \frac{2}{ \big\vert {\cal M}_{qg\to q}^{\rm vac} \big\vert^2 } 
\ \Cnkq{1}{1} \ ,  
\eea
where the small circular blob outside the blue rectangle indicates that the final emission vertex now depends only on color indices, and we also anticipate that $\Gamma_\ns$ depends only on $\bm k$. 
$\Gamma_\ns(\bm k)$ can be obtained recursively as follows. First, in absence of rescatterings we find: 
\be
\label{initial-condition-Gamma}
\Gamma_0(\bm k) = \frac{2}{ \big\vert {\cal M}_{qg \to q}^{\rm vac} \big\vert^2 } 
\ \GammazeroNorescatt{0}{1} \ = \ \frac{2}{\vacuum{}{0.8}} \ \psi_0(\bm k)  \ \GammazeroNorescatt{1}{1} \ \  .
\ee
In the latter graph the only remaining Lorentz dependence is the factor $V(\bm q)$, which cancels with the same factor contained in $\big\vert {\cal M}_{qg\to q}^{\rm vac} \big\vert^2$. Thus $\Gamma_0(\bm k) $ is independent of $\bm q$ and reads
\be
\label{Gamma0}
\Gamma_0(\bm k) = \psi_0(\bm k) \, 2\, T_1 T_3 =  \psi_0(\bm k) \, (C_1+C_3-C_2 ) \, ,
\ee
where $T_1$ and $C_1=T_1^2$ (resp.~$T_3$ and~$C_3=T_3^2$) denote the $\sun$ generator and Casimir of the incoming (resp.~outgoing) parton, and $C_2 = (T_1 - T_3)^2$ the Casimir of the color exchange in the $t$-channel. (In the particular case $qg \to q$, we have $C_1=C_3=C_F$ and $C_2=\Nc$.)
Then, supposing $\Gamma_\ns(\bm k)$ is known, $\Gamma_{\ns+1}(\bm k)$ can be obtained by inserting an additional rescattering just after the last one in the graphical representation \eq{Gamma-n} of $\Gamma_\ns(\bm k)$, namely, by inserting the blob: 
\be
\label{blob}
\raisebox{-1.8mm}{\blueblob{1}} \ \ \equiv 
\rescatta{\bm{\scriptstyle{k}-\scriptstyle{\ell}}}{\bm{\scriptstyle{k}}}{1} + 
\rescattb{\bm{\scriptstyle{k}-\scriptstyle{\ell}}}{\bm{\scriptstyle{k}}}{1} + 
\rescattc{\bm{\scriptstyle{k}}}{\bm{\scriptstyle{k}}}{1} +  
\rescattd{\bm{\scriptstyle{k}}}{\bm{\scriptstyle{k}}}{1} + 
\rescatte{\bm{\scriptstyle{k}}}{\bm{\scriptstyle{k}}}{1} + 
\rescattf{\bm{\scriptstyle{k}}}{\bm{\scriptstyle{k}}}{1} \ . 
\ee
Here the graphs only represent color connections, but the momenta of the radiated gluon entering and exiting the blog are also indicated, to properly keep track of the transverse momentum flow when inserting \eq{blob} in \eq{Gamma-n}. The last soft exchange $\bm \ell_{\ns+1} \equiv \bm \ell$ is attached to the blue cross which denotes the last scattering center. The diagrams where this attachment occurs either only above the cross (in the amplitude) or only below the cross (in the conjugate amplitude) correspond to the so-called virtual corrections in the multiple scattering process. Inserting \eq{blob} in \eq{Gamma-n} we obtain
\be
\Gamma_{\ns+1}({\bm k}) = \Nc \int \dd^2 \bm \ell \, V(\bm \ell) 
\big[ \Gamma_{\ns}({\bm k - \bm \ell}) - \Gamma_{\ns}({\bm k }) \big] \, ,
\label{recurrenceG}
\ee
where on the r.h.s.~the first term arises from the first two terms of \eq{blob} and the second term arises from the last four terms of \eq{blob}. To reach \eq{recurrenceG} we used the fact that at any given time (specified by a vertical line) between two successive rescatterings in the blue rectangle of \eq{Gamma-n}, the overall $q\bar{q}g$ state is color singlet, $T_g + \sum_{i} T_i = 0$, where $T_g$ and $T_i$ denote the $\sun$ generators of the gluon and of all other partons (here $i=q, \bar{q}$), respectively. Thus, the first two terms of \eq{blob} sum up to the color factor (recall our conventions in footnote~\ref{foot:birdtrack-rules})
\be
\label{colfact1}
\rescatta{}{}{1} + \rescattb{}{}{1} \ = \ - T_g \cdot \sum_{i} T_i 
\ =\  T_g^2 
\ =\  \Nc \, , 
\ee
and the last four terms of \eq{blob} sum up to 
\bea
\label{colfact2}
\rescattc{}{}{1} + \rescattd{}{}{1} + \rescatte{}{}{1} + \rescattf{}{}{1}
&=& \ -\frac{1}{2} \left[ T_g^2 +   \sum_{i} T_i^2  \right] -  \sum_{i \neq j} T_i T_j  \nn\\
&=& \  -\frac{1}{2}   \left( T_g + \sum_{i} T_i \right)^2 + T_g \cdot \sum_{i} T_i  \nn\\[.4cm]
&=& \  - T_g^2 \ =\  - \Nc \, .  
\eea

Using \eq{Gamma0} and \eq{recurrenceG}, it is clear that $\Gamma_\ns(\bm k)$ is of the form
\be
\label{Gamma-psi-n}
\Gamma_\ns(\bm k)  = 2\, T_1 T_3 \, \Nc^\ns \, \psi_\ns(\bm k) \, , 
\ee
where $\psi_\ns(\bm k)$ satisfies the recurrence relation 
\be 
\psi_{\ns+1}({\bm k}) = \int \dd^2 \bm \ell \, V(\bm \ell) 
\big[ \psi_{\ns}({\bm k - \bm \ell}) - \psi_{\ns}({\bm k }) \big] \, ,
\label{recurrenceF}
\ee
with initial condition $\psi_0(\bm k)$. 
Equation~\eq{recurrenceF} can be solved exactly by going to impact parameter space to obtain:
\be
\label{psin-k}
\psi_{\ns}(\bm k) =  \int \!\! \frac{\dd^2 \bm b}{(2 \pi)^2} \, \psi_{0}(\bm b) \, [V(\bm b) -1 ]^\ns \, e^{i \footnotesize{\bm k \cdot \bm b}} \, , 
\ee
where $V(\bm b) = b \mu \, {\rm K}_1(b \mu)$ and $\psi_{0}(\bm b)$ is given by
\be
\label{psi0-b}
\psi_{0}(\bm b) = \int \dd^2 \bm k \, \psi_0(\bm k) \, e^{-i \footnotesize{\bm k \cdot \bm b}} = -2i\pi \frac{\bm b \cdot \bm \varepsilon}{b^2} \, x m b \, {\rm K}_1(x m b)  \, .
\ee

Using \eq{CnGamma}, \eq{Gamma-psi-n} and \eq{psin-k} in the expression \eq{spec-order-n} of the spectrum at order $\ns$ in opacity, and then summing over $\ns \geq 1$, we obtain for the medium-induced spectrum (with $r \equiv L/\lambda_g$),
\be
\label{arb order opacity}
x \frac{\dd I}{\dd x} = 
\frac{\alpha_s}{\pi^2} \, 2 T_1T_3  \int \!\! \frac{\dd^2 \bm b}{(2 \pi)^2} \, e^{i  x \footnotesize{\bm q \cdot \bm b}} \, \psi_{0}(\bm b)^* \psi_{0}(\bm b) \, \left[ 1 -e^{- r \, \footnotesize{\bm(1-V(b)\bm)} }\right] \, . 
\ee
The summation carried out above is valid for all $r$, which supports item (2) on page~\pageref{bullet2}. 
This is very different from the case of parton energy loss in the LPM regime, where the opacity expansion may not converge at large-$L$~\cite{Arnold:2008iy,Isaksen:2022pkj}. Using now \eq{psi0-b}, summing over the radiated gluon polarization, and attributing masses $m_1$ and $m_3$ to the incoming and outgoing quark, we get
\be
\label{spectrum-1to1}
x \frac{\dd I}{\dd x} = \frac{\alpha_s}{\pi} \, \big( C_3 + C_1 - C_2 \big) \, \sigma( \tilde{q}_\perp, \tilde{m}_1, \tilde{m}_3, r ) \ ; \ \ \ \ 
\tilde{q}_\perp \equiv \frac{x q_\perp}{\mu} \, ; \ \tilde{m}_i \equiv \frac{x m_i}{\mu} \, , 
\ee
where the function $\sigma(\tilde{q}_\perp, \tilde{m}_1, \tilde{m}_3, r)$ is defined as 
\be
\label{sig-fun}
\sigma(\tilde{q}_\perp, \tilde{m}_1, \tilde{m}_3, r) = \int_0^{\infty} \frac{\dd B^2}{B^2} \, {\rm J}_0(\tilde{q}_\perp B) \, \tilde{m}_1 B \, {\rm K}_1(\tilde{m}_1 B) \,  \tilde{m}_3 B \, {\rm K}_1(\tilde{m}_3 B)  \, \left[ 1 -e^{- r \, (1-B \, {\rm K}_1(B) )}  \right] \, . 
\ee
In the case of $qg \to q$ scattering, we have $m_1=m_3$. 
But it should be clear from the derivation that the result \eq{spectrum-1to1}--\eq{sig-fun} 
holds for any $2 \to 1$ scattering 
(including cases where the initial and final partons have different masses). 
In particular, using $\tilde{m}_i B \, {\rm K}_1(\tilde{m}_i B) \to 1$ 
when $m_i \to 0$, we check that \eq{spectrum-1to1}--\eq{sig-fun} 
encompasses the results of ref.~\cite{Peigne:2014uha} for $qg \to q$ ($m_1 =m_2= m_3 =0$) and $gg \to Q$ ($m_1 = m_2= 0$ and $m_3 \neq 0$).

\subsection{Generalization to $2 \to \np$ processes}  
\label{app:master-spectrum}

Here we derive the FCEL spectrum for $\np \geq 2$ particles in the final state, confirming a conjecture in ref.~\cite{Peigne:2014rka}. 
For illustration we consider the same $2\to 2$ process as in ref.~\cite{Peigne:2014rka} (called $1\to 2$ forward scattering in that reference), namely, $qg \to qg$ scattering, however without restricting to the large $\Nc$ limit, and in a way that makes obvious the generalization to any $2\to 2$ partonic process, as well as to $2 \to \np$ processes with $\np >2$. 

The picture for $qg \to qg$ scattering simply follows from the model for $qg \to q$ scattering considered in section~\ref{app:qtoq}, by promoting the recoil gluon drawn in Eq.~\eq{2to1vs2to2} (and carrying the transverse momentum $-{\bm q}$) to an {\it energetic} gluon. 
The $qg \to qg$ amplitude is of the form 
\be
\label{qtoqg:amp}
{\cal M}_{qg \to qg}^{\rm vac} \ \equiv \ {\cal M} \ = \ 
\qtoqgtchannel{1} \ + \ \qtoqguchannel{1} \ +\ \qtoqgschannel{1} \ \ , 
\ee
where the final quark and gluon transverse momenta are denoted as $\bm{p_3} \equiv -\bm{K}$ and $\bm{p_4} \simeq {\bm K}$ ($K_\perp$ playing the role of the hard scale), and their longitudinal momentum fractions as $\xi_3 = 1-\xi  \equiv \bar \xi$, $\xi_4 \equiv \xi$ (in conformity with the notations of section~\ref{sec:setup}, see figure~\ref{fig:2to2}). 
In Eq.~\eq{qtoqg:amp} (as in Eq.~\eq{2to1vs2to2}), the gluon in black is the incoming gluon from the target. As mentioned earlier, when $\xi \to 0$ the first ($t$-channel) term of \eq{qtoqg:amp} dominates, and \eq{qtoqg:amp} then coincides with the ``$2 \to 1$ amplitude''~\eq{2to1vs2to2}.

The square of \eq{qtoqg:amp} is given by the sum of graphs
\bea
\vert {\cal M} \vert^2 \ =\ \HardRescatta{1} \ + \ \HardRescattb{1} \ + \ \HardRescattc{1} \hskip 7mm &&  \nn \\[2mm]
+ \ 2 \left( \HardRescattd{1} \ + \ \HardRescatte{1} \ + \ \HardRescattf{1} \right) \ . \hskip 1mm  && 
\label{qtoqg:vac}
\eea

Similarly to \eq{2to1:med}, in a medium the hard process is supplemented by additional rescatterings, 
denoted by a blue rectangle. The rescatterings $\bm \ell_i$ [averaged with the Coulomb potential $V(\bm \ell_i)$] being soft compared to the inverse transverse size of the $qg$ pair, $|\bm \ell_i| \sim \mu \ll K_\perp \sim 1/r_\perp$, the pair behaves as a pointlike object w.r.t.~soft rescatterings. 
The derivation made in appendix A of ref.~\cite{Peigne:2014uha} for $2 \to 1$ processes also applies 
to $2 \to 2$ (and $2 \to \np$) processes, resulting in an expression for the medium-induced spectrum similar to \eq{spec-order-n}--\eq{eq:Cn}, up to the following modifications: the graph in the denominator of \eq{eq:Cn} is now replaced by the sum of graphs \eq{qtoqg:vac}, and in the numerator of \eq{eq:Cn} the additional soft radiated gluon $k$ (with $x \equiv k^+/p_1^+ \ll 1$) can be added to any of the diagrams contributing to the hard process. 
Equation~\eq{eq:Cn} thus becomes: 
\be
\label{eq:Cnbis}
C_\ns(\bm k, \bm K) = \, \frac{2}{\vert {\cal M} \vert^2} 
\ \left\{ \ \Cnbisellipse{1} \ \ + \ \ \ldots \ \ \right\}  \ \ ,
\ee
where inside the bracket we have drawn the contribution associated to the last diagram of \eq{qtoqg:vac}, and the dots stand for the contributions arising from the other diagrams. The small ellipse denotes the two possible attachments of the soft gluon in the conjugate amplitude, either to the hard gluon or to the hard quark. Finally, as in section~\ref{app:qtoq}, all connections of rescattering gluons $\bm \ell_i$ to the hard partons and to the soft radiated gluon are implicitly accounted for in the blue rectangle. 

Let us now focus on the first term in the bracket of \eq{eq:Cnbis} and its contribution to $C_\ns(\bm k, \bm K)$, and see how the hard part of this diagram [given by the last diagram of \eq{qtoqg:vac}] is `dressed' by multiple soft rescatterings. We first consider the attachment of the soft gluon to the final hard gluon in the conjugate amplitude. Factoring out the final emission vertex in the conjugate amplitude and using the notational convention of section~\ref{app:qtoq}, this specific contribution to  \eq{eq:Cnbis} reads
\be
- \Gamma_\ns(\bm k, \bm K) \cdot \psi_0^g\Big(\bm k - \frac{x}{\xi} \bm K\Big)^*  \, , 
\ee
where the upper script $g$ on $\psi_0^g$ indicates that we should use here the $g \to gg$ wavefunction, given by \eq{psi0-k} for $m=0$, and $\Gamma_\ns(\bm k, \bm K)$ is defined  by 
\be
\Gamma_\ns(\bm k, \bm K) = \, \frac{2}{ \vert {\cal M} \vert^2 } 
\ \GammanKK{1} \ \ ,
\ee
where the circular blob represents the final emission vertex as in the $2\to 1$ processes. One can show similarly to section~\ref{app:qtoq} that $\Gamma_\ns(\bm k, \bm K)$ obeys the recurrence relation \eq{recurrenceG}, however with the initial condition
\be
\Gamma_0(\bm k, \bm K) = \psi_0(\bm k) \, \frac{2}{\vert {\cal M} \vert^2} 
\ \Gammazerobis{0.85} \ \ ,
\ee
where $\psi_0(\bm k)$ is the {\it initial} $q \to qg$ wavefunction, obtained also from \eq{psi0-k} by setting $m=0$.

The crucial point in proving that $\Gamma_\ns(\bm k, \bm K)$ obeys \eq{recurrenceG} is that the systematics of eqs.~\eq{colfact1} and \eq{colfact2} holds independently of the number and type of hard partons accompanying the soft radiated gluon. Each rescattering will always provide a factor $\Nc$ in the recurrence relation \eq{recurrenceG}. 
This is a consequence of color conservation: the energetic parton system in the amplitude and that in the conjugate amplitude, together with the radiated gluon $\bm k$, always form an overall color singlet system. 

Since $\Gamma_\ns(\bm k, \bm K)$ satisfies \eq{recurrenceG} we define, analogously to \eq{Gamma-psi-n}, 
\be
\Gamma_\ns(\bm k, \bm K) = \Nc^\ns \, \psi_\ns(\bm k) \, \frac{2}{\vert {\cal M} \vert^2} 
\ \Gammazerobis{0.85} \ \ ,
\ee
and observe that $\psi_\ns(\bm k)$ obeys \eq{recurrenceF} with initial condition $\psi_0(\bm k)$, and is thus given by \eq{psin-k}. The specific contribution under consideration thus contributes to the spectrum as
\be
\label{specific-cont} \frac{\alpha_s}{\pi} \, \sigma\left( \frac{x K_\perp}{\xi \mu}, \frac{x m_1}{\mu}, \frac{x m_4}{\xi \mu}, r \right) \, \frac{2}{\vert {\cal M} \vert^2} \ \Gammazerobis{0.85} \ \ .
\ee
Here $m_4 =0$, but we keep it to make the generalization to any $2 \to 2$ scattering transparent. Let us emphasize that the variable $x$ appearing in the final $g \to gg$ wavefunction $\psi_0 (\bm k - x \bm K)$, with $\psi_0$ defined by \eq{psi0-k}, should be interpreted here as $x/\xi$. Thus, $\psi_0(\bm k - x \bm K) \to \psi_0(\bm k - x \bm K /\xi )$ {\it and} $x \to x/\xi$ in the denominator of \eq{psi0-k}. Thus, when $m_4 \neq 0$ both $K_\perp$ and $m_4$ are rescaled by $\xi$ in \eq{specific-cont}. 

Adding now to \eq{specific-cont} similar contributions (with the soft gluon still attached to the final hard gluon in the conjugate amplitude) arising from the other terms of \eq{eq:Cnbis} (represented by the dots), we obtain 
\be
\frac{\alpha_s}{\pi} \, \sigma\left( \frac{x K_\perp}{\xi \mu}, \frac{x m_1}{\mu}, \frac{x m_4}{\xi \mu}, r \right)  \, \frac{2}{\vert {\cal M} \vert^2} \ \MMgluon{0.9} \ \ ,
\ee
where the opaque grey rectangles stand for the complete hard amplitude \eq{qtoqg:amp}. 

The calculation of the contribution to the spectrum arising from the attachment of the soft gluon $\bm k$ to the final quark (in the conjugate amplitude) is completely analogous, up to the change $-\psi_0^g(\bm k - x \bm K / \xi )^* \to -\psi_0(\bm k + x\bm K/ \bar{\xi} )^*$ for the final emission vertex. Adding this to the previous contribution we obtain (for a final quark mass $m_3 \neq 0$):
\bea
x \frac{\dd I}{\dd x} = 
\frac{\alpha_s}{\pi} \, \frac{2}{ \big\vert {\cal M} \big\vert^2} 
\left\{ \ \ \sigma\left( \frac{x K_\perp}{\xi \mu}, \frac{x m_1}{\mu}, \frac{x m_4}{\xi \mu}, r \right)  
\ \MMgluon{0.9} \right. \hskip 1.4cm &&  \nn \\[2mm] 
\left. + \sigma\left(\frac{x K_\perp}{\bar{\xi} \mu}, \frac{x m_1}{\mu}, \frac{x m_3}{\bar{\xi} \mu}, r \right) 
\ \MMquark{0.9} \ \ \right\} \ \ . \hskip 4mm && 
\label{eq:final-specqtoqg}
\eea

Clearly, the derivation of the spectrum \eq{eq:final-specqtoqg} associated to $qg \to qg$ would be exactly the same for any other type of $2 \to 2$ process. The spectrum associated to a generic 
$2 \to 2$ partonic process thus reads 
\bea
x \frac{\dd I}{\dd x} = 
\frac{\alpha_s}{\pi} \, \frac{2}{ \big\vert {\cal M} \big\vert^2} 
\left\{ \ \ \sigma\left( \frac{x K_\perp}{\xi \mu}, \frac{x m_1}{\mu}, \frac{x m_4}{\xi \mu}, r \right)  
\ \xibird(0.5,-0.8) \right. \hskip 1.4cm &&  \nn \\[1mm] 
\left. + \sigma\left(\frac{x K_\perp}{\bar{\xi} \mu}, \frac{x m_1}{\mu}, \frac{x m_3}{\bar{\xi} \mu}, r \right)
\ \xibird(0.5,-0.2) \ \ \right\} \ \ , \hskip 4mm && 
\label{eq:final-spec2to2}
\eea
where parton 2 from the target is now drawn entering from the left, to emphasize that the spectrum, although derived in the target rest frame, is actually frame independent and can be used \eg\ in the c.m.~frame of the collision. It is also understood that in \eq{eq:final-spec2to2}, the soft radiated gluon line carries only color indices, all the Lorentz (kinematical, spin) dependence being confined to the amplitude ${\cal M}$. 
Setting 
$m_1 \to 0$ and $m_3=m_4=m$ 
in \eq{eq:final-spec2to2} 
leads to the initial expression of the spectrum used in our study, eqs.~\eq{master-spectrum}--\eq{TuTt}. 

\begin{figure}[t]
\begin{center}
\includegraphics[scale=0.9]{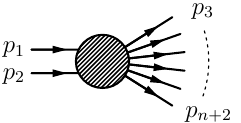}  
\end{center}
\vskip -5mm
\caption{
  \label{fig:1tom}
  Labelling convention for $2 \to \np$ processes. 
  }
\end{figure} 

Finally, the generalization of \eq{eq:final-spec2to2} to any $2 \to \np$ process with $\np > 2$ (as depicted in figure~\ref{fig:1tom}) is straightforward, 
\be
\label{eq:spec1tom}
x \frac{\dd I}{\dd x} \bigg|_{2\to \np} =
\frac{\alpha_s}{\pi} \, \sum_{i=3}^{\np+2} \ 
\sigma\left( \frac{x p_{i\perp}}{\xi_i \mu}, \frac{x m_1}{\mu}, \frac{x m_i}{\xi_i \mu}, r \right)  \ \T_i  \ , 
\ee 
where $\xi_i \equiv p_i^+/p_1^+$ and $\bm p_i$ are the longitudinal momentum fraction and transverse momentum of parton $i$ in the multi-parton final state (with $\sum_{i=3}^{\np+2} \xi_i = 1$ and $\sum_{i=3}^{\np+2} \bm p_i = \bm 0$), and $\T_i$ is a graph analogous to the graphs appearing in \eq{eq:final-specqtoqg} (multiplied by $2/\vert {\cal M} \big\vert^2$, with ${\cal M}$ the $2 \to \np$ amplitude), with the line of the soft radiated gluon attached to parton $i$ in the conjugate amplitude. 
Equation~\eq{eq:spec1tom} shows how the 
Lorentz structure 
of the soft radiation can be `factorized' from the hard scattering, provided that the hard scattering amplitude (embedded in $\Theta_i$ or $\Phi$) 
is known. 
We stress, however, that evaluating $\Theta_i\,$ is itself a non-trivial task in general.

In this respect, let us mention that for $2 \to \np$ processes with $\np \geq 2$, the Lorentz dependence does not fully cancel between the numerator of $\T_i$ and its denominator $|{\cal M}|^2$. This is because already for $2 \to 2$ processes, \eq{qtoqg:vac} contains several terms [in contrast with \eq{2to1:vac}], which have different Lorentz dependences. For $2 \to 2$ processes, the Lorentz dependence can be fully encompassed by one scale (\eg\ $p_{1\perp}$) and one dimensionless parameter ($\xi$), and the scale dependence does cancel between each graph of \eq{eq:final-specqtoqg} and the denominator $|{\cal M}|^2$. This is why in the present study, the color density matrix $\Phi$ defined by \eq{Phi-def} may only depend on the kinematical variable $\xi$ (see appendix~\ref{app:Phi-matrix}). However, for $2 \to \np$ processes with $\np \geq 3$, the partonic final state depends on at least two independent transverse momentum scales, and we expect the generalization of the $\Phi$-matrices to depend not only on the $\xi_i$'s, but also on ratios like $p_{2\perp}/p_{1\perp}$, for instance. 

\section{Color matrices $\B_{\pm}$} 
\label{app:BBbar}

Since $\B$ and $\Bbar$ are pure color matrices, the calculation of $\B_\pm \equiv \B \pm \Bbar$ only requires specifying the type of process under consideration, namely, $q q \to q q$, $q \bar{q} \to q \bar{q}$, $q g \to q g$, $gg  \to gg$ or $g g \to q \bar{q}$, irrespectively of quark flavors. In addition, $\B$ and $\Bbar$ are symmetric, real matrices, and the expressions of $\B_{\pm}$ for $\bar{q} \bar{q} \to \bar{q} \bar{q}$, $\bar{q} g \to \bar{q} g$, $q \bar{q} \to gg$, are easily shown to be the same as for ${q q \to q q}$, $q g \to q g$, $g g \to q \bar{q}$, respectively.\footnote{More precisely: $\B_{\al\beta}^{qq \to qq} = \B_{\bar{\al} \bar{\beta}}^{\bar{q}\bar{q} \to \bar{q}\bar{q}}$ (with $\al, \beta \in \{\bm{\overline{3}}, \bm{6} \}$ and $\bar{\al}, \bar{\beta} \in \{ \bm{3}, \bm{\overline{6}} \}$); $\B_{\al\beta}^{qg \to qg} = \B_{\bar{\al} \bar{\beta}}^{\bar{q}g \to \bar{q}g}$ (with $\al, \beta \in \{\bm{3}, \bm{\overline{6}}, \bm{15} \}$ and $\bar{\al}, \bar{\beta} \in \{ \bm{\overline{3}}, \bm{6}, \bm{\overline{15}} \}$);  $\B_{\al\beta}^{gg \to q\bar{q}} = \B_{\al \beta}^{q\bar{q} \to gg}$ (with $\al, \beta \in \{\bm{1}, \bm{8}_{\rm a}, \bm{8}_{\rm s}\}$).} 

We first recall that $\B_+$ is a diagonal matrix for all channels, see eq.~\eq{Bplus}, where $C_\al$ may be copied from table~\ref{tab:irreps}. For convenience, we explicitly list them here:
\bea
\label{BpBbar matrix: qq} 
\B^{q q \to q q}_{+} &=&
\diag \big( \,
2 \CF - \tfrac{\Nc+1}{\Nc} \, ,\ 
2 \CF + \tfrac{\Nc-1}{\Nc} \,
\big) \, , \\[1mm]
\label{BpBbar matrix: qqbar} 
\B^{q \bar q \to q \bar q}_{+} &=&
\diag \big( \,
0 \, ,\ \Nc\,
\big) \, , \\[1mm]
\label{BpBbar matrix: qg} 
\B^{q g \to q g}_{+} &=&
\diag \big( \,
2\CF - \Nc \, ,\ 2\CF - 1 \, ,\ 2 \CF + 1
\big) \, , \\[1mm]
\label{BpBbar matrix: gg} 
\B^{g g \to g g}_{+} &=&
\diag \big( \,
\Nc \, ,\ 2\Nc \, ,\ 0\, , \ \Nc \, , \ 2(\Nc+1) \, , \ 2(\Nc-1) \,
\big) \, , \\[1mm]
\label{BpBbar matrix: ggqqbar} 
\B^{g g \to q \bar q}_{+} &=&
\diag \big( \, 0 \, ,\ \Nc \, ,\ \Nc \,  \big) \, ,  
\eea
with the ordering of irreps according to table~\ref{tab:channels}.

Note that the process $g q \to g q$, corresponding in our notational convention to an  incoming gluon from the projectile proton, is associated to a matrix $\B_{+}$ which is different from the process $q g \to q g$ (where the incoming gluon is from the target nucleus), because the incoming partons have different Casimir charges. We thus complement the above list with the matrix $\B_{+}$ for this process: 
\be
\label{BpBbar matrix: gq} 
\B^{g q \to g q}_{+} = \diag \big( \, \Nc \, ,\ 2\Nc - 1 \, ,\ 2 \Nc + 1 \big) \, . 
\ee
 
As for $\B_{-}$, its expression following from \eq{eq:B}--\eq{eq:Bbar}, for each $2\to 2$ partonic process, can be obtained using the birdtrack pictorial technique~\cite{Cvitanovic:2008zz,Dokshitzer:1995fv,Keppeler:2017kwt,Peigne:2023iwm}, or alternatively using {\sc form}~\cite{Kuipers:2012rf}. 
(Note that the birdtracks \eq{eq:B}--\eq{eq:Bbar} also appear in ref.~\cite{Cougoulic:2017ust}, from which the expression of $\B_{-}$ can be borrowed for most partonic channels.) 
The matrices $\B_{-}$ are listed in section~\ref{app:Bminus}, and we recall the relation of $\B_{-}$ to the soft anomalous dimension matrix in section~\ref{app:BBbar-softADM}. 

\subsection{$\B_{-}$ for all partonic channels}
\label{app:Bminus}

We list below the matrices $\B_{-}$ for each process (recall that $U_k \equiv \sqrt{\Nc +k}\,$, $D_k \equiv \sqrt{\Nc -k}$ and $K_A \equiv \Nc^2 -1$), with an ordering in $\al$ that follows the layout of table~\ref{tab:channels}.
\bea
\label{BmBbar matrix: qq} 
\B^{q q \to q q}_{-} 
&=&  \left( 
\begin{matrix}   0  & \displaystyle - \sqrt{K_A}\  \\ \displaystyle - \sqrt{K_A} \ &  0 \end{matrix}  
\right)  \, , \\[2mm] 
\label{BmBbar matrix: qqbar}
\B^{q \bar{q} \to q\bar{q}}_{-} 
&=& -\,\frac{1}{\Nc} \left( 
\begin{matrix}   0  & \displaystyle 2 \sqrt{K_A}\  \\ \displaystyle 2 \sqrt{K_A} \ & \displaystyle  \Nc^2-4 \end{matrix}  
\right)  \, ,  \\[2mm]
\label{BmBbar matrix: qg}
\B^{q g \to qg}_{-} 
&=& \left( 
\begin{matrix}   
  - \frac{\Nc^2+1}{\Nc K_A} \quad
  & 
  -\frac{\sqrt{2}  \Nc U_1 D_2}{K_A} \quad
  & 
   -\frac{\sqrt{2} \Nc D_1 U_2}{K_A}
  \\[2mm]
   -\frac{\sqrt{2} \Nc U_1 D_2}{K_A} \ 
  &  \frac{2\Nc -1}{\Nc(\Nc-1)}
  &  -\frac{\Nc U_2 D_2}{U_1 D_1} \\[2mm]
   -\frac{\sqrt{2} \Nc D_1 U_2}{K_A}
  &  -\frac{\Nc U_2 D_2}{U_1 D_1}
  &  \frac{2\Nc +1}{\Nc(\Nc+1)}
\end{matrix}  
\right)  \, , 
\hskip 1cm 
\eea 
\bea 
\label{BmBbar matrix: gg}
\B^{g g \to gg}_{-} 
&=& -2 \left( 
\begin{matrix}   
  0 \quad
  & 
   0 \quad
  & 
  \frac{2 \Nc}{U_1 D_1} \quad
  &
  \frac{\Nc}{2} \quad
  &
  \frac{U_3 }{U_1}  \quad
  &
    \frac{D_3 }{D_1} \quad
  \\[1mm]
   0 \quad
  &
   0 \quad
  &
  0 \quad
  &
  \frac{\sqrt{2} \Nc}{U_2 D_2} \quad
  &
  \frac{U_1 D_2 U_3}{U_2 \sqrt{2} } \quad
  &
  \frac{ D_1 U_2 D_3}{D_2 \sqrt{2} }
    \\[1mm]
  \frac{2 \Nc}{U_1 D_1} \quad
  &
  0 \quad
  &
  0 \quad
  &
   0 \quad
  &
   0 \quad
  &
   0 \quad
  \\[1mm]
   \frac{\Nc}{2}
  & 
    \frac{\sqrt{2} \Nc}{U_2 D_2} \quad
  &
   0 \quad
  &
   0 \quad
  &
   0 \quad
  &
   0 \quad
  \\[1mm]
   \frac{U_3 }{U_1} \quad
  & 
   \frac{U_1 D_2 U_3}{U_2 \sqrt{2} } \quad
  &
   0 \quad
  &
   0 \quad
  &
    0 \quad
  &
   0 \quad
  \\[1mm]
   \frac{D_3}{D_1}
  &
   \frac{D_1 U_2 D_3}{D_2 \sqrt{2} } \quad
  &
   0 \quad
  &
   0 \quad
  &
   0 \quad
  &
    0 \quad
\end{matrix}  
\right)  \, , \\[4mm]
\label{BmBbar matrix: ggqqbar}
\B^{g g \to q\bar{q}}_{-} 
&=& \left( 
\begin{matrix}   
  \displaystyle 0
  & 
  \displaystyle  -2 \sqrt{2} 
  & 
  \displaystyle 0
  \\[2mm]
  \displaystyle   -2 \sqrt{2} \quad
  & \displaystyle 0
  & \displaystyle - \sqrt{\Nc^2-4}
  \\[2mm]
  \displaystyle   0 
  & \displaystyle - \sqrt{\Nc^2-4} \quad
  & \displaystyle 0
\end{matrix}  
\right)  \, . 
\eea
Finally, as far as the process $g q \to g q$ is concerned, by examining the birdtracks involved one easily finds that its matrix $\B \equiv \frac{1}{2} (\B_{+} +  \B_{-})$ is the same as for the process $q g \to q g$. Thus, its matrix $\B_{-}$ directly follows from \eq{BpBbar matrix: qg}, \eq{BmBbar matrix: qg} and \eq{BpBbar matrix: gq}: 
\be
\B^{g q \to g q}_{-} = 
\B^{q g \to q g}_{+}  
+ \B^{q g \to q g}_{-}  - \B^{g q \to g q}_{+}   \, . 
\ee

\subsection{Correspondence of $\B_{-}$ with the soft anomalous dimension matrix} 
\label{app:BBbar-softADM}

For a given $2\to 2$ process, the matrix $\B_{-}$ is directly related to the soft anomalous dimension matrix~\cite{Botts:1989kf,Sotiropoulos:1993rd,Contopanagos:1996nh,Kidonakis:1998nf, Oderda:1999kr,Bonciani:2003nt,Appleby:2003hp,Banfi:2004yd,Kyrieleis:2005dt,Dokshitzer:2005ek,Dokshitzer:2005ig,Sjodahl:2008fz,Forshaw:2008cq}, which reads~\cite{Dokshitzer:2005ig}
\be \label{Q-def}
\ADM  = \frac{1}{2\Nc} \left[\, T_t^2 + T_u^2  + b\, (T_t^2-T_u^2)  \, \right]  \, ,
\ee
where $T_t \equiv T_1-T_3$ and $T_u \equiv T_1-T_4$ are the color exchanges in the $t$-channel and $u$-channel, respectively, and $b$ some ratio of logarithms~\cite{Dokshitzer:2005ig}. Indeed, using 
\bea
T_t^2 &=& T_1^2 + T_3^2 - 2 T_1 T_3 = C_1 + C_3 - \Bbar \ , \\ 
T_u^2 &=& T_1^2 + T_4^2 - 2 T_1 T_4 = C_1 + C_4 - \B \ , 
\eea
and the relation \eq{Bplus}, we obtain (in the orthonormal $s$-basis $\ket{\al}$) 
\be \label{Q-expr}
\ADM_{\al\beta} \equiv \bra{\al} \ADM \ket{\beta} = \frac{1}{2\Nc} \left\{ b \left( \B_{-} \, \right)_{\al\beta} + \delta_{\al\beta} \left[ C_1+C_2+(1+b)C_3+(1-b)C_4 -C_\al \right] 
 \right\} \, .
\ee
Thus, using the expressions of $\B_{-}$ listed in section~\ref{app:Bminus}, one can recover the soft anomalous dimension matrix for any $2 \to 2$ partonic process. (For instance, for the $gg \to gg$ process, using $C_i=\Nc$ for $i=1 \ldots 4$ and $\B_{-}$ given by \eq{BmBbar matrix: gg}, the expression \eq{Q-expr} coincides with the $\ADM$-matrix derived in ref.~\cite{Dokshitzer:2005ig} for this process.)
Note that since $\B_{-}$ is a symmetric matrix, the matrix ${\cal Q}$ defined by \eq{Q-expr} is also symmetric, as must be the case for anomalous dimension matrices when expressed in an orthonormal basis~\cite{Seymour:2005ze,Seymour:2008xr}.

\bibliography{mybib}
\bibliographystyle{JHEP}

\end{document}